\documentclass{elsart} 
\usepackage{ifpdf,graphicx, amsfonts, amsopn, amsbsy, amssymb, natbib}
\journal{Physica D}

\renewcommand{\cite}{\citep}

\newcommand{\beq}{\begin{equation}}
\newcommand{\eeq}{\end{equation}}
\newcommand{\beqa}{\begin{eqnarray}}
\newcommand{\eeqa}{\end{eqnarray}}

\renewcommand{\vec}[1]{{\bf{#1}}}
\renewcommand{\Re}{{\mathfrak{Re}}}

\newcommand{\ave}[1]{\left\langle #1\right\rangle}

\newcommand{\mean}[1]{\overline{#1}}

\newcommand{\sgn}{\mathrm{sgn}}
\newcommand{\tr}{\operatorname{tr}}
\newcommand{\dd}{\mathrm{d}}
\newcommand{\ii}{\mathrm{i}}

\newcommand{\mod}{\operatorname{mod}}

\newcommand{\diag}{\operatorname{diag}}

\newcommand{\const}{\operatorname{const}}
\newcommand{\id}{\operatorname{id}}

\newcommand{\bR}{\mathbb{R}}
\newcommand{\bN}{\mathbb{N}}
\newcommand{\bQ}{\mathbb{Q}}
\newcommand{\bZ}{\mathbb{Z}}
\newcommand{\bT}{\mathbb{T}}
\newcommand{\bC}{\mathbb{C}}

\newtheorem{theorem}{Theorem}%[section]

\newtheorem{corollary}[theorem]{Corollary}
\newtheorem{lemma}[theorem]{Lemma}

\newtheorem{conjecture}[theorem]{Conjecture}

\newcommand{\cA}{{\cal A}}
\newcommand{\cB}{{\cal B}}

\newcommand{\cD}{{\cal D}}

\newcommand{\cM}{{\cal M}}

\newcommand{\cP}{{\cal P}}
\newcommand{\cQ}{{\cal Q}}

\newcommand{\La}{\Lambda}
\newcommand{\om}{\omega}

\makeatletter
\renewenvironment{figure}[1][]%
    {%
        \@float{figure}%
        \vspace{3mm}\centering%
  }{%
    \end@float
  }
\makeatother

\begin{document}

% ============================================================================
%   Title
% ============================================================================

\begin{frontmatter}

\title{On ergodic and mixing properties of the triangle map}

\author[label1,label2]{Martin Horvat\corauthref{cor1}},
\ead{martin.horvat@fmf.uni-lj.si}
\ead[url]{http://chaos.fmf.uni-lj.si}
\author[label2]{Mirko Degli Esposti},
\ead{desposti@dm.unibo.it}
\author[label3]{Stefano Isola},
\ead{stefano.isola@unicam.it}
\author[label1]{Toma\v z Prosen}\and
\ead{tomaz.prosen@fmf.uni-lj.si}
\author[label4]{Leonid Bunimovich}
\ead{bunimovh@math.gatech.edu}

\corauth[cor1]{corresponding author}

\address[label1]{Department of Physics, Faculty of Mathematics and Physics,
    University of Ljubljana, Slovenia}

\address[label2]{Dipartimento di Matematica, Universit\`a di Bologna, Italy}

\address[label3]{Dipartimento di Matematica e Informatica, Universit\`a di Camerino, Italy}

\address[label4]{School of Mathematics, Georgia Institute of Technology, Atlanta, USA}

\begin{abstract}
In this paper, we study in detail, both analytically and numerically, the dynamical properties of the triangle map, a piecewise parabolic automorphism of the two-dimensional torus, for different values of the two independent parameters defining the map. The dynamics is studied numerically by means of two different symbolic encoding schemes, both relying on the fact that it maps polygons to polygons: in the first scheme we consider dynamically generated partitions made out of suitable sets of disjoint polygons, in the second we consider the standard binary partition of the torus induced by the discontinuity set. These encoding schemes are studied in detail and shown to be compatible, although not equivalent. The ergodic properties of the triangle map are then investigated in terms of the Markov transition matrices associated to the above schemes and furthermore compared to the spectral properties of the Koopman operator in $L^2(\bT^2)$. Finally, a stochastic version of the triangle map is introduced and studied. A simple heuristic analysis of the latter yields the correct statistical and scaling behaviours of the correlation functions of the original map.

\end{abstract}


\end{frontmatter}
% ============================================================================
%   Main
% ============================================================================
%
\section{Introduction}
The research of conservative dynamical systems has a long and fruitful history. From the discovery of the exponential sensitivity on initial conditions by Poincar\'e the main part of research directed its attention to such {\em chaotic} situations. The Kolmogorov-Sinai entropy is one of the basic invariants of dynamical systems, which divides them into two classes: (i) the chaotic or hyperbolic systems, with positive dynamical entropy, and (ii) non-hyperbolic systems, with zero dynamical entropy. A particular subclass of non-hyperbolic systems has been discovered recently, exhibiting properties common mostly to chaotic systems such as diffusion, ergodicity and mixing. The study of such systems is important in order to understand the origin of stochastic-like-dynamics in non-hyperbolic dynamical systems - and in particular in non-equilibrium statistical mechanics - and the effects on their quantum counterparts. A physically relevant example in this class is the triangle map \cite{casati:prl:00}. As we will see below, this is a two parameter $(\alpha, \beta)$ family of two dimensional maps on the torus. They are related to the returning map of a point particle moving inside an elongated rectangular triangle billiard, namely for the correspondence to be accurate one of the angles of the triangular billiard has to be small.
Roughly speaking $\alpha$ is related to one angle in the triangle and $\beta$ is introduced in order to generalize the map and to make the dynamics richer. For the detailed construction of the map and its relation to the triangular billiards we refer to original articles \cite{casati:prl:99,casati:prl:00}. For a more complete description of the dynamics in polygonal billiards see e.g. \cite{masur:handbook:02, tabachnikov:book:95}. Some known generic properties of dynamics in polygonally shaped billiards are reviewed in \cite{gutkin:jsp:96}. The triangle map $\phi:\bT^2\to\bT^2$ is a one-to-one transformation (automorphism) of a two-dimensional torus $\bT^2 = [0,1)^2$ written in additive notation (mod 1) as
\beq
  \phi(q,p) = (q + p + \alpha \theta(q)+\beta,
               p + \alpha \theta(q) + \beta)\>,
  \label{eq:trm_map}
\eeq
where
$$
  \theta(q) = \left \{
    \begin{array}{lll}
    1 &:& q\in [0,{1\over 2}) \cr
    -1 &:& {\rm otherwise}
    \end{array}\right. \>
$$
with two parameters $(\alpha,\beta)\in\bR^2$. The Jacobian matrix of the map reads
$$
  J =\frac{\partial \phi(q,p)}{\partial (q,p)}
    = \left(
      \begin{array}{cc}
        1 - 2\alpha\{\delta(q)+\delta(q-\half)\} & 1 \cr
        -2\alpha\{\delta(q)+\delta(q-\half)\}    & 1
      \end{array}
      \right) \>
$$
so that $\det J = 1$ and $\tr J = 2[1- \alpha\{\delta(q)+\delta(q-\half)\}]$. The map is area-preserving and piecewise parabolic. However it is not continuous, the `discontinuity set' $\cD$ is the following codimension one manifold
$$
  \cD =\{\, (0,p)\, | \, p\in [0,1)\, \} \cup
       \{\,(\half,p)\, | \, p\in [0,1)\, \}
      =: v(0)\cup v(\half)\>.
$$
The action of the map $\phi$ can be decomposed into three simple transformations:
\begin{enumerate}
\item cut of the torus along two vertical lines $v(0)$ and $v({1\over 2})$ and translation of the two resulting pieces in opposite directions parallel to the cuts: $q\to q$ and $p\to p+\alpha \theta (q)$;
\item rigid translation by $\beta$ along the $p$-direction: $q\to q$ and $p\to p +\beta$;
\item parabolic skew translation: $q\to q+p$ and $p\to p$.
\end{enumerate}
The triangle map defines a time-discrete dynamical system $\phi^t$ for $t\in\bZ$ by a recursion relation
\beq
  {\bf x}_{t} = \phi({\bf x}_{t-1}) = \phi^t ({\bf x}_0)\>,\qquad
  {\bf x}_t = (q_t, p_t) \>.
  \label{eq:trm_dyn}
\eeq
where the $t$-th iteration of the map can be explicitly written as
\beqa
  q_t(q_0,p_0) &=& q_0 + p_0\, t +
          \frac{\beta}{2}\, t(t+1) +
          \alpha \sum_{k=1}^{t-1} S_k\>,\nonumber\\
  p_t(q_0,p_0) &=& p_0 + \beta\, t + \alpha\, S_t\>,
  \label{eq:trm_dyn_t}
\eeqa
where
\beq
  S_t:=\sum_{k=0}^{t-1}\theta (q_k)\quad
  \hbox{satisfies}\quad S_{t+1}=S_t \pm 1 \>.
\eeq
Note that $S_t\in \bZ$ and $|S_t| \le t$.

Using (\ref{eq:trm_dyn_t}) and the fact that the function $\theta(q)$ is locally constant, it is easy to see that the dynamics transforms a horizontal segment to a finite number of horizontal segments. On the other hand, for each $t$, the image of a given vertical line $v(q_0)=\{(q_0,p)\, , p\in (0,1]\}$ under the map $\phi^t$ is a family of segments having a slope $1/t$ and both the length and the position are determined by the pseudo-random series $\sum_{k=1}^{t-1}k\,(\beta+\alpha\,\theta (q_k))$.

 Away from the singular set $\cD$, the map $\phi$ acts locally as a linear stretching: a small parallelogram $P$, which is bounded by two horizontal sides and two non-horizontal sides forming an angle $\gamma$ with the horizontal axis, is mapped to another parallelogram $P'$. $P'$ has the same area as $P$ and is again bounded by two horizontal sides of unchanged length, whereas the other two sides are rotated clockwise forming  an angle $\gamma' = \gamma/(1+\gamma)$ with the horizontal axis, so that they get both stretched by a factor $\sin \gamma \cdot \sqrt{1+(1+\cot \gamma)^2}$.\\
Two nearby points $(q,p)$ and $(q+\Delta q,p+\Delta p)$ on the same
side of discontinuity are evolving so that
\beq
  \Delta(t) = \|\phi^t(q+\Delta q, p + \Delta p) -  \phi^t(q, p)\|
  =
  \sqrt{\Delta p^2 + (\Delta q + t \Delta p)^2}\>.
\eeq
Notice that the distance between points lying on a horizontal segment ($\Delta p = 0$) does not grow with time, $\Delta(t) = |\Delta q|$, whereas for points on a vertical segment ($\Delta q=0$) the distance grows as $\Delta(t)=|\Delta p|\sqrt{1+t^2}$. This implies that two arbitrary points on the same horizontal segment can be separated only by the cutting mechanism on the singular line. This implies that the points in a horizontal line are not separated by stretching, but rather by chance that they fall with time on opposite sides of discontinuity, which is guaranteed by the stochastic like behaviour which shall be discussed in the sequel.\par
The triangle map has been further explored and applied in investigations of fundamental properties of
statistical mechanics in several recent papers \cite{tsallis,jinghua,querios}.
In the following text we study,  mainly numerically, ergodic properties of the triangle map using two different ways to symbolically encode the dynamics, see e.g \cite{christiansen:non:96}, called the polygonal and the binary description. We establish some common properties of the two descriptions as well as some their specific features. In the frame of a given description we study a finite state Markov chain corresponding to the triangle map and its spectral gap. We also analyze certain interesting scaling relations of the spectrum of the Koopman operator in the truncated Fourier basis. Additionally we draw parallels between the triangle map and its stochastic version called random triangle map. The later possesses similar properties as the deterministic version of the triangle map, but enables analytical predictions thanks to the possibility of averaging over different realizations.\par
\section{Some further properties of the triangle map}

The ergodic properties of the triangle map system strongly depend on the arithmetic properties of the parameters $\alpha$ and $\beta$. In the following we list all the analyzed cases and present the results.

\subsection{The case $\alpha =0$}

If $\alpha =0$ the map reduces to the skew translation $\phi(q,p) = (q+p+\beta, p+\beta)$. As is well known, for $\beta$ irrational the map is (uniquely) ergodic but not even weakly mixing (see reference e.g. \cite{katok:book:95} and below)

\subsection{The case $\alpha$ and $\beta$ rational}

If $\alpha$ and $\beta$ are rational numbers then the dynamics is {\em pseudo-integrable} \cite{berry:phys2D:81}. For example, if $\alpha = n/m$, and $\beta = r/m$, for some integers $n,m,r$ such that ${\rm gcd}(n,m,r)=1$, then the phase space $\bT^2$ is foliated into {\em invariant curves} $\{(q,p + k/m) | q \in [0,1), k \in \{0,1,\ldots, m-1\}\}$. It is easy to see that similar foliations exist also for arbitrary rational values of the parameters and the dynamics $\phi$ restricted on each set of invariant curves is just an {\em interval exchange transformation} (IET). See Lemma \ref{lemma:IET} below for a more precise statement.

\subsection{The generic case}

The case in which both parameters are non zero and irrational, i.e. $0\not=\alpha, \beta\in \bR\backslash\bQ $, with $\beta \notin \bZ +\alpha \bZ$, will be referred to as the {\it generic case}. We
have the following
\begin{lemma}\label{nopo}
In the generic case $\phi$ has no periodic orbits.
\end{lemma}
\noindent
{\sl Proof.} According to (\ref{eq:trm_dyn_t}), a necessary condition for $\phi$ to have a periodic point $(q_0,p_0)$ of period $t\geq 1$ is that $\beta\, t+\alpha\,S_t=0 \, (\mod \, 1)$. But this is clearly impossible under the assumption of the lemma.$\qed$\par
\begin{figure}[!htb]
\includegraphics[width=0.5\textwidth]{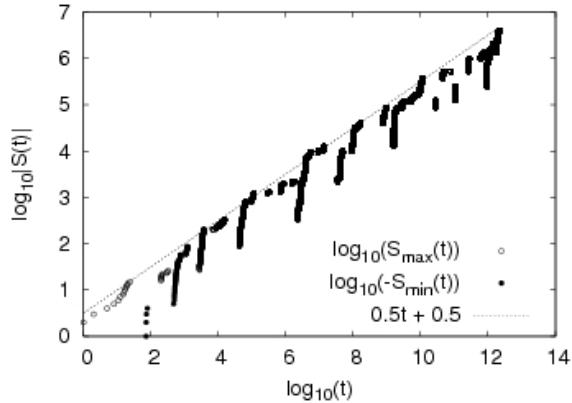}
\caption{The extreme values of the sum $S_t$ in the generic case $\alpha=1/e$ and $\beta=(\sqrt{5}-1)/2$ at $q_0=0.1$ and $p_0=0.2$.}
\label{fig:erg_gen}
\end{figure}

We are now tempted to argue that in the generic case every orbit $(q_t,p_t)_{t\geq 0}$ is uniformly distributed (u.d.) mod 1 in $\bT^2$. This means that for any lattice point $(r,s)\in\bZ^2\setminus \{0,0\}$ the sequence $(rq_t + s p_t)_{t\geq 0}$ is u.d. mod 1 in $\bT$. Note that if this is the case then, arguing for example using the Weyl criterion (see \cite{kuipers:book:74}, Chap. 1.2),  the system would be {\it strictly ergodic}, i.e. {\sl minimal} and {\sl uniquely ergodic}. To be more specific, consider a fixed observable $f(x,y)$ of the form $f(x,y)= e_{r,s}(x,y)$, where $e_{r,s}(q,p):=e^{2\,\pi\, i\,(\,r\,q+s\, p\,)}$. Assuming the uniform distribution of {\it all} orbits, together with the Weyl criterion, one gets for {\it any} $\phi$-invariant measure $\nu$:
$$
\nu (f)=\nu \left({1\over n} \sum_{t=0}^{n-1}f\circ \phi^t\right)={1\over n}
\sum_{t=0}^{n-1}e_{r,s}(q_t,p_t)\to 0\> \quad \textrm{when}\quad
n\to\infty \>,
$$
namely $\nu (e_{r,s})=0$ for all $(r,s)\in \bZ^2\setminus \{0,0\}$. For an arbitrary $f\in L^2(\bT^2)$ we write $f =\sum_{r,s} c_{\,r,s} \,e_{r,s}$ so that $\nu(f)=c_{0,0}=\mu(f)$. Since a complete proof of the uniform distribution of all orbits is still missing (see however below for a semi-rigorous argument) we can only state the following:
\begin{conjecture} \label{ergo}
In the generic case the system $(\bT^2, \phi, \mu)$ is strictly
ergodic.
\end{conjecture}
\noindent We now sketch an argument, half-rigorous, half-heuristic, to support the above conjecture.\\
The $\phi$-orbit $(q_t,p_t)_{t\geq 0}$ is u.d. mod 1 in $\bT^2$  if  for any lattice point $(r,s)\in \bZ^2\setminus \{0,0\}$ the sequence $rq_t + s p_t$ is u.d. mod 1 in $\bT$. From classical difference theorems of the theory of uniform distribution (see \cite{kuipers:book:74}, Chap. 1.3) we know that $rq_t + s p_t$ is u.d. mod 1 iff for every positive integer $h$ the sequence $u_{t}^{(h)}:=r(q_{t+h}-q_t)+s(p_{t+h}-p_t)$ has this property (see \cite{kuipers:book:74}, p.26). Note that
\begin{eqnarray*}
 u_{t}^{(h)}
       &=&\, r\left[ hp_0 +h\beta (t+1)+\alpha h\, S_{t+1} + h(h-1){\beta \over 2} +  \alpha \sum_{k=1}^{h-1}(h-k)\theta(q_{k+t})\right]\\
      &&   +\, s\left[ h \beta +\alpha  \sum_{k=0}^{h-1}\theta(q_{k+t}) \right]\>,\\
      &=&r\, h \,p_{t+1}  +  \alpha \left[ r {\tilde S}_h +s  S_h\,\right]\circ \phi^t + \beta \left[r \frac{ h(h-1)} 2 +s h\right] \>, 
\end{eqnarray*}
where
$$
  {\tilde S}_h:=\sum_{k=1}^{h-1}(h-k)\theta(q_{k}) \>.
$$
Whence, the time dependence of $u_{t}^{(h)}$ is that of $r\, h \,p_{t+1}$ plus a term which is bounded above by $(|\alpha| +|\beta|)(|r|h(h-1)/2 +|s| h)$. Therefore, since $p_t=p_0 +\beta t + \alpha S_t$ we have
$$
  \lim_{t \to \infty} \frac {  u_{t}^{(h)}} t = r\, h\, \left( \beta +
  \alpha \lim_{t \to \infty} \frac {S_t} t \right) \>.
$$
Now, by the ergodic theorem $S_t/ t$ converges a.e. to an integrable function $f^*(q,p)$ which is $\phi$-invariant and $\int_{\bT^2} f^*(q,p)\,\dd q\, \dd p =0$ and thus $u_{t}^{(h)}/t$  converges a.e. to $\gamma = rh(\beta +f^* \alpha)$. Up to now the argument is rigorous. To complete it heuristically, one could use Lemma
\ref{nopo} and the invertibility of $\phi$ to argue that $\phi$ cannot have eventually periodic orbits and hence  that $\gamma$ is irrational. Then, stepping backward deduce from this that $(u_{t}^{(h)})_{t\geq 0}$ is u.d. mod 1 in $\bT$ and thereby $(q_t,p_t)_{t\geq 0}$ is u.d. mod 1 in $\bT^2$.\\
We have numerically investigated this behaviour and the results are reported in Figure \ref{fig:erg_gen}, where, in order to simplify the representation, we just recorded the extreme points:
\beq
   S_{\rm max} (t) = \max_{\tau \in [0,t]} \{ S_\tau \}\>,
   \qquad
   S_{\rm min} (t) = \min_{\tau \in [0,t]} \{ S_\tau \}\>.
\eeq
It turns out that in the generic case {\sl for all} tested initial points  the sum $S_t$  yields the same behaviour as for the standard symmetric random walks:
\beq
  S_{\rm max}\>, -S_{\rm min} \sim t^{\half}\>,
\eeq
The above result is an indication of the stochastic like behaviour of this non hyperbolic dynamical system. We will soon discuss the case $\beta=0$, where a different asymptotic behaviour of the sums $S_t$ will emerge.\vspace{2mm}\\
We now turn to mixing properties, providing a strong  numerical evidence for the following
\begin{conjecture} \label{mix}
In the generic case the system $(\bT^2, \phi, \mu)$ is mixing.
\end{conjecture}

\begin{figure}[!htb]
  \includegraphics[width=0.49\textwidth]{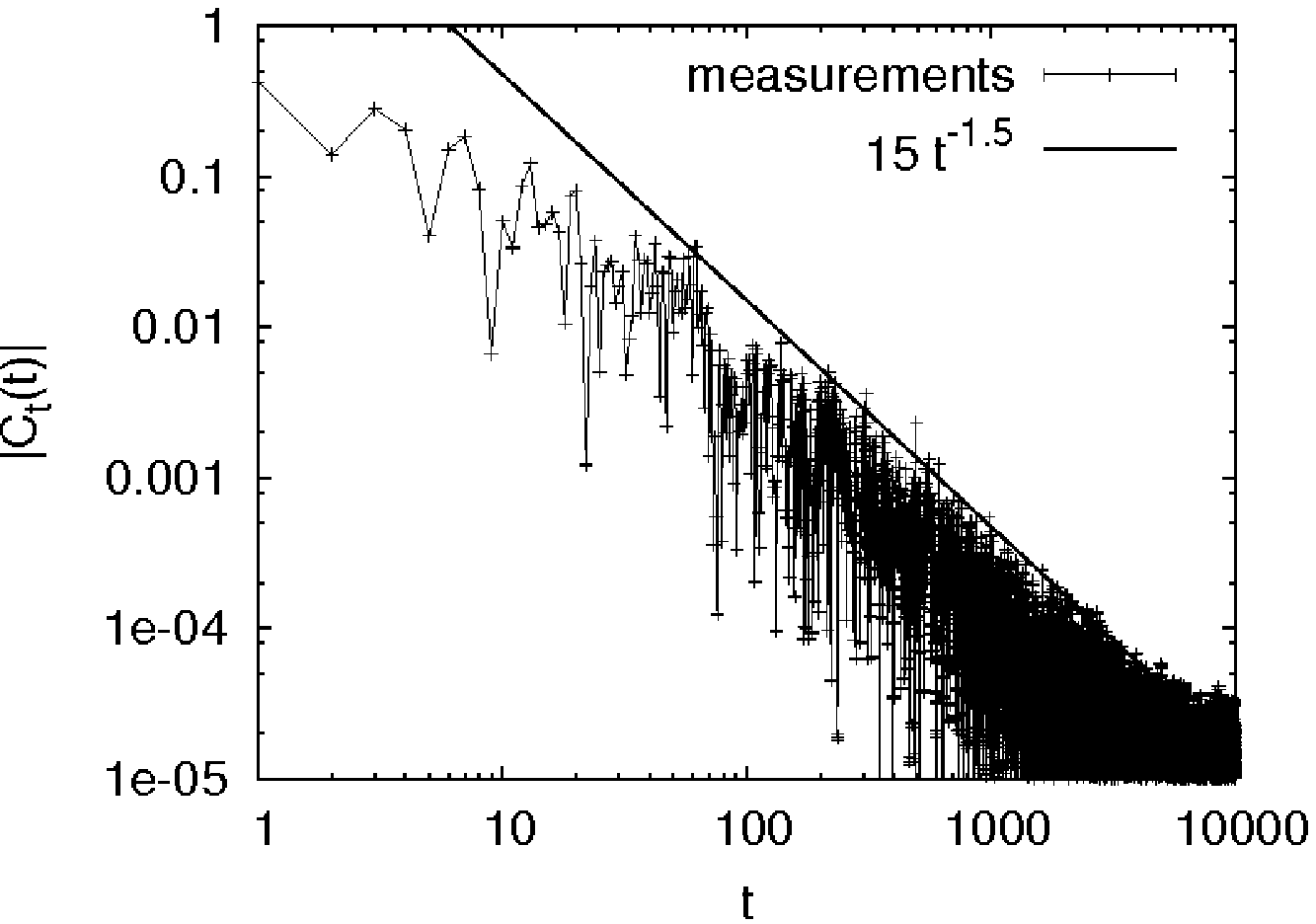}\hskip4pt%
  \includegraphics[width=0.49\textwidth]{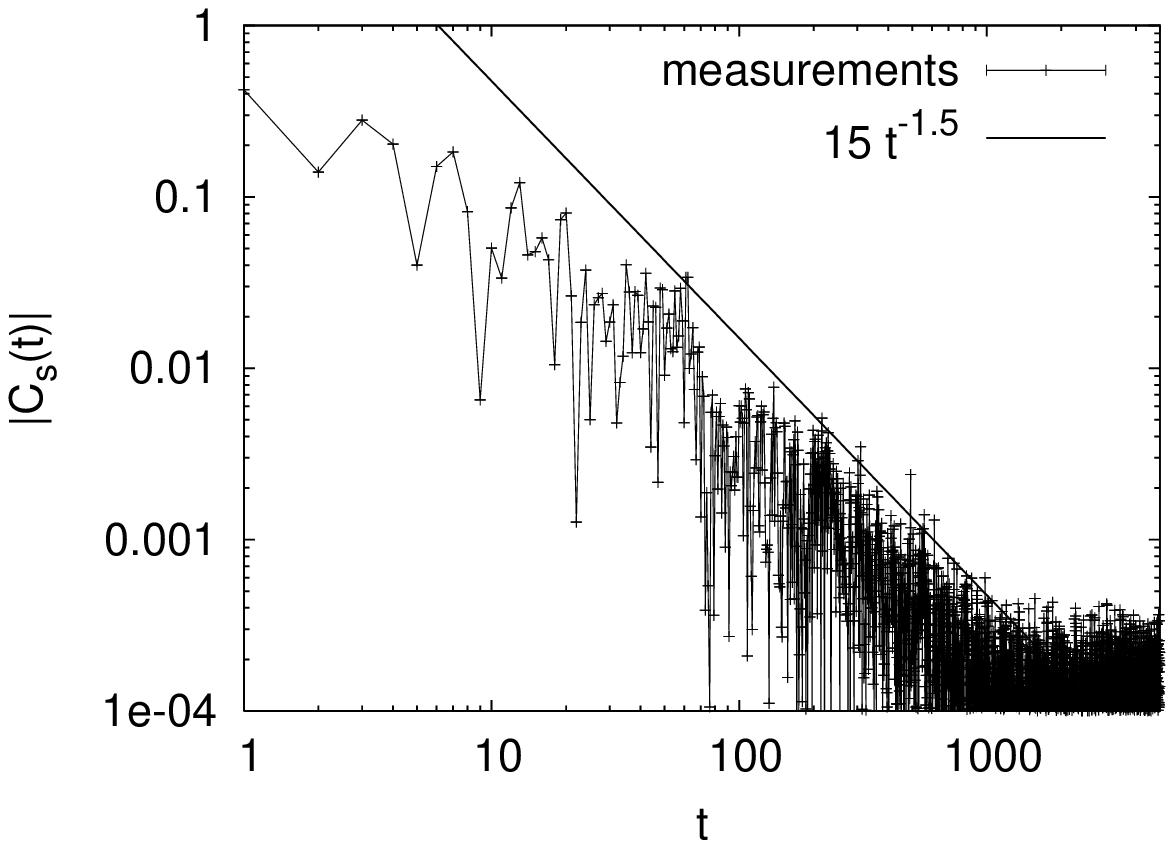}
  \hbox to \textwidth{\small\hfil(a)\hfil\hfil(b)\hfil}
\caption{The auto-correlations $C_{\rm t}(t)$ (a) and $C_{\rm s}(t)$ (b) using observable $f({\bf x})=\sin(2\pi q) + \sin(2\pi p)$ at $\alpha = 1/e$ and $\beta=(\sqrt{5}-1)/2$. In (a) we took $N=2^{15}$ and averaged the result over $m=4\cdot 10^6$ initial points and in (b) we use a Simpson integration scheme with $N_x=500, N_y\doteq 5\cdot 10^4$.}
\label{fig:C_gen_det}
\end{figure}

\noindent
In the  next experiments we numerically compute both the auto-correlation of a given observable $f(q,p)$  as a time average (for a fixed $x_0=(q_0,p_0)\in\bT^2$)
\beq
  C_{\rm t}(t)[f]
   = \lim_{N\to\infty}
     \frac{1}{N-t}\sum_{k=0}^{N-t} f(\phi^k(x_0)) f(\phi^{k+t}(x_0))\>,
  \label{eq:C_point}
\eeq
and the auto-correlation as a phase space average, i.e.
\beq
  C_{\rm s}(t)[f]
  =  \mu(f \circ \phi^t \cdot  f )\>,
  \label{eq:C_space}
\eeq
where we consider smooth observable $f$ with zero mean $\mu(f)= \int_{\bT^2} f(q,p)\, \dd q\, \dd p = 0$. Whenever the system is ergodic the expressions (\ref{eq:C_point}) and (\ref{eq:C_space}) are clearly equivalent, but since for some values of the parameters ergodicity is still questionable, we separately calculate and compare both quantities (see in particular the  next subsection). In agreement with both of the above conjectures in the {\it generic} case the numerical results in Figure \ref{fig:C_gen_det} show the same results  as  already reported in \cite{casati:prl:00}, i.e. an  algebraic decay of correlations with the law
\beq
  C_a(t) = O(t^{-\frac{3}{2}})\>,\qquad a= {\rm t}, {\rm s}\>.
\eeq
This indicates that $(\bT^2, \phi, \mu)$ is strongly mixing with polynomial rate of decay of correlations. In numerical implementation of $C_{\rm s}$ we use a Simpson integration scheme \cite{stoer:book:02} with additional extrapolation between results obtained on grids $N_x \times N_y$ and $2N_x \times 2N_y$. Because the observed scaling laws of the map, discussed in the next section, are directly related to the geometrical properties of the map, it is preferable to consider  $N_x \ge N_y$. Actually, in the following sections we will see that the optimal choice is $N_y/N_x\propto t$ and $N_x\propto t$, but this is usually not possible, because of the large memory consumption. The possible mechanism of mixing should be similar to that occurring for general polygonal billiards with irrational angles (see \cite{casati:prl:99}), and the model we are facing here clearly shows that not only mixing can appear even when hyperbolicity is absent, but furthermore it could emerge just by composing two simple piecewise parabolic linear maps.
\subsection{The case $\beta =0$} If $\beta =0$, the map becomes $\phi(q,p) = (q+ p+\alpha \,\theta (q), p+\alpha \, \theta (q))$, and for rational $\alpha =n/m$ we have the following result.
\begin{lemma}\label{lemma:IET}
Let $\beta =0$ and $\alpha =n/m$ with $n,m\in \bN$, ${\rm gcd}(n,m)=1$. For $p_0\in \bT$ set
$$
  \La (\alpha,p_0) := \bigcup_{k=0}^{m-1} h((p_0+k\alpha) \mod\, 1)\>
$$
with $h(p_0)=\{(q,p_0)\, , q\in \bT \}$ being a `horizontal foliation' of the phase space. Let moreover $\cQ(\alpha,p_0)=(Q_1,\dots, Q_r)$ be the partition of $\La(\alpha,p_0)$ into disjoint semi-intervals whose endpoints form the set $\La(\alpha,p_0)\cap \phi^{-1}\cD$. The map $\phi$ restricted to $\La(\alpha,p_0)$ acts as an interval exchange transformation on the elements of the partition $\cQ(\alpha,p_0)$. Furthermore, if $p_0\in \bQ$ then there exists an integer $k=k(\alpha, p_0)$ such that $\phi^k=\id$, otherwise $\phi$ acts as an aperiodic IET.
\end{lemma}
\noindent
{\sl Proof.} The proof is based on the following observation: setting
\beq
  O(\alpha, p_0)
  = \left( (p_0 + \alpha\, t)\, \mod\, 1 \right)_{t\in \bZ}\>,
\eeq
we have $|O(\alpha,p_0)|=m$ for all $p_0\in [0,1)$, so that the set
$O(\alpha,p_0) \times [0,1)$ is made of the $m$ `levels'
$h((p_0+k\alpha) \mod\, 1)$, $k=0,\dots, m-1$, i.e. coincides with
$\La (\alpha,p_0)$. Moreover, the set
\beq
  I(\alpha,q,p_0)
  = \left( \left(p_0+\alpha\, S_t(q,p) \right) \mod\, 1 \right)_{t\in \bZ^*}
\eeq
satisfies $I(\alpha,q,p_0)\subseteq O(\alpha,p_0)$.  The assertion can now be readily checked.$\qed$\vspace{2mm}\\
The case $\alpha$ irrational and $\beta=0$ represents a situation quite similar to the one observed in the free motion of a ball inside a triangular billiard \cite{casati:prl:99}, and furthermore the most delicate and less explored situation, from both the mathematical and the numerical point of view. We start with a simple remark concerning periodic orbits and then we turn to ergodicity and decay of correlations, where new numerical experiments presented here yield evidence of somewhat unexpected an behaviour.\par
\begin{figure}[!htb]
  \includegraphics[width=0.49\textwidth]{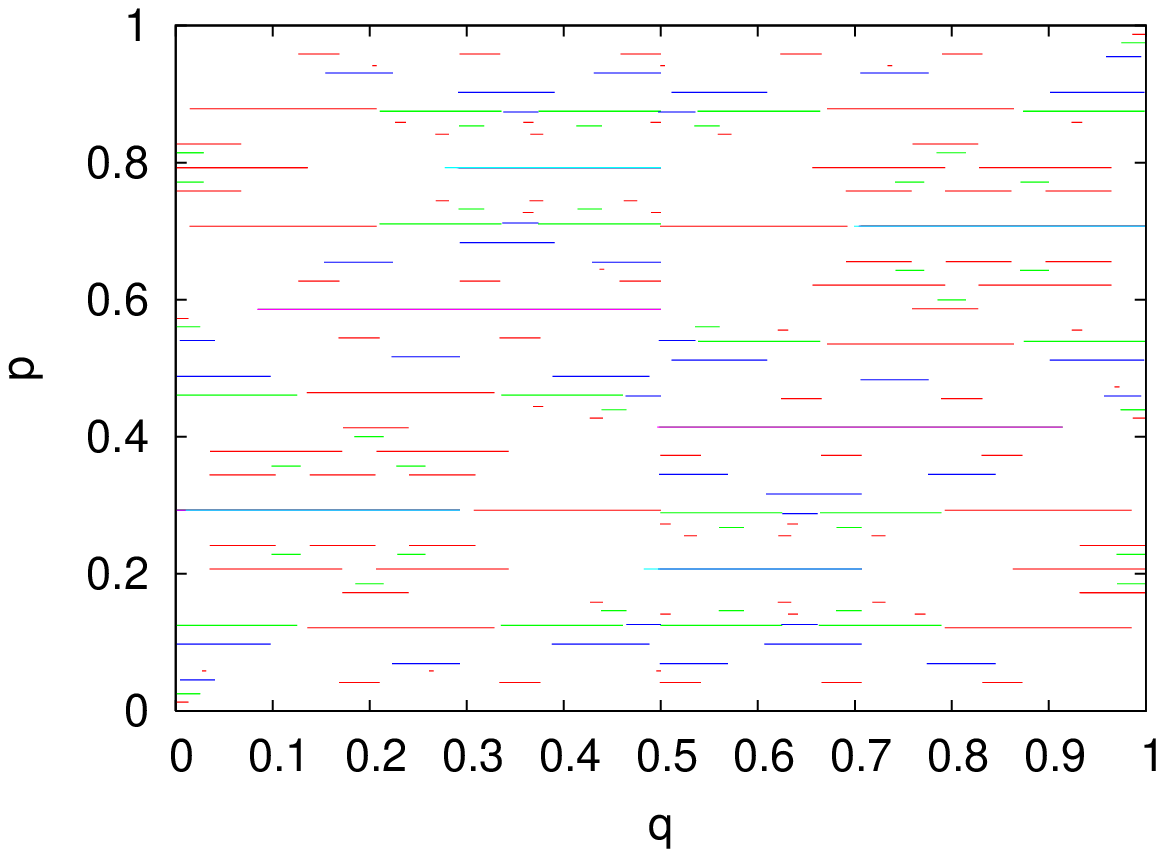}\hskip4pt%
  \includegraphics[width=0.49\textwidth]{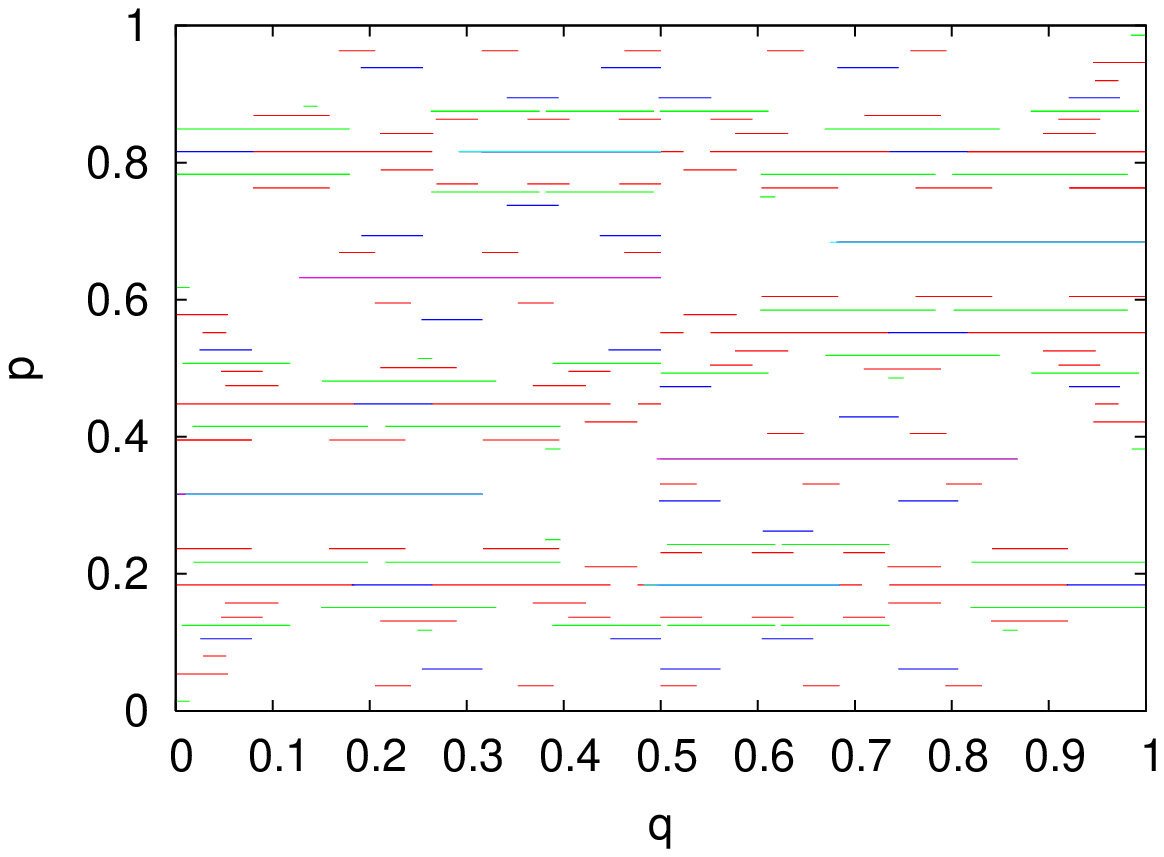}\\
  \includegraphics[width=0.8\textwidth,clip=]{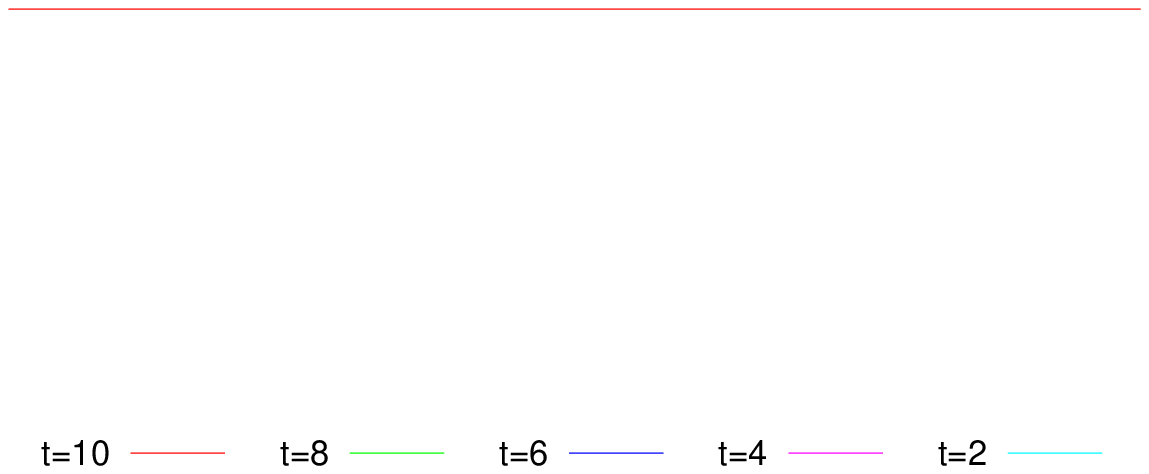}
  \hbox to\textwidth{\small\hfil(a)\hfil\hfil(b)\hfil}
\caption{The periodic orbits of triangle map at $\alpha=\sqrt{2}-1,e^{-1}$ (a,b) and $\beta=0$ for period lengths $t$ indicated in the figure.}
\label{fig:per_nongen}
\end{figure}
\begin{figure}[!htb]
  \includegraphics[width=0.49\textwidth]{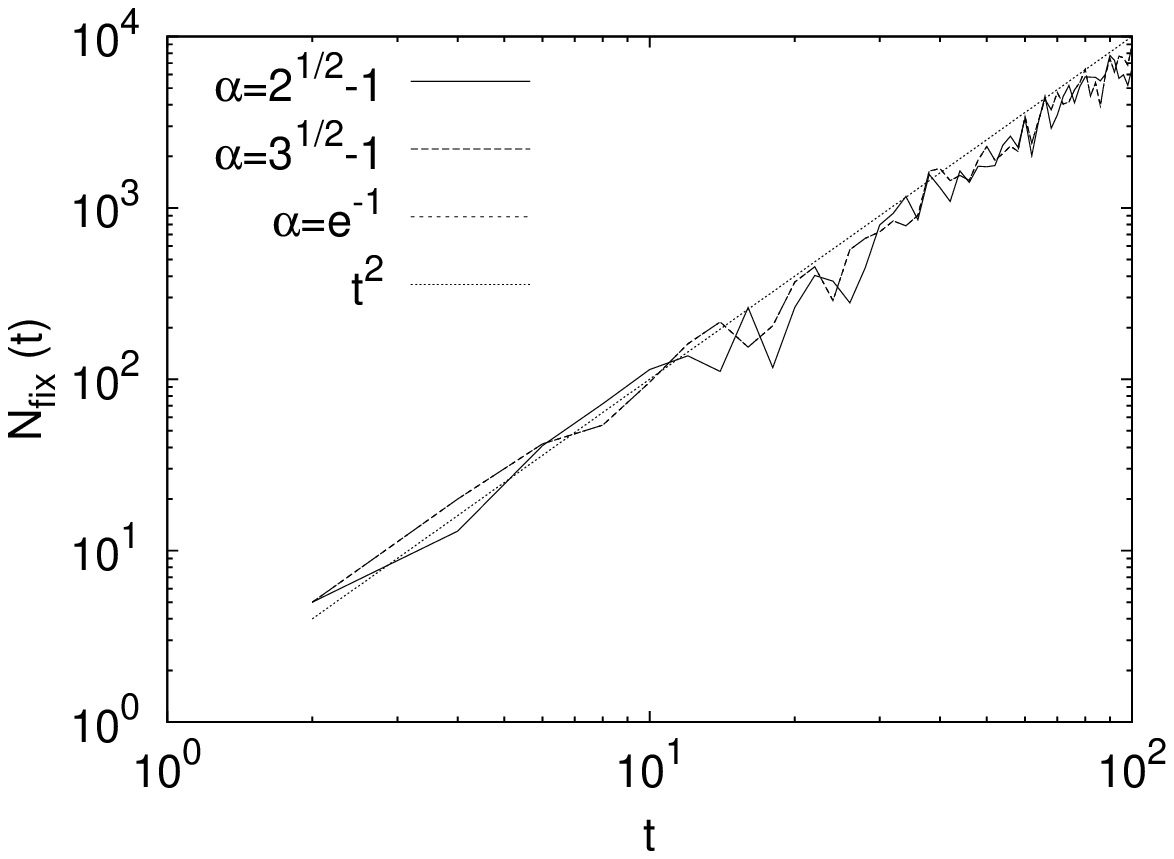}\hskip4pt%
  \includegraphics[width=0.49\textwidth]{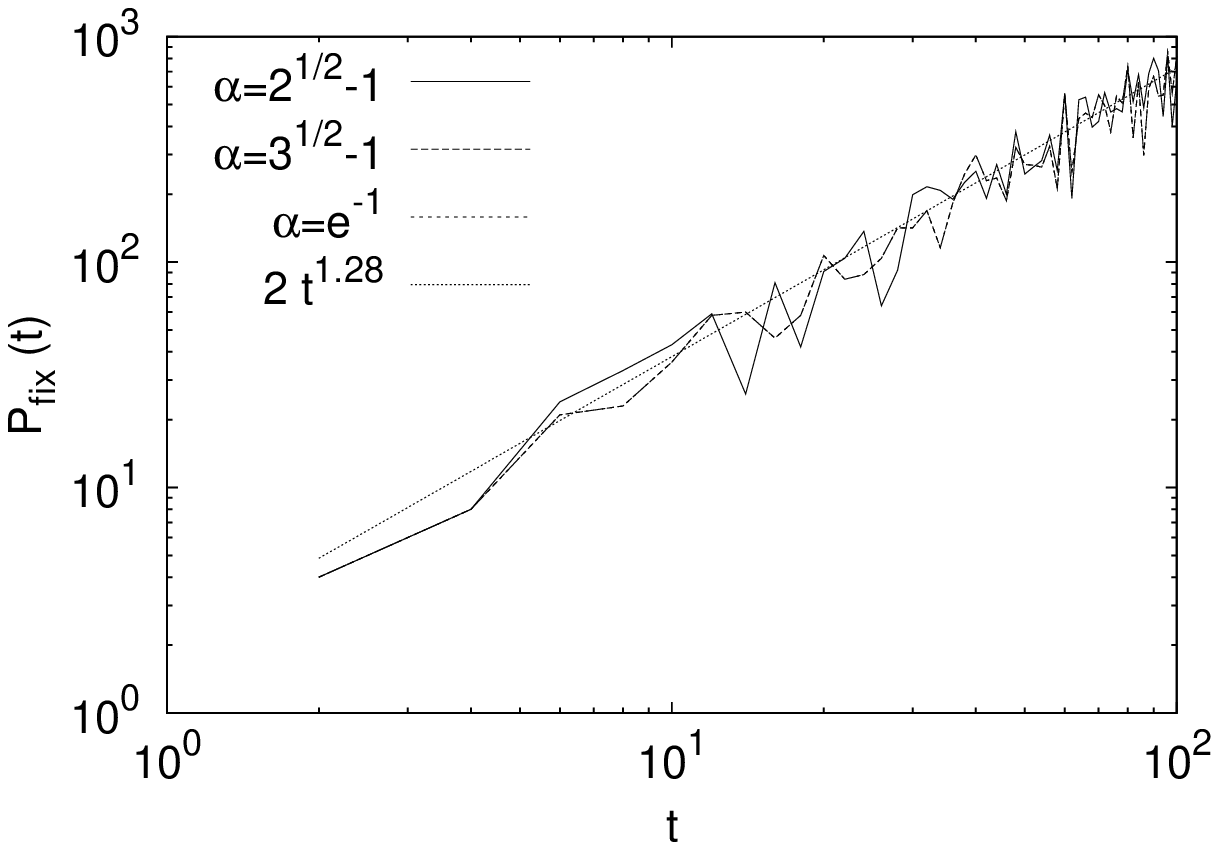}
  \hbox to \textwidth{\small\hfil(a)\hfil\hfil(b)\hfil}
\caption{The number of invariant intervals $N_{\rm fix}$ (a) and the number in invariant intervals visited momentum $N_{\rm fix}$ (b) as the function of time $t$ in non-generic case for different values of $\alpha$.}
\label{fig:trm_fix}
\end{figure}
\begin{figure}[!htb]
\includegraphics[width=0.5\textwidth]{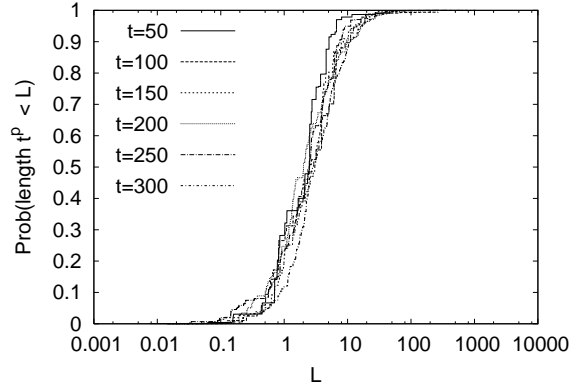}
\caption{The cumulative distribution of rescaled lengths $L=t^{1.28} l$ of invariant intervals calculated for $\alpha=1/e$ and at various times as denoted in the figure.}
\label{fig:trm_fix_len}
\end{figure}

We point out that for $\alpha$ irrational (and $\beta =0$) we
observe the existence of  periodic orbits. More precisely, according
to (\ref{eq:trm_dyn_t}), a point $(q_0,p_0)$ with either $p_0=0$ or
$p_0\in \bR\backslash\bQ$ is periodic of even period $t$ provided
the conditions $ p_0 t + \alpha \sum_{k=0}^{t-1}S_k =0$ and $S_t=0$
are both satisfied. The set of periodic orbits of length $t$ is
characterised by the invariant set $d_t(q,p)=0$, where
$(q_t,p_t)=\phi^t(q,p)$ and
\beq
  d_t(q, p) =
  \sqrt{(\min\{|q - q_t|,1-|q -q_t|\})^2 + (\min\{|p-p_t|,1-|p -p_t|\})^2}\>.
\eeq
Formula (\ref{eq:trm_dyn_t}) and the fact that the function $\theta(q)$ is locally constant immediately implies that periodic orbits of given length come in families forming horizontal segments. This structure and the statistics of segment lengths does not depends on $\alpha$, but the individual lentgh of the segments does. This is shown in Figure \ref{fig:per_nongen}, where we draw (primitive) periodic orbits for $\alpha=\sqrt{2}-1, e^{-1}$, up to period $t=10$. The intervals of periodic orbits are found first by randomly sampling the torus and then by searching for nearby zeros of $d_t(q, p)$.
The number of intervals of periodic orbits is growing  and their length is decreasing with the increasing period $t$. Finding compact intervals of primitive periodic orbits can be quite time demanding. Therefore we instead search for a related object i.e. invariant intervals  on $\phi^t$, which represent the set for all periodic points of length $t'$ such that $t'| t$.  From numerical results, shown in Figure \ref{fig:trm_fix}a,  we find that the number of invariant intervals $N_{\rm fix}$ increases with increasing $t$ as
\beq
  N_{\rm fix} \asymp t^2\>.
\eeq
In addition to that we check the number of different values of momentum $p$ in invariant intervals $P_{\rm fix}$. The results are presented in Figure \ref{fig:trm_fix}b and they fit well to a power law asymptotics
\beq
  P_{\rm fix} \asymp t^\zeta\>, \qquad \textrm{where}\quad \zeta = 1.28 \pm 0.02\>.
\eeq
The number $P_{\rm fix}$ increases asymptotically slower than $N_{\rm fix}$, which indicates that invariant lines at a given momentum $p$ are with increasing $t$ cut in more and more pieces. Their number is on average given by the ratio $N_{\rm fix}/P_{\rm fix}\sim t^{2-\zeta}$.
In addition we check the statistics of lengths $l$ of the invariant horizontal intervals at some $t$. We find that lengths are decreasing with increasing $t$ proportionally to $t^{-\zeta}$, with the same $\zeta$ as before. This behaviour is observed using the cumulative distribution of rescaled lengths $t^\zeta\,l$ which does not change with increasing $t$ , see Figure \ref{fig:trm_fix_len}. The observed scaling reflects the natural simmetry between the {\it diffusion} process in the momentum and the {\it cutting} process in $q$ induced by the discontinuity line.\par
\begin{figure}[!htb]
  \includegraphics[width=0.49\textwidth]{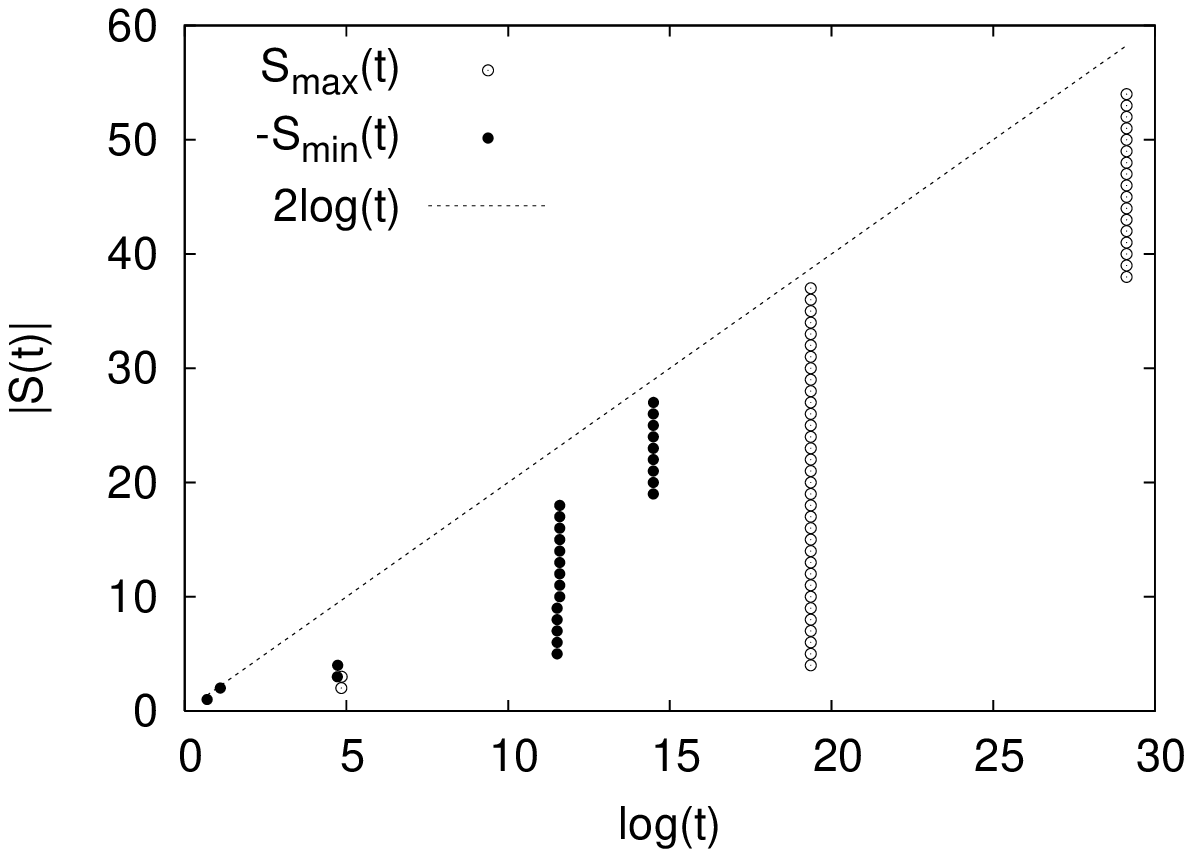}\hskip4pt%
  \includegraphics[width=0.49\textwidth]{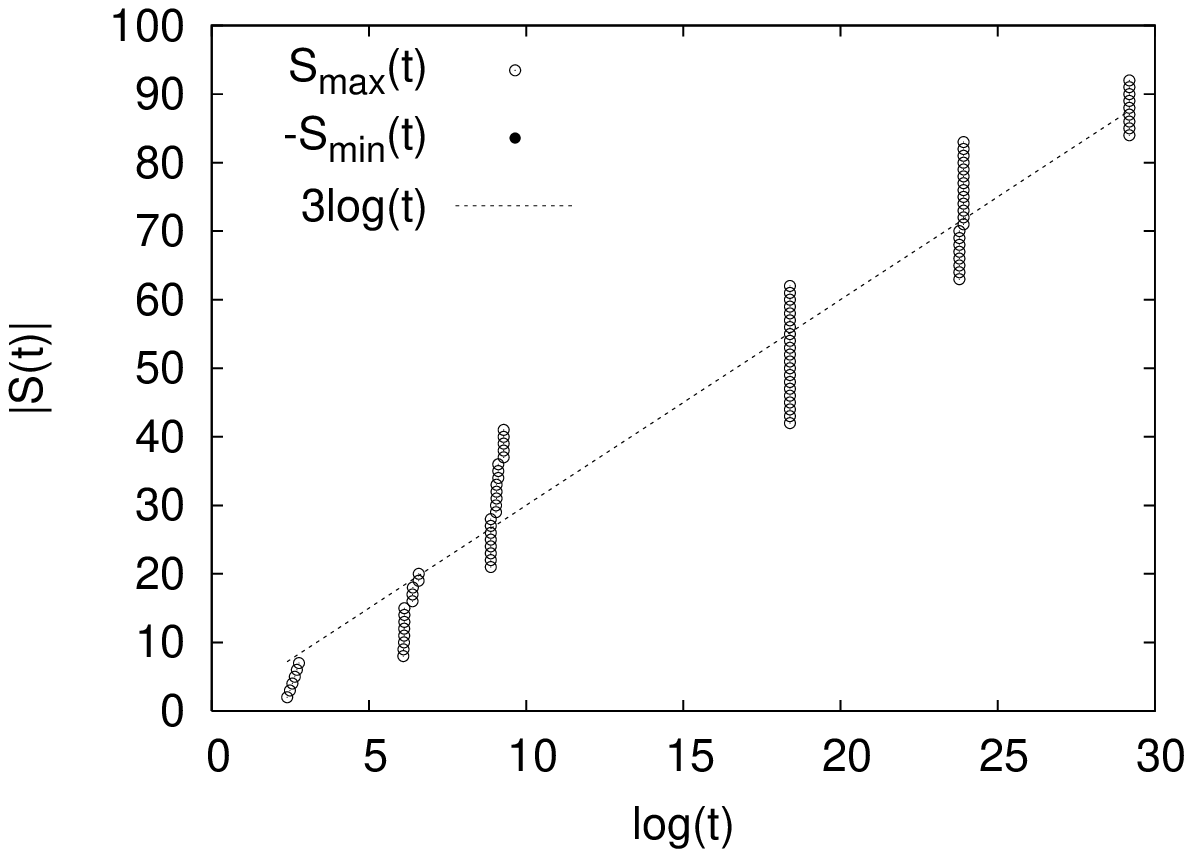}
  \hbox to  \textwidth{\small\hfil(a)\hfil\hfil(b)\hfil}
\caption{The points of extreme values $S_{\rm max}(t)$ and $S_{\rm min}(t)$ of the sum $S_t$ for $\alpha=\sqrt{2}-1, (\sqrt{5}-1)/2$ (a,b) at $q_0=0.1$ and $p_0=0.2$. }
\label{fig:erg_S}
\end{figure}

Regarding the rate of growth of the sums  $S_t$, we find interesting behaviours closely related to similar phenomena that one can find already in the case of irrational rotations of the circle \cite{isola:ijmms:06}.  In particular, we find evidence that for $\alpha$ being a quadratic irrational, the extreme values of the sum $S_t$ ($S_{\rm max}$ and $-S_{\rm min}$) increase logarithmically, see Figure \ref{fig:erg_S}. We note that to obtain these results we had to rely to high (more than double) precision floating point arithmetic \cite{demmel:siam:03}.
Moreover, similarly to the case of rotations discussed in \cite{isola:ijmms:06}, we expect a significant dependence of the growth behaviour of $S_t$ on the arithmetic type of $\alpha$. \par
\begin{figure}[!htb]
  \includegraphics[width=0.49\textwidth]{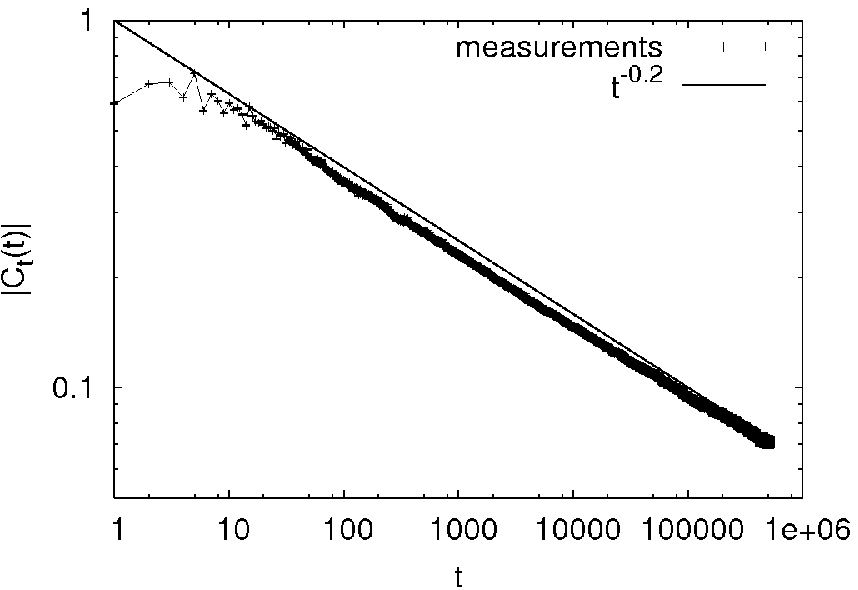}\hskip4pt%
  \includegraphics[width=0.49\textwidth]{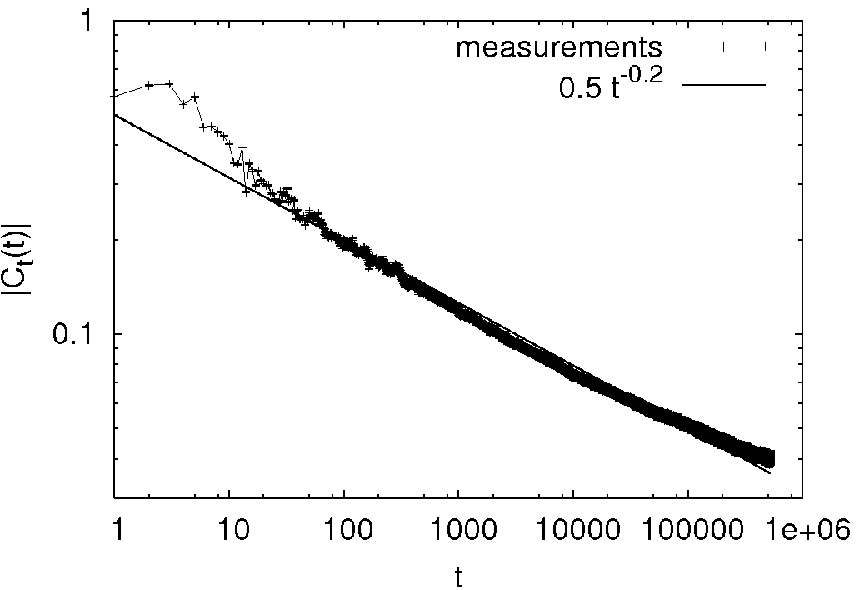}
  \hbox to \textwidth{\small\hfil(a)\hfil\hfil(b)\hfil}
\caption{The auto-correlation $C_{\rm t} (t)$ using observable $f(q,p)=\sin(2\pi q) + \sin(2\pi p)$ at $\beta=0$ and $\alpha=\sqrt{2}-1,e^{-1}$ (a,b), where we take $N=2^{19}$ and average the result over $m=10^4$ realisations.}
\label{fig:C_beta0_det}
\end{figure}
\begin{figure}[!htb]
  \includegraphics[width=0.49\textwidth]{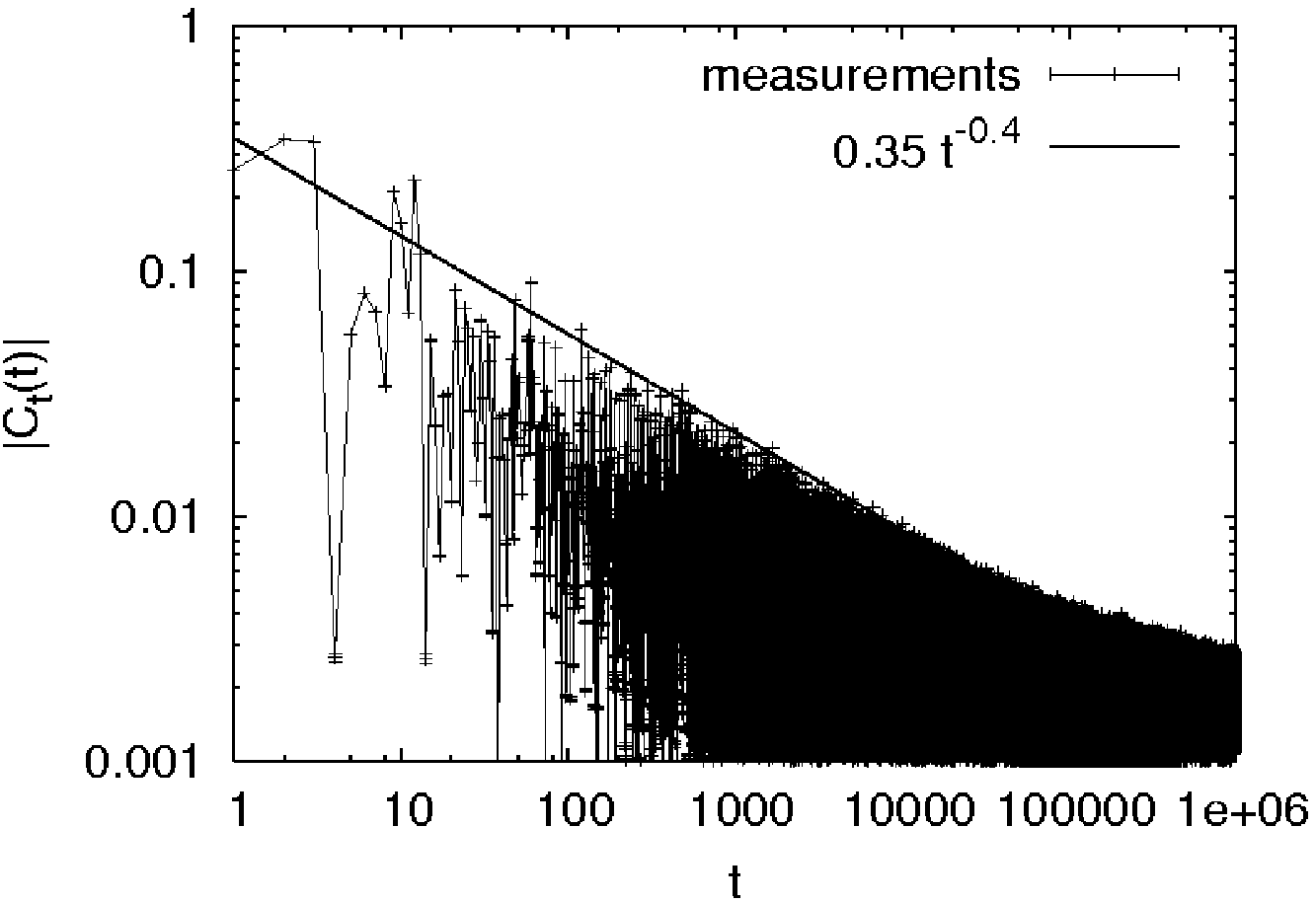}%
  \includegraphics[width=0.49\textwidth]{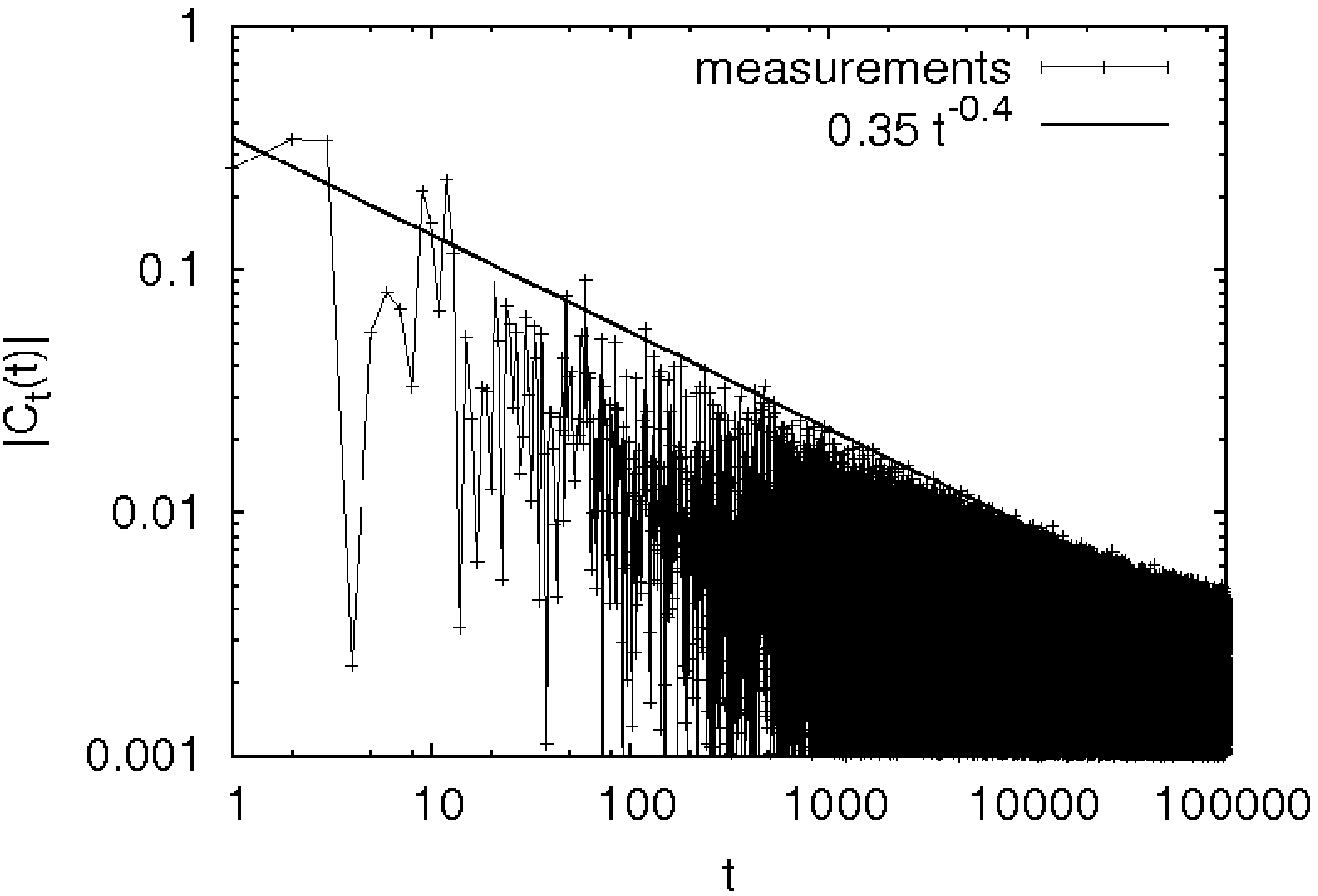}
    \hbox to\textwidth{\small\hfil(a)\hfil\hfil(b)\hfil}
\caption{The auto-correlations $C_{\rm t}(t)$ (a) and $C_{\rm s}(t)$ (b) of the observable $f({\bf x})=\sin(2\pi q) + \sin(2\pi p)$ at $\beta=0$ and $\alpha=\sqrt{3}-1$. In (a) we take $N=2^{20}$ and average the results over $m=4\cdot 10^4$ realisations, and in (b) we use the Simpson integration scheme with $N_x=200, N_y\doteq 6\cdot 10^5$}
\label{fig:C_beta0_anomaly_det}
\end{figure}

On the basis of preliminary numerical experiments performed in \cite{casati:prl:00}, it was conjectured that for $\alpha$ irrational and $\beta =0$ the system $(\bT^2, \phi, \mu)$ was just ergodic, and perhaps only weakly-mixing with no decay of correlations. Further  and more detailed numerical experiments reported here indicate that certain polynomial decay of correlations can instead be present. This behaviour strongly depends on the values of $\alpha$, and probably also on the specific choice of the observable but nevertheless it appears as a solid evidence of mixing.  This is yet another indication of the multifaced behaviour of the map \ref{eq:trm_map}. More precisely, in most cases of irrational $\alpha$ we find numerically non-exponential mixing with the auto-correlation approximately described by
\beq
  C(t) = O(t^{-\omega})\>,\qquad\textrm{where}\quad \omega = 0.2 \pm 0.02\>.
  \label{eq:corr_nongen}
\eeq
In Figure \ref{fig:C_beta0_det} we show the auto-correlation $C_{\rm t}(t)$ for two representative examples $\alpha=\sqrt{2}-1,e^{-1}$ using observable $f(q,p)=\sin(2\pi q) + \sin(2\pi p)$. We observe a very smooth decay. Basically the same results are obtained also for $\alpha=(\sqrt{5}-1)/2$ and by using other observables e.g. $f(q,p)=\sin(2\pi q) + \sin(2\pi p) + \sin(2\pi q) \sin(2\pi p)$, or $f(q,p)=\theta(q)$. The auto-correlation $C_{\rm s}(t)$ is not shown as it is numerically indistinguishable from $C_{\rm t}(t)$.
We also found an exception to the given approximation e.g. for $\alpha=\sqrt{3}-1$, where we find for certain observables, e.g. $\sin(2\pi q) + \sin(2\pi p)$ and $\sin(2\pi q) + \sin(2\pi p) + \sin(2\pi q)\sin(2\pi p)$, that the correlations could decay a bit faster, approximately as
\beq
  C(t) = O(t^{\omega})\>,\qquad\textrm{where}\quad \omega = 0.4 \pm 0.05\>,
\eeq
whereas for other observables (e.g step function) the law of decay is (\ref{eq:corr_nongen}). The anomalous situation in presented in Figure \ref{fig:C_beta0_anomaly_det}, where we show results for both definitions of auto-correlations $C_{\rm t} (t)$ and $C_{\rm s} (t)$. It is not surprising that in this situation the  connection between $\alpha$, the choice of the observable  and the power of the correlation decay $\omega$ can be quite complicated. Notice that due to a slow diffusion in momentum space, the calculated correlation decay in the presented time windows is actually taking place on a small subset of the phase space i.e. possibly on only few horizontal intervals per initial condition.

\section{The Koopman operator}

Some further particular cases can be discussed in terms of the spectral properties of the unitary operator $U$ on $H=L^2(\bT^2,d\mu)$, where $d\mu=dq\,dp$, defined by
\beq
  (U f)(q,p)=f(\phi (q,p))\>.
  \label{eq:koop}
\eeq
Consider the orthogonal decomposition $H=H_1\oplus H_2$, where $H_2$ is the subspace of functions depending only on the second coordinate, i.e. the subspace generated by $\{e_{0,s}\}_{s\in\bZ}$ where $e_{r,s}(q,p)=e^{2\,\pi\,\ii\,(\,r\,q+s\,p\,)}$. By the relation (\ref{eq:koop}),
\beq \label{gen}
  U^t\,e_{r,s}\,
  = e_{r,s+rt} \, e^{2\,\pi\,\ii\,\sum_{k=0}^{t-1}\bigl((t-k)r+s\bigr)\bigl(\,\alpha\,\theta (q_k)+\beta\bigr)}\>,
\eeq
which for $t=1$ writes
\beq\label{gen1}
  U\,e_{r,s}\,=e_{r,s+r} \, e^{ 2\,\pi\,\ii\,(s+r)(\alpha\,\theta(q)+\beta)}\>.
\eeq
On the other hand, a simple calculation yields
\beq
  e^{2\,\pi\,\ii\,\ell \, (\,\alpha\, \theta(q) +\beta)} =
  e^{2\,\pi\,\ii\,\ell \,\beta}\,\left( \cos{(2\pi\, \ell\,\alpha)}+ {2\over \pi} \,\sin{(2\pi \,\ell\,\alpha)}\,\sum_{k\;\; {\rm odd}}\;{e_{k,0}\over k}\, \right)
\eeq
and therefore

\beqa\label{carattere}
  U\,e_{r,s}\,
  &=& e^{2\,\pi\,\ii\,(s+r)\,\beta}\,
      \biggl(\cos{[2\pi (s+r)\alpha]}\,\, e_{r,s+r}\nonumber \\
  && \qquad +\;{2\over \pi} \,\sin[2\pi (s+r)\alpha]\,
      \sum_{k\;\; {\rm odd}} \;{e_{r+k,s+r}\over k}\, \biggr)
\eeqa
From this expression we immediately deduce the following
\begin{lemma} \label{autof}
If $\alpha = n/m$ and  $\beta \in \bR$ then $U$ has  eigenfunction $e_{0,m}$, with eigenvalue $e^{2\,\pi\,\ii \,m\,\beta}$.
\end{lemma}
\noindent
{\sl Proof.} Writing (\ref{carattere}) for $\alpha =\frac{n}{m}$, $r=0$ and $s=m$ we get $U\,e_{0,m}\,= e^{2\,\pi\,\ii\,m\,\beta}\, e_{0,m}$.$\qed$

\begin{corollary}
Let $\mu$ be the two dimensional Lebesgue measure. If $\alpha=n/m$ and $\beta \in \bR$  then $(\bT^2,\phi ,\mu)$ is not weakly-mixing. If furthermore $\beta \in \bZ +\alpha \bZ$ then the system is non-ergodic.
%the constant function and $e_{0,m}$ being two invariant functions in $L^2(\bT^2)$.
\end{corollary}

\noindent
Now, given a function $f \in L^2(\bT^2)$ let us expand it as
\beq\label{phi}
  f = \sum_{r,s} c_{\,r,s} \,e_{r,s} \quad\hbox{with}\quad
  \sum_{r,s}|c_{\,r,s}|^2 < \infty
\eeq
so that by (\ref{carattere}) we have
\beqa\label{Uphi}
  U\,f &=&  \sum_{r,s} c_{\,r,s}\, e^{2\,\pi\,\ii\,(s+r) \,\beta}\,  \biggl( \cos{[2\pi (s+r)\alpha]}\,\, e_{r,s+r}\nonumber\\
  &&\qquad +\;{2\over \pi} \,\sin[2\pi (s+r)\alpha]\,
            \sum_{k\;\;{\rm odd}} \;{e_{r+k,s+r}\over k}\, \biggr)\,\>.
\eeqa
Suppose now that there is $\lambda \ne 0$ with $|\lambda|=1$ such that $Uf=\lambda\, f$. By (\ref{phi}) and (\ref{Uphi}) the eigenvalue equation becomes
\beq\label{coeffi}
  \lambda \,c_{\,r,s}
  = e^{2\pi\,\ii\, s \beta}\,
  \left(
  \cos (2\pi s\alpha)\,\, c_{\,r,s-r}+
        {2\over \pi} \,\sin{(2\pi s \alpha \,)}\,
        \sum_{k\;\; {\rm odd}} \;{c_{\,r-k\,,\,s-r}\over k}
  \right)\>.
\eeq
For $\alpha=0$ equation (\ref{coeffi}) becomes
\beq
 \lambda \,c_{\,r,s}=  e^{2\,\pi\,\ii s \beta}\, c_{\,r,s-r}\>.
\eeq
Thus, when restricted to the (invariant) space $H_2$, $U$ multiplies
the Fourier coefficients by $e^{2\,\pi\,\ii\, s \beta}$ and
therefore has a discrete spectrum. Instead, on $H_1$ (where $r\ne
0$) $U$ generates an infinite family of trajectories with infinite
cardinality and therefore has a Lebesgue spectrum. In summary, for
$\alpha =0$ and $\beta$ irrational $(F,\bT^2,\mu)$ is ergodic but
not weak mixing (although for functions belonging to $H_1$ one may
have arbitrarily
fast decay of correlations \cite{courbage:etds:97}).\\
For $\alpha \ne 0$ the space $H_2$ is  not invariant. An
eigenfunction $f\in H_2$ would give
$$
  \lambda \, c_{0,s} =  e^{2\,\pi\,\ii\,s\,\beta}\,
                        \cos (2\pi s\alpha)\, c_{0,s}
$$
which is impossible for $\alpha$ irrational, since it would be $|\lambda|<1$ (unless $f=c_{0,0}\not=0$ and $\lambda=1$). Moreover, for $s=0$ we get $\lambda \,c_{r,0}=c_{r,-r}$ so that $|c_{r,0}|=|c_{r,-r}|$. Since $\sum_{r,s}|c_{\,r,s}|^2<\infty$ we have $c_{r,0}=c_{r,-r}=0$ for all $r\ne 0$. Using $c_{r,0}=0$ and (\ref{coeffi}) with $r=s$ we see that $c_{r,r}=0$ for all $r\ne 0$ as well. Therefore, if $c_{0,0}=0$ (so that $\lambda \ne 1$) then $f$ must be of the form
\beq\label{dive}
  f = {\sum_{|r|\ne |s|\ne 0}} c_{\,r,s} \,e_{r,s},
\eeq
where the sum has at least one term with $r\ne 0$.

\subsection{Fidelity of the truncated dynamics}

Having no further analytical results concerning the spectral properties of the  operator $U$, we now present some interesting scaling laws observed in the numerical investigation of a suitable truncated Koopman operator $U_{\rm tr}$ on the finite dimensional space $H_{\rm tr}$ spanned by the Fourier basis $\{e_{r,s} \}_{|r|\le N,|s|\le M}$. Even if at the moment we do not have a clever theoretical understanding of these results, their evidence is so robust that we believe it is worth mentioning.\par
\begin{figure}[!htb]
  \includegraphics[width=0.49\textwidth]{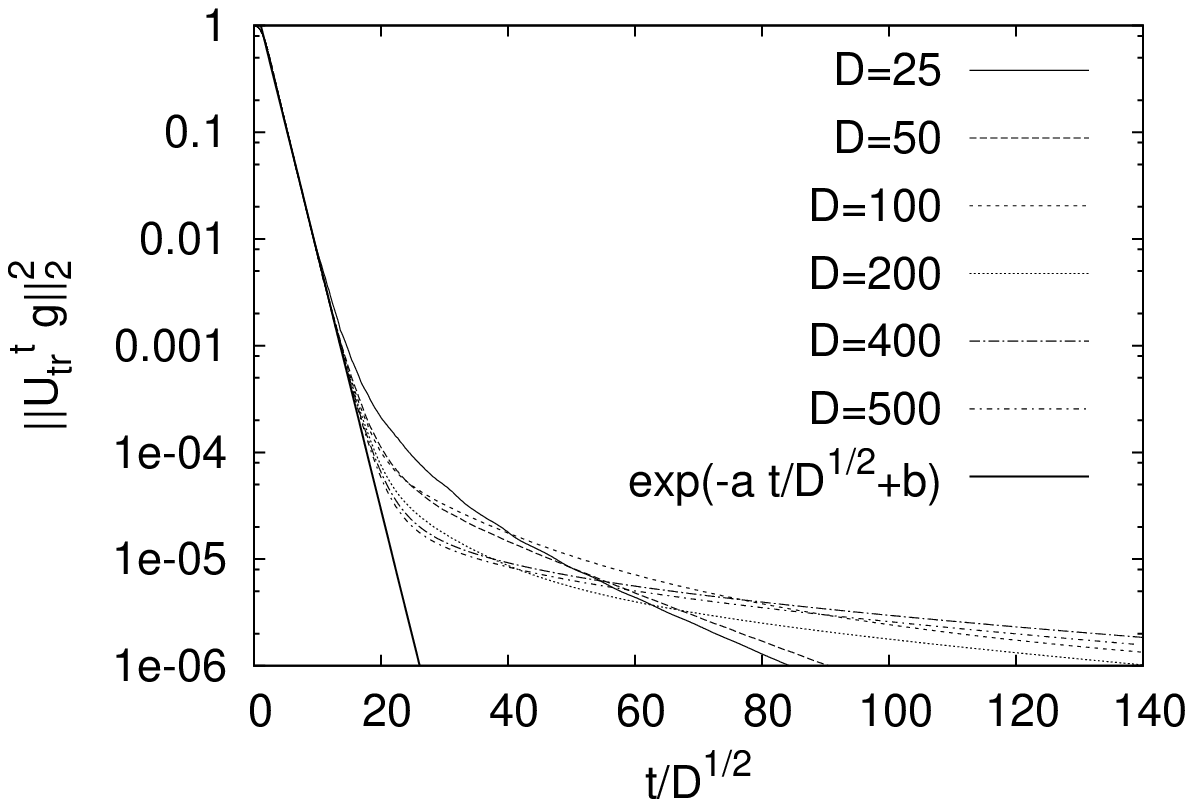}\hskip4pt%
  \includegraphics[width=0.49\textwidth]{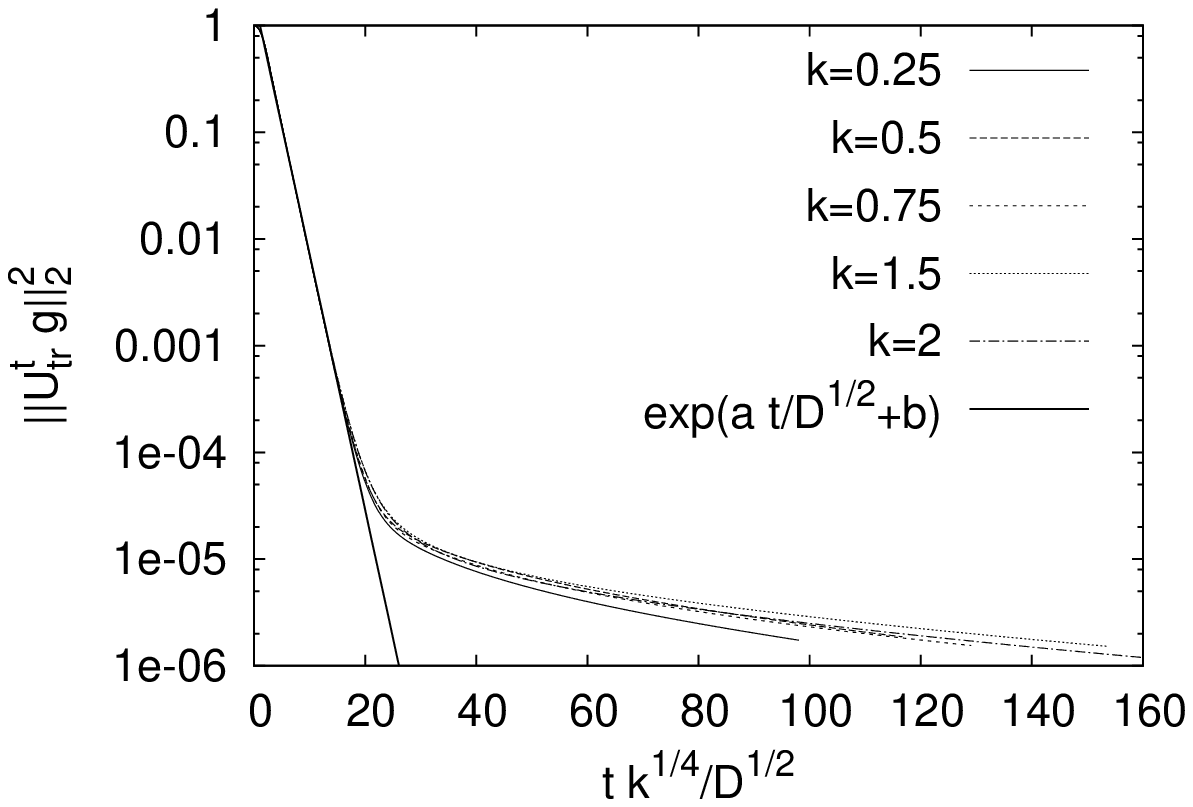}
  \hbox to\textwidth{\small\hfil(a)\hfil\hfil(b)\hfil}
  \caption{The evolution of $F_{\rm u}(t)$ for a Gaussian packet $g$ ($s=10,q_0=0.3,p_0=0.6$) at $\alpha=1/e$, $\beta=(\sqrt{5}-1)/2$, where in (a) it is calcuated at $\kappa=1$ and for different $D$ (a), and in (b) at fixed $D=500$ and for various $\kappa$ (b) as denoted in the figure. The fitted constants are $a=0.549624\pm 0.0008$ and $b=0.578\pm 0.007$.}
  \label{fig:trm_ueer}
\end{figure}
\begin{figure}[!htb]
  \includegraphics[width=0.5\textwidth]{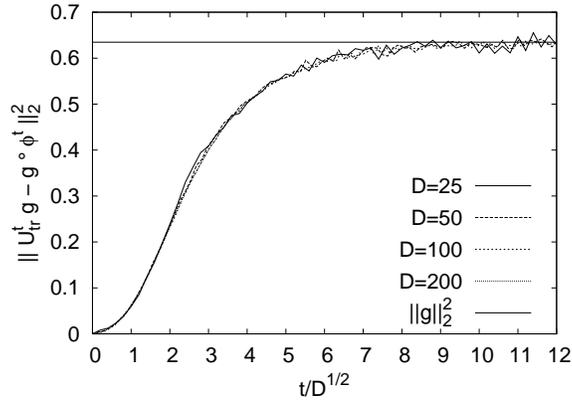}
  \caption{The $L_2$ norm of the difference between a Gaussian packet $g$ ($s=10,q_0=0.3,p_0=0.6$) propagated by the truncated Koopman operator and by exact triangle map calculated for  $\alpha=1/e$, $\beta=(\sqrt{5}-1)/2$ with $k=1$ and for various $D$.}
  \label{fig:trm_deer}
\end{figure}

The operator $U_{\rm tr}$ is no longer unitary. The Fourier basis is chosen so that $U_{\rm tr}$, as previously $U$, maps real functions into real functions satisfying the identity
\beq
   K U_{\rm tr} K = U_{\rm tr}\>,
\eeq
where $K$ is the operator of the complex conjugation. Thereby complex eigenvalues of $U_{\rm tr}$ come in complex conjugated pairs. The parameters $N$ and $M$ determine the dimension of $H_{\rm tr}$ and we express them by using new parameters $D$ and $\kappa$ as
\beq
  N = ((2 D+1) \kappa^{\frac{1}{2}} -1)/2\quad\textrm{and}\quad
  M = ((2 D+1) \kappa^{-\frac{1}{2}} -1)/2\>,
\eeq
so that $(2D+1)^2 = (2 N + 1)(2 M +1)$ and $\kappa =(2N+1)/(2M+1)$. We see that $D$ represents the geometric mean and $\kappa$ the ratio between the numbers of Fourier modes chosen along $q$ and $p$ axis. We quantify the {\it fidelity} of the truncated dynamics generated by the
$U_{\rm tr}$ by
\beq
  F_{\rm u}(t) = \| U_{\rm tr}^t g \|_2\>,\qquad
  F_{\rm d}(t) = \| g\circ \phi^t - U_{\rm tr}^t g \|_2\>,
\eeq
applied to an observable $g$ (and its truncation in $H_{\rm tr}$), with $\int_{\bT^2} \dd q\, \dd p\, g(q,p)=0$. The first quantity, $F_{\rm u}(t)$, measures the deviation from unitarity, whereas the second, $F_{\rm d}(t)$, captures  the distance from the dynamics generated by the triangle map. Here we use an observable $g(q,p)$, which is constructed by projecting a two dimensional Gaussian function, centered in $(q_0,p_0)$ and of variance $s^{-1}$, onto the torus and subtracting its mean. Thereby we obtain
\beq
  g(q, p) = g_s(q\!-\!q_0) g_s(p\!-\!p_0) - 1 \>,\;\;
  g_s (x) = \sqrt{\frac{s}{\pi}} \sum_{n\in\bZ}
  \exp\left(-s(x\!-\!n)^2\right)\>.
\eeq
The $L_2$ norm of the Gaussian packet is $\|g(q,p)\|_2^2 = \theta_{\rm jac}(2\pi^2/s)^2$, where $\theta_{\rm jac}( x)=\sum_{n\in\bZ} \exp(-x\, n^2)$ is connected to the Jacobi theta functions. Nevertheless, in the following, presented results are similar for all sufficiently smooth functions i.e. those with the Fourier expansion decaying sufficiently fast. In Figure \ref{fig:trm_ueer} we show the time evolution of $F_{\rm u}(t)$ for one Gaussian packet.
The analysis of results reveals a remarkable scaling property of $F_{\rm u}(t)$ w.r.t. the parameters $D$ and $\kappa$ valid in a certain time window $0\ll t < O(\sqrt{D})$, and reads
\beq
  \log(F_{\rm u}(t)) = -a \frac{\kappa^{\frac{1}{4}}}{\sqrt{D}} t + b
\qquad \textrm{for}\quad D \gg 1\>.
\eeq
where $a$ and $b$ are constants (independent of $t,D,\kappa$). Surprisingly, the constant $a$ is not strongly dependent on observables and its value is around $\half$ for various observables that have been tested. The obtained scaling indicates that the property of unitarity decays exponentially approximately with rescaled time $t/{\rm dim}(H_{\rm tr})^{\frac{1}{4}}\approx t/D^{1/2}\approx$ and the conservation of the norm is improving with decreasing $\kappa$. A similar scaling law is observed in the quantity $F_{\rm d}(t)$, which at $t=0$ starts from zero and then with increasing time saturates to a plateau given by $\|g\circ \phi^t\|_2=\|g\|_2$, see Figure \ref{fig:trm_deer}.\par
\subsection{The spectrum of the truncated Koopman operator}
\begin{figure}[!htb]
  \includegraphics[width=0.46\textwidth]{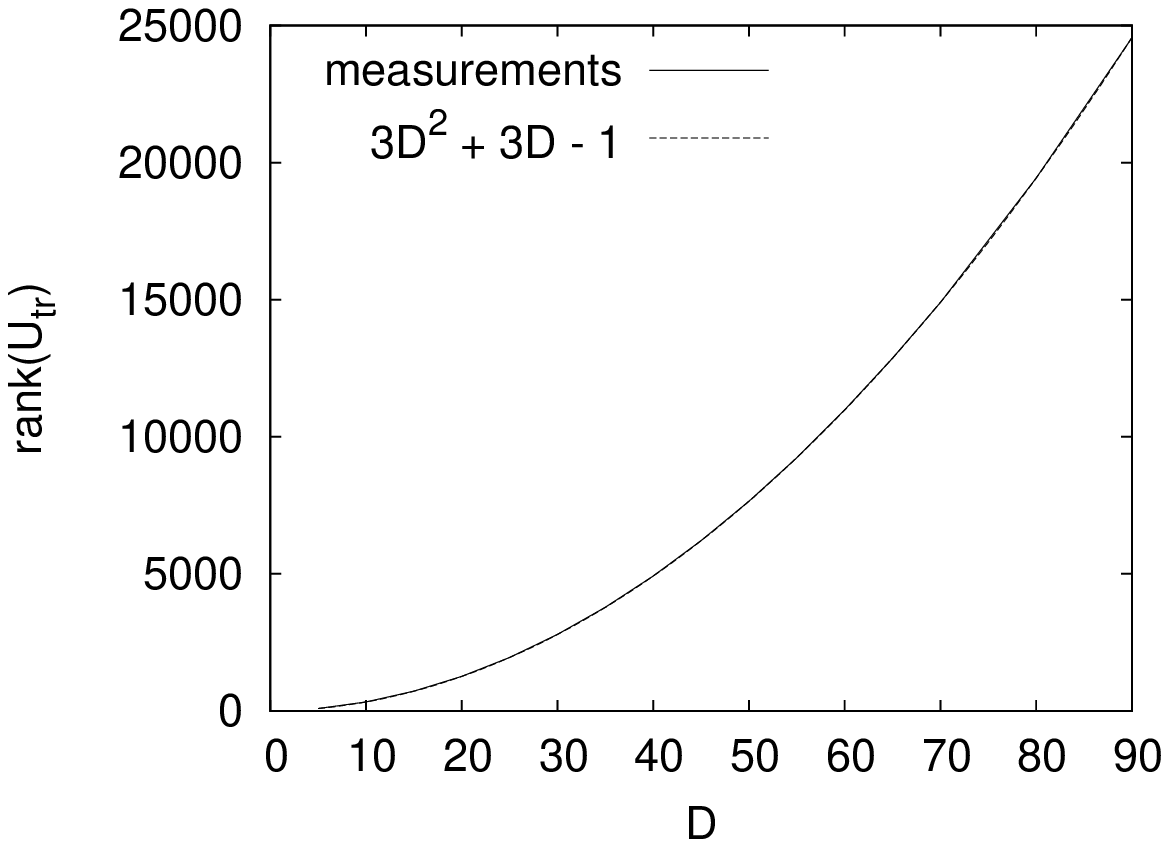}\hskip4pt%
  \includegraphics[width=0.52\textwidth]{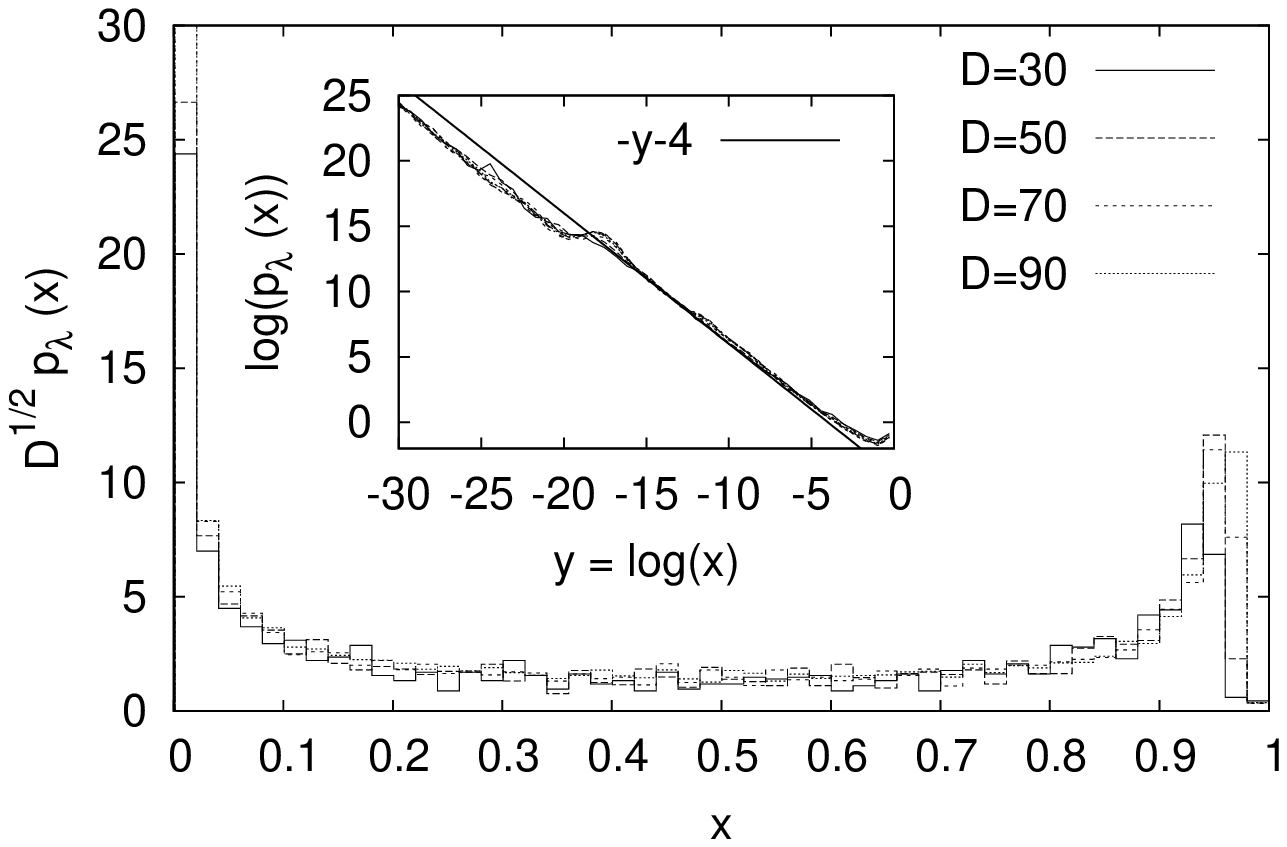}
  \hbox to\textwidth{\small\hfil(a)\hfil\hfil(b)\hfil}
  \caption{The rank of the truncated Koopman operator as a function of $D$ (a) and the distribution of the modulus of eigenvalues $p_{\lambda}(t)$ for different $D$ (b) calculated at $\alpha=1/e$ and $\beta=(\sqrt{5}-1)/2$.}
  \label{fig:koop_lam}
\end{figure}

\begin{figure}[!htb]
\includegraphics[width=0.49\textwidth]{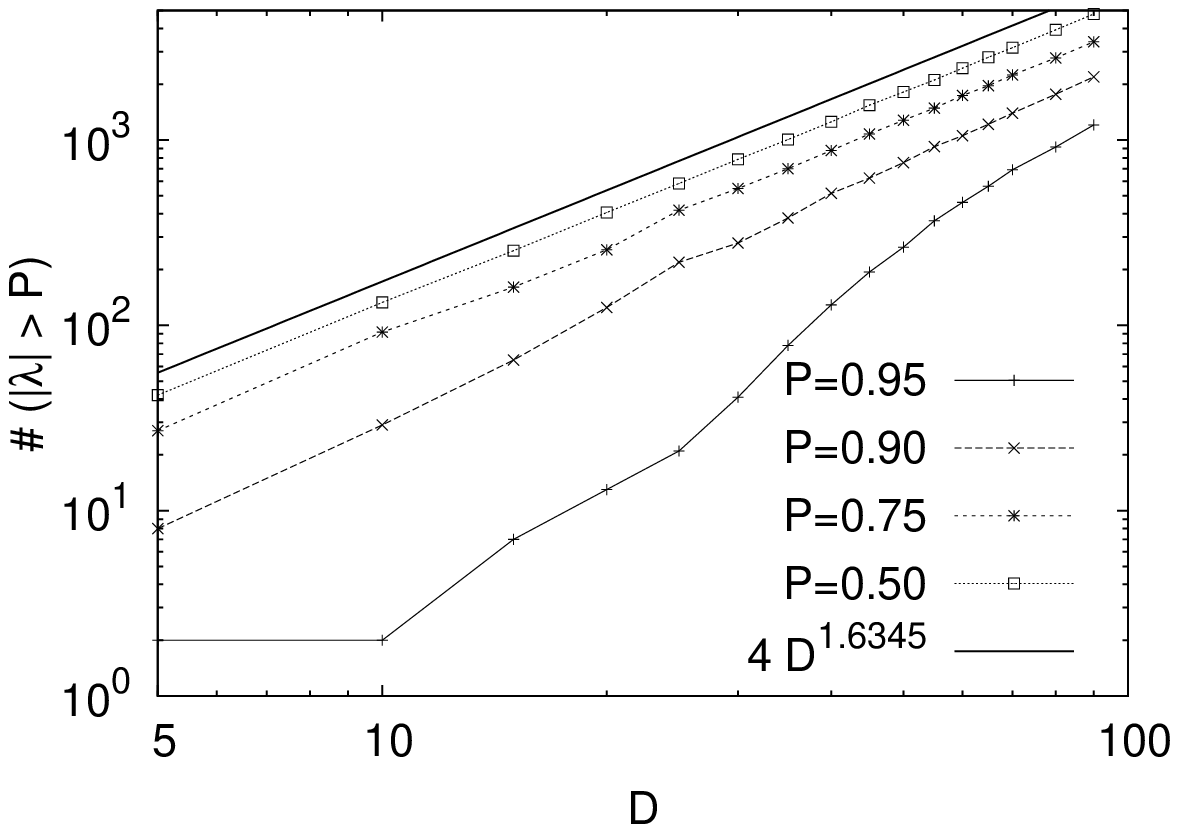}\hskip4pt%
\includegraphics[width=0.49\textwidth]{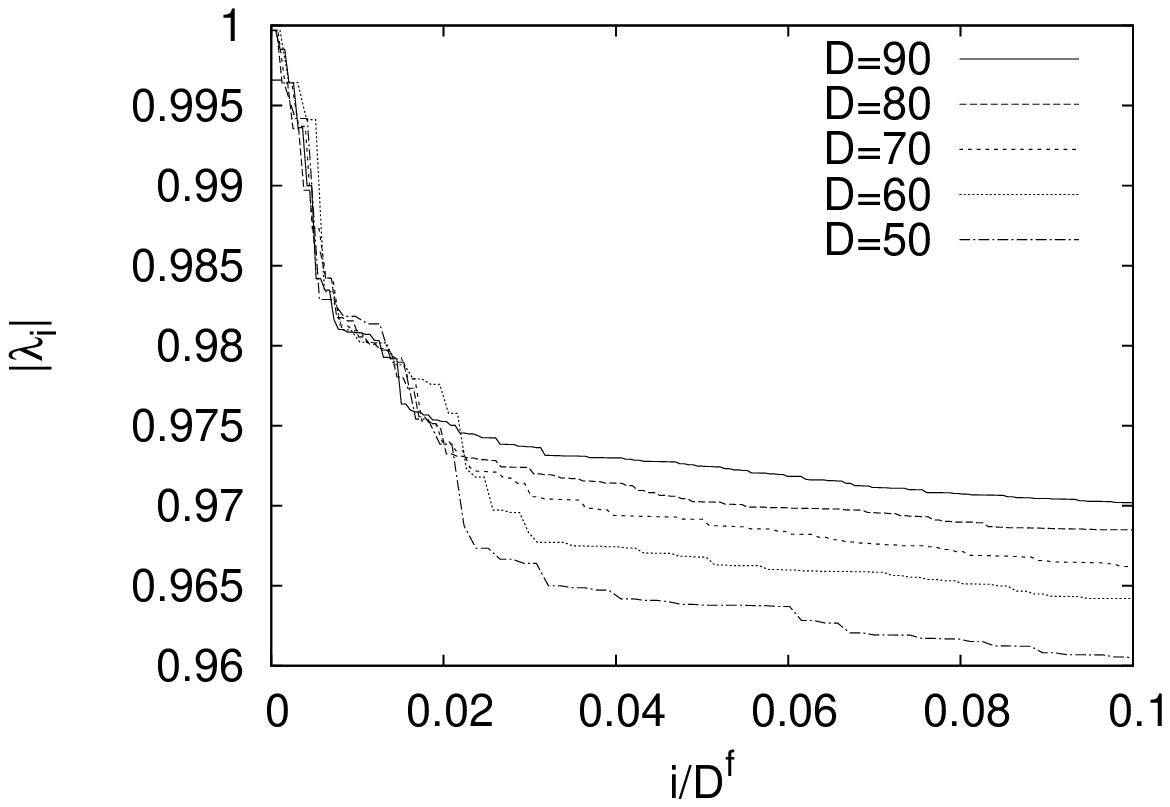}
\hbox to\textwidth{\small\hfil(a)\hfil\hfil(b)\hfil}
\caption{The largest eigenvalues of the truncated Koopman operator for different $D$ (a) and the number of eigenvalues larger then some $p$ as a function of $D$ (b) both calculated at $\alpha=1/e$ and $\beta=(\sqrt{5}-1)/2$.}
\label{fig:koop_bump}
\end{figure}

The spectrum $\{\lambda : \det (U_{\rm tr}  - \lambda \id) = 0 \}$ of the truncated Koopman operator on this finite functional space of dimension $(2D+1)^2$ is inside the unit-circle on the complex plane. The side effect of the truncation is that $U_{\rm tr}$ is not invertible, having the rank smaller than the dimension, see Figure \ref{fig:koop_lam}a. Notice that only non-zero eigenvalues are far enough from the unit circle participate in decay of $F_{\rm u}(t)$ from the start. We studied  the distribution of modulus of eigenvalues denoted by $p_\lambda (x)$ and have found that the probability scales like $O(D^{-\frac{1}{2}})$, see Figure \ref{fig:koop_lam}b. On the other hand the distribution of eigenvalue modules has singularity at zero of the form $p_\lambda (x)= O(x^{-1})$.
It is worthwhile to mention that eigenvalues have higher density near the unit circle. The peak of eigenvalues is with increasing $D$ slowly moving to the unit-circle and thereby we observe the squeezing of the spectral gap defined as $\Delta = 1-\max\{|\lambda|< 1\}$. The gap in our case is a non-smooth function of $D$. However, due to the condensation, the gap does not have an important dynamical meaning. The number of eigenvalues by magnitude larger than some value $0 < x \ll 1$ is increasing as $O(D^\nu)$, where fitted value of the exponent is $\nu = 1.634\pm 0.002$, see Figure \ref{fig:koop_bump}a.  This scaling can be also applied to eigenvalues near the unit-circle so that behaviour of eigenvalue modules near the circle align for different $D$ as we can see in Figure \ref{fig:koop_bump}b.
Although the connection between the fidelity $F_{\rm u}(t)$ and the spectrum of the truncated Koopman operator is clear, we find that the behaviour of eigenvalue modulus with $D$ is insufficient to explain the scaling properties of the fidelity.

\section{Dynamics of polygons}\label{sec:poly}

In this section we consider a heuristic but quite fruitful geometric approach, namely we study the dynamics of the map $\phi$ acting on particular simple subsets of phase space $\cA \subset \bT^2$, the polygons. It is easy to verify that all the ingredients of the dynamics $\phi$ preserve polygons, namely $\phi$ maps a polygon to a polygon, and hence its iterates $\phi^n$ map a polygon to  a finite set of several polygons. Furthermore it is possible to make an efficient and accurate numerical algorithm for performing dynamics of polygons by representing them - or their boundaries - in terms of sets of lines.

In the following subsection we describe and apply such explicit construction and the related numerical procedure in order to study the statistical properties of transformed polygons as it emerges from iterating the dynamics for sufficiently long times.

\subsection{Statistical properties of polygons}
We now consider an arbitrary initial polygon $\cA\subset\bT^2$ and evolve it with time $\cA_t := \phi^t(\cA)$, $t\in \bZ$. After some time $t$, the set $\cA_t$ is composed of many, say $N(t)$, {\em pieces}. The following main questions are addressed in our numerical experiment:
\begin{itemize}
\item Do there exist well defined statistical distributions of geometric properties of these little polygonal pieces, such as {\em area}, diameter in $x$-direction, diameter in $y$-direction etc., after long time $t$, and do these distributions possess properly defined averages and variances?
\item What are the scalings of geometric properties of little pieces as $t$ grows? Are asymptotic, $t\to\infty$, statistical distributions universal, i.e. independent of the geometry and volume of the initial set $\cA$?
\item Are these pieces distributed uniformly in phase space $\bT^2$, i.e.  that the number of pieces inside an arbitrary (`window') set  $\cB\subset \bT^2$ with area $\mu(\cB)$ is, after long time $t$, equal to $\mu(\cB)N(t)$?

\end{itemize}
A computer program has been written which computes an exact evolution of an arbitrary initial polygon $\cA$ in terms of a set of polygons $\cA_t$. Affirmative answers to the above questions (for generic, irrational values of parameters $\alpha,\beta$) are strongly suggested by the numerical results which are summarised below. It can be easily shown that any connected piece, say $\cP$, of $\cA_t$ has the following properties (which are preserved under time evolution):
\footnote{Provided such conditions are imposed on the initial polygon $\cA_0=\cA$. Otherwise, these properties hold for ever increasing fraction of its little pieces, namely for those whose boundary no longer contains the image of any part of the boundary of the initial set $\cA$.}
\begin{enumerate}
\item There exist left-lower and right-upper corner points, $C_1,C_2\in \cP$, with coordinates $(q_1,p_1)$ and $(q_2,p_2)$, respectively, such that for arbitrary $(q,p)\in{\mathcal P}$: $q_1 \le q \le q_2$, $p_1 \le p \le p_2$.
\item The boundary of the polygon ${\mathcal P}$ can be written in terms of two piece-wise linear curves (monotonically increasing in $q-p$ plane) going from $C_1$ to $C_2$, namely lower $\cB_{\rm l}$ and upper $\cB_{\rm u}$ boundary. Going from $C_1$ to $C_2$, the slopes of the straight segments of $\cB_{\rm u}$ in $q-p$ plane can be written in terms of an increasing sequence of positive integers $1\le n_1 < n_2\ldots$, as $1/n_1 > 1/n_2 > \ldots$, while the slopes of the straight lines of ${\mathcal B}_l$ are given by a decreasing sequence of positive integers $m_1 > m_2 \ldots \ge 1$, as $1/m_1 < 1/m_2 < \ldots$
\end{enumerate}

\begin{figure}[!htb]
  \includegraphics[width=0.8\textwidth]{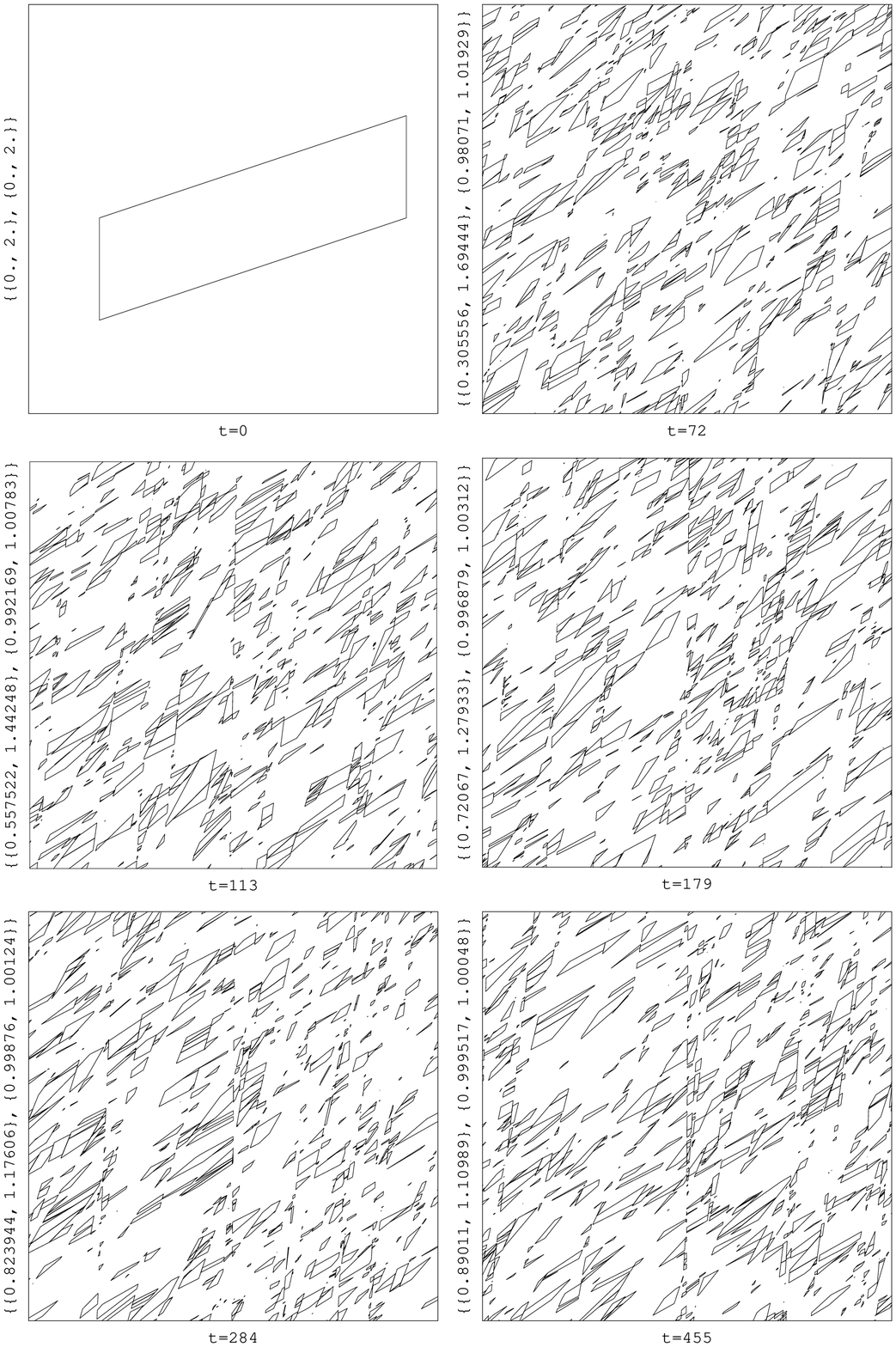}
\caption{Renormalised (zoomed) snapshots around some point in phase space $(q_c=1,p_c=1)$ at different times on a rescaled total phase space $[0,2)\times [0,2)$, for generic case $\alpha=e^{-1}, \beta=(\sqrt{5}-1)/2$. Vertical labelling to the left of each plot indicates the coordinates $\{\{q_1,q_2\},\{p_1,p_2\}\}$ of the window shown.}
\label{fig:poly_gen}
\end{figure}

In Figure \ref{fig:poly_gen} we show the image of the initial polygon (first plot) in terms of five snapshots (additional five plots) at generic irrational values, $\alpha=e^{-1}, \beta=(\sqrt{5}-1)/2$, where integer times $t$ shown are roughly by a constant factor apart.
Each snapshot is plotted in a renormalised window, around certain reference
point $(q_c,p_c)=(1,1)$ (in the center of the phase space, which is here scaled to $[0,2)\times [0,2)$), of $q-$width $\delta_q/t$ and $p-$width $\delta_p/t^2$ in order to suggest the scaling. 

Observe the statistical similarity of consecutive snapshots!

\begin{figure}[!htb]
  \includegraphics[width=0.8\textwidth]{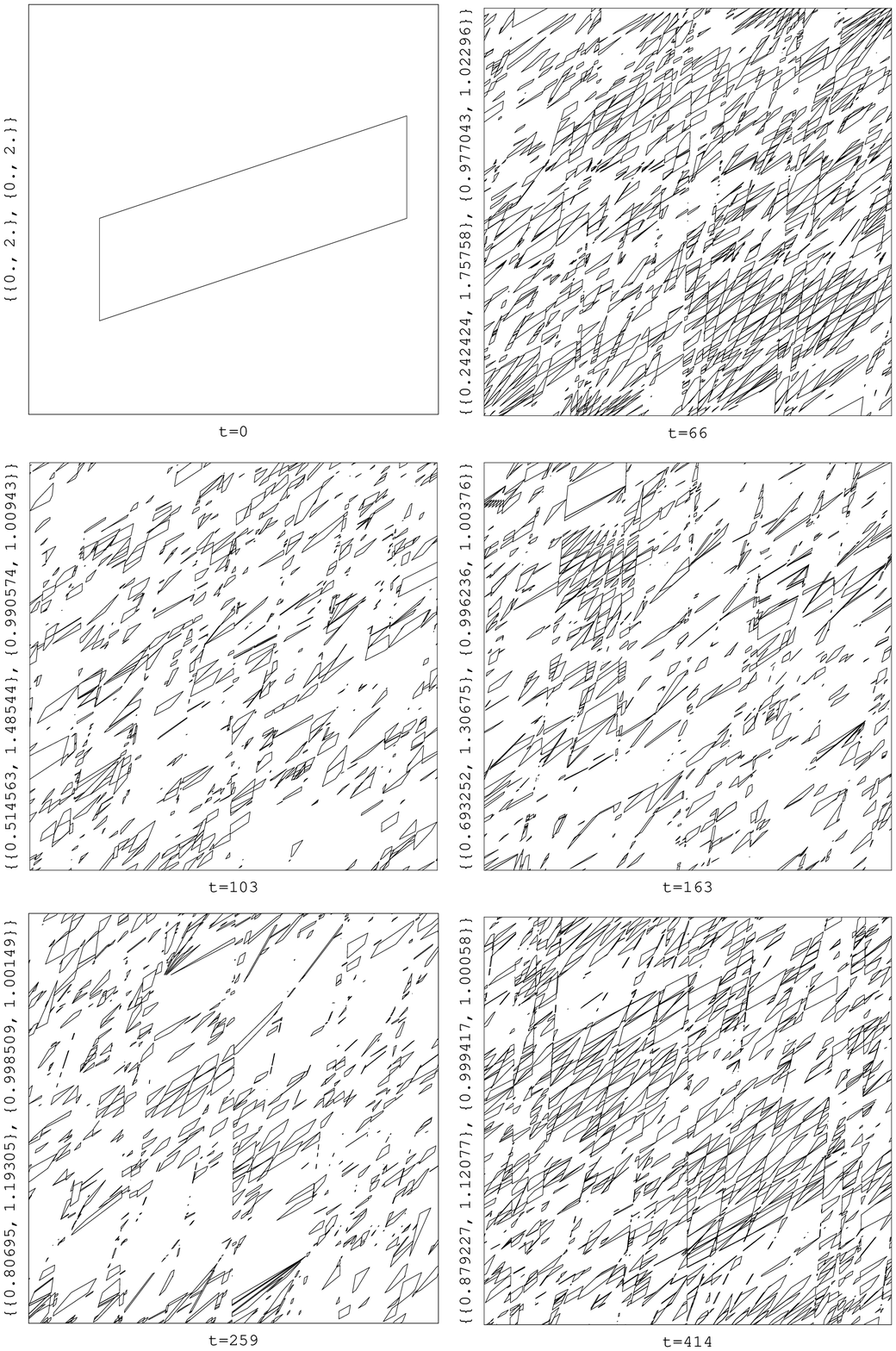}
\caption{Renormalised snapshots in non-generic case $\alpha=e^{-1}$,
$\beta=0$. See also caption of fig. \ref{fig:poly_gen}.}
\label{fig:poly_weak}
\end{figure}

By repeating the experiment for a smaller initial set the
statistical self-similarity persists just the density of little
pieces reduces proportionally to the area of the initial set. Figure
\ref{fig:poly_weak} presents computation for the non-generic
(`weakly-ergodic') case of $\beta=0$ (such map has been studied
already in \cite{kaplan:physD:98}).
We see large temporal and spatial fluctuations in the density of little pieces which may complicate the mechanism of dynamical mixing in this situation. Clearly, general heuristic arguments for the mixing mechanism would not hold for this case. In the other non-generic case $\alpha\in\bQ$ and $\beta\in\bR\backslash\bQ$ we get similar nonuniform structure with disappearance of some stripes in initial transient time.\par
\begin{figure}[!htb]
  \includegraphics[width=0.5\textwidth]{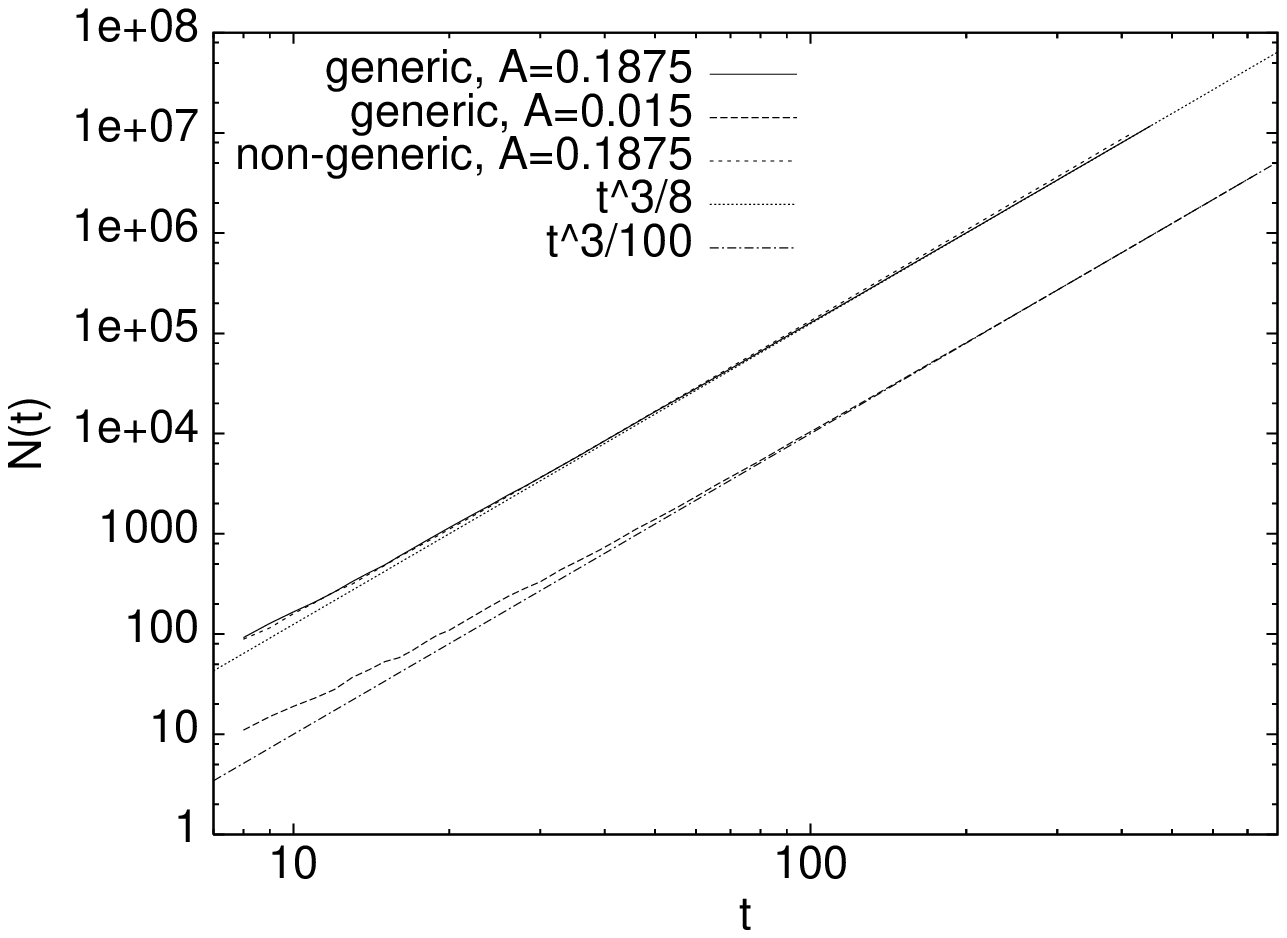}
\caption{The number of pieces $N(t)$ as a function of iteration-time $t$ for three different cases of initial set sizes and their estimated asymptiotics $\widetilde C t^3$.}
\label{fig:np}
\end{figure}
\begin{figure}[!htb]
  \begin{tabular}{@{}c@{}cc}
 {\small(a)\hskip-5mm}&\includegraphics[width=0.47\textwidth]{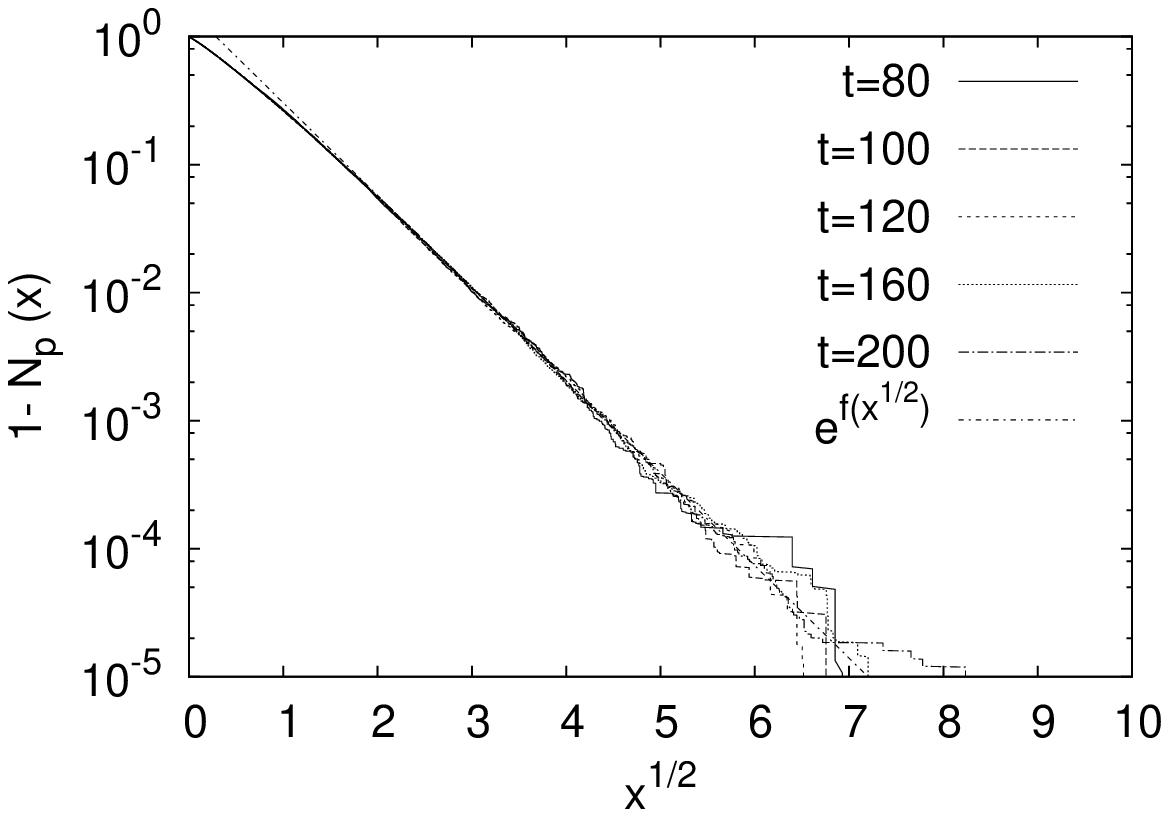}&
    \includegraphics[width=0.47\textwidth]{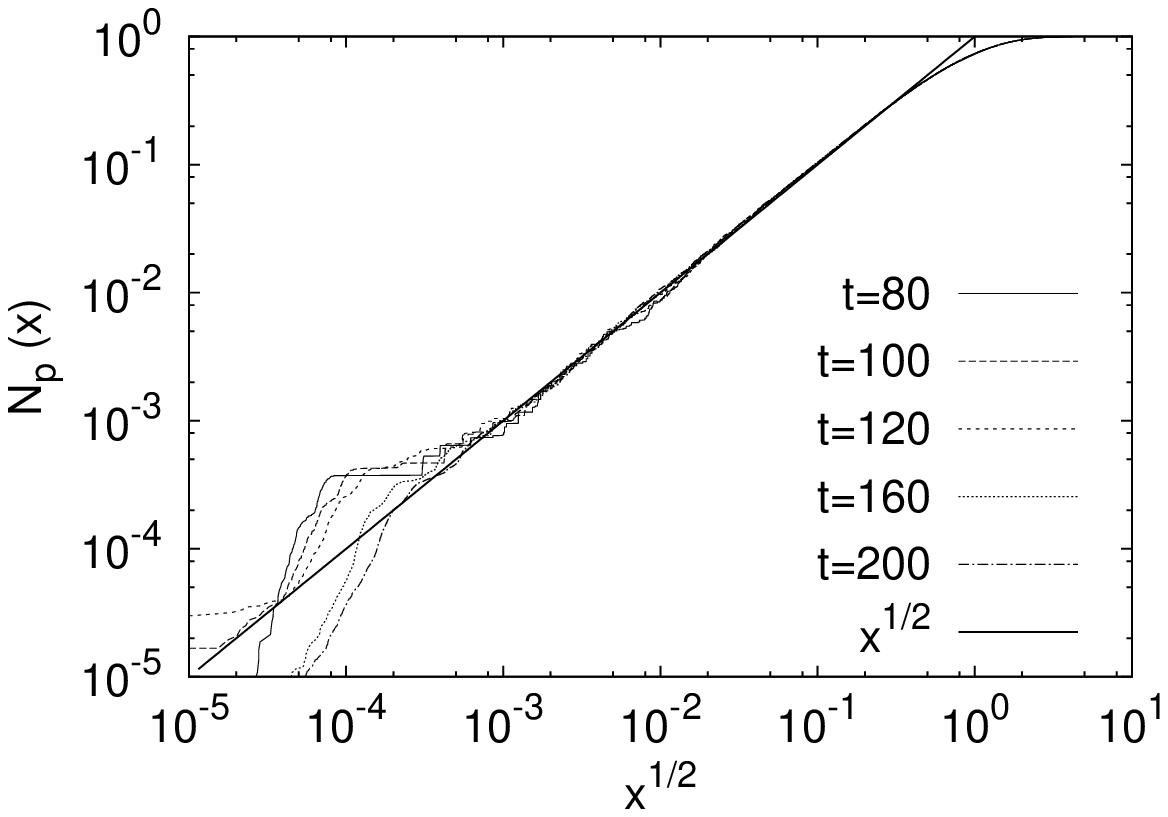}\cr
 {\small(b)\hskip-5mm}&\includegraphics[width=0.47\textwidth]{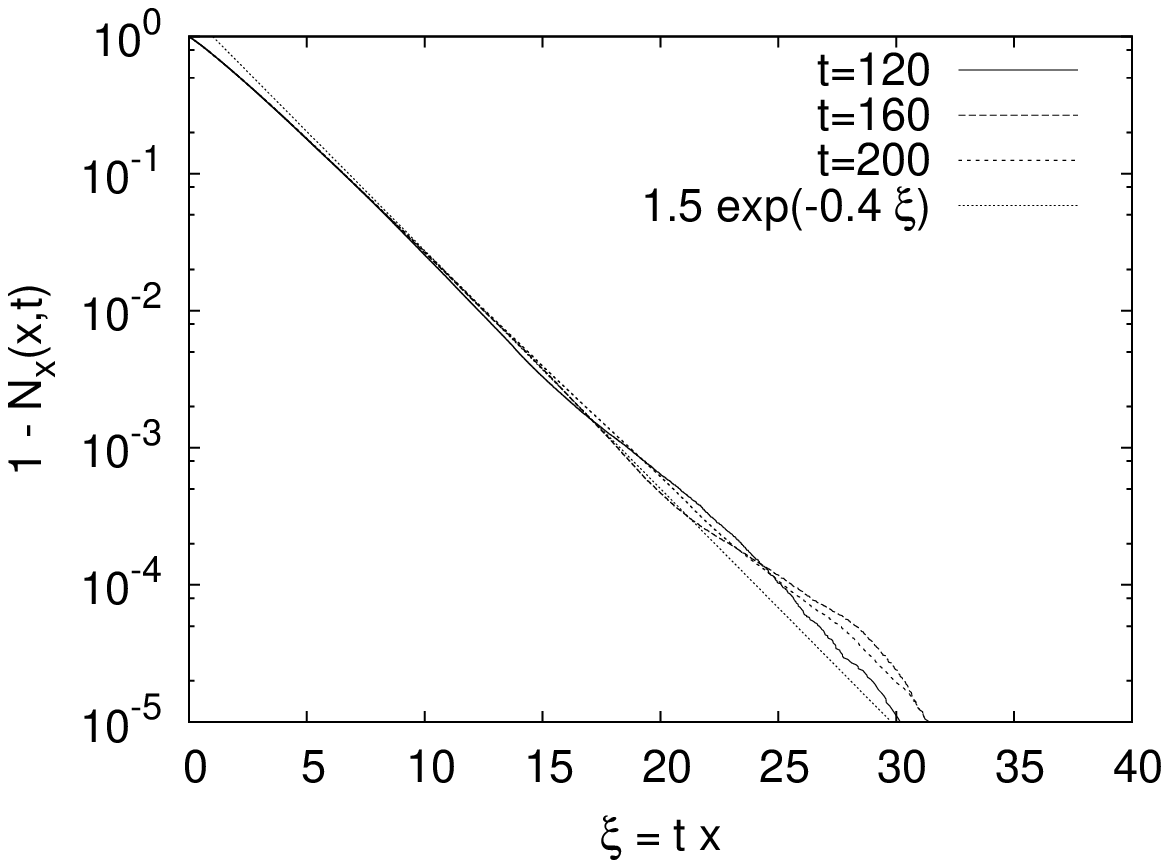}&
    \includegraphics[width=0.47\textwidth]{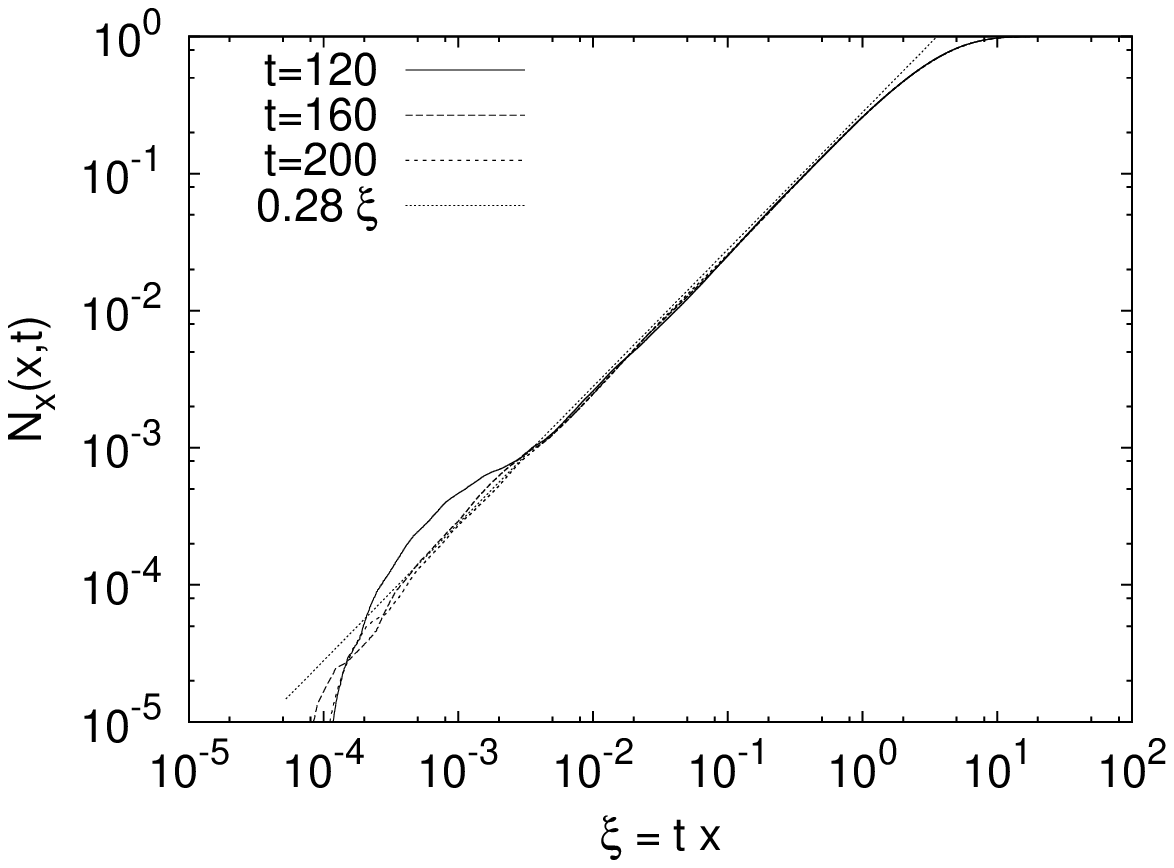}\cr
 {\small(c)\hskip-5mm}&\includegraphics[width=0.47\textwidth]{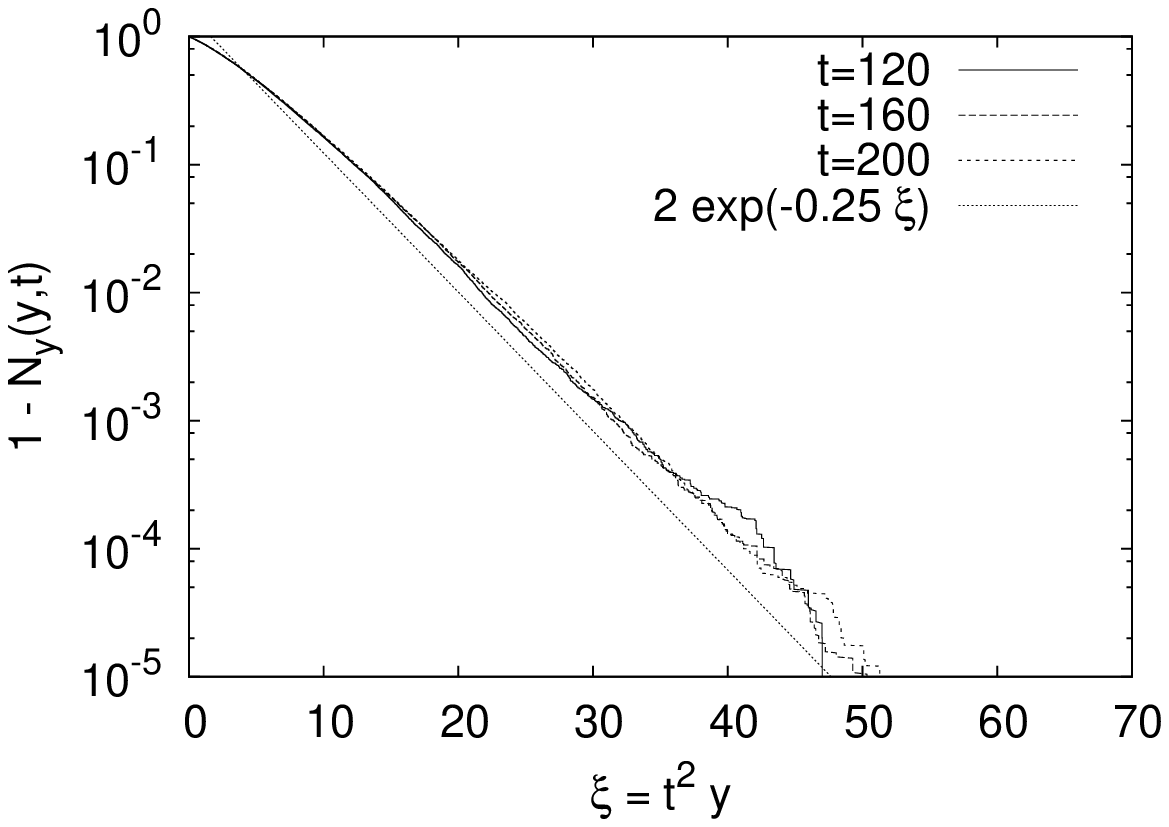}&
    \includegraphics[width=0.47\textwidth]{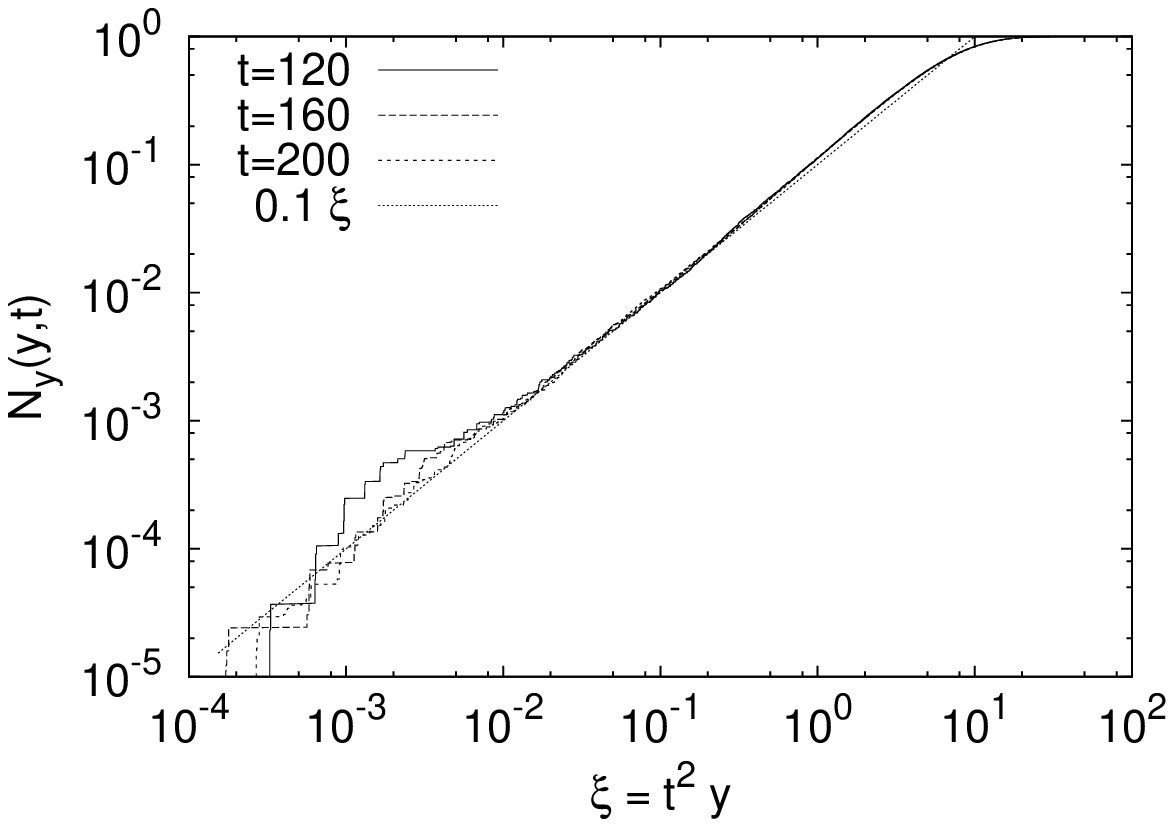}
  \end{tabular}
  \caption{The cumulative number distribution of the polygon areas $N_a(x,t)$ (a), polygon horizontal dimension $N_x(x,t)$ (b), and polygon vertical dimension $N_y(x,t)$ (c), for different times $t$ in the generic case at $\alpha=1/e$ and $\beta=(\sqrt{5}-1)/2$. The fitted curve is $f(x)=-1.6644\, x + 0.475028$.}
  \label{fig:tp_part_cum}
 \label{fig:nl}
\end{figure}

We count the number of the little pieces $N(t)$ obtained by evolving initial set up to time $t$. In Fig.\ref{fig:np} we show $N(t)$ for the two generic cases with different initial sets and for a weakly ergodic case.
Always, the growth of $N(t)$ is found to be cubic in time after some transient period, with the proportionality constant factor proportional to the area of the initial set
\beq
  N(t) \sim \widetilde C t^3\>,\qquad
  \widetilde C\propto \mu(\cA)\>.
\eeq
We analyse the statistical distributions of the following geometric measures of little pieces: (i) the area $a:=\mu(\cP)$, (ii) the $q$-diameter $\delta_q:=q_2-q_1$, (iii) the $p$-diameter $\delta_p:=p_2-p_1$, and (iv) the average slope of the piece $s:=\delta_p/\delta_q$.
%
%In figures \ref{fig:ave_ar_sl} and \ref{fig:ave_dx_dy}) we plot the averages of these four quantities (with statistical weights given by the probability measures (areas) $\mu(\cP)$) as a function of time $t$. For generic irrational parameters, the long-time averages and also full distributions do not depend on the size of the initial set, nor on particular values of generic irrationals $\alpha,\beta$. Suggested numerically determined scaling laws for the averages and for the standard deviations $\sigma_{(\cdot)}:=\sqrt{\ave{(\cdot)^2}-\ave{\cdot}^2}$ are:
%
\beqa
  &\ave{a} = 30.0/t^3\>,\qquad
  &\sigma_a = 1.12 \ave{a}\>, \label{eq:scala}\\
  &\ave{\delta_q} = 7.2/t\>,\qquad
  &\sigma_{\delta_q} = 0.59 \ave{\delta_q}\>,\\
  &\ave{\delta_p} = 14.4/t^2\>,\qquad
  &\sigma_{\delta_p} = 0.52 \ave{\delta_p}\>,\\
  &\ave{s} = 2.2/t\>,\qquad
  &\sigma_s = 0.37 \ave{s}\>. \label{eq:scalz}
\eeqa
This quantitative results and self-similarity of pieces on the phase space suggest a kind of statistical scaling invariance under the transformation $q'=q t$ and $p'=p t^2$. Therefore, let us introduce $t$-invariant quantities
\beq
  \chi_a = t^3 a\>,\quad \chi_q = t \delta_q\>,\quad
  \chi_p = t^2 \delta_p\>,\quad \chi_s = t s\>.
\eeq
Next we investigate the full distributions of scaled quantities $\chi_{a,q,p,s}$. This is done by calculating two different distributions:
\begin{enumerate}
\item[(i)] {\em Cumulative number distributions}, equation for the quantity $Q$ which is one of $\chi_a,\chi_x,\chi_y,\chi_s$
\beq
  N_Q (\chi) = \frac{\#\{Q({\mathcal P}) \le \chi\}}{N(t)}\>.
\eeq
\item[(ii)] {\em Cumulative probability distributions},
\beq
  W_Q (\chi)
  = \frac{1}{\mu({\mathcal A})}
    \sum_{\mathcal P} \mu({\mathcal P})H(\chi-Q({\mathcal P}))\>.
\eeq
where $H(x)=\{0: x< 0; 1: x\ge 0\}$ is the step function.
\end{enumerate}
We note that the qualitative features of the number distributions $N_Q$ and
probability (measure) distributions $W_Q$ are essentially the same, so we shall
in the following only report numerical results on the former.
It has been checked with great numerical accuracy that distributions $N_Q(\chi)$ and $W_Q(\chi)$ are asymptotically (for large $t$) independent:
(i) of time $t$,
(ii) of the size and shape of the initial set $\cA$ and
(iii) of the particular values of generic irrationals $\alpha,\beta$.
In Figure \ref{fig:nl} we show numerical results on these distributions which suggest several small and large argument asymptotics which are indicated within figures' labels. It should be noted that the numerical data in Figure \ref{fig:nl} can be described even globally quite well by an exponential fit $N_{x,y}(\chi) = 1 - \exp(-\gamma_{x,y} \chi)$ with some exponents $\gamma_x,\gamma_y$.

We conjecture and show later in subsection \ref{xy} for a simplified special case that these distributions (and the scaling) can be derived from the fixed point condition for certain dynamic renormalisation group equations. The later is found for the random triangle model or more precisely for its one-dimensional simplification.

\subsection{A Markov approximation}

Let us define a finite Markov approximation of the dynamics by the following construction of a transition matrix. We start by the set of two vertical discontinuity lines $\cD$. They can be viewed as boundaries of two elementary polygons - vertical stripes - which divide the entire torus $\bT$ in two pieces and have no corners. This is just the initial partition that we will be using later to construct a suitable  binary coding of the dynamics. Here we discuss only the generic case at parameters $\alpha=1/e$, $\beta=(\sqrt{5}-1)/2$.\par
\begin{figure}[!htb]
  \includegraphics[width=0.49\textwidth]{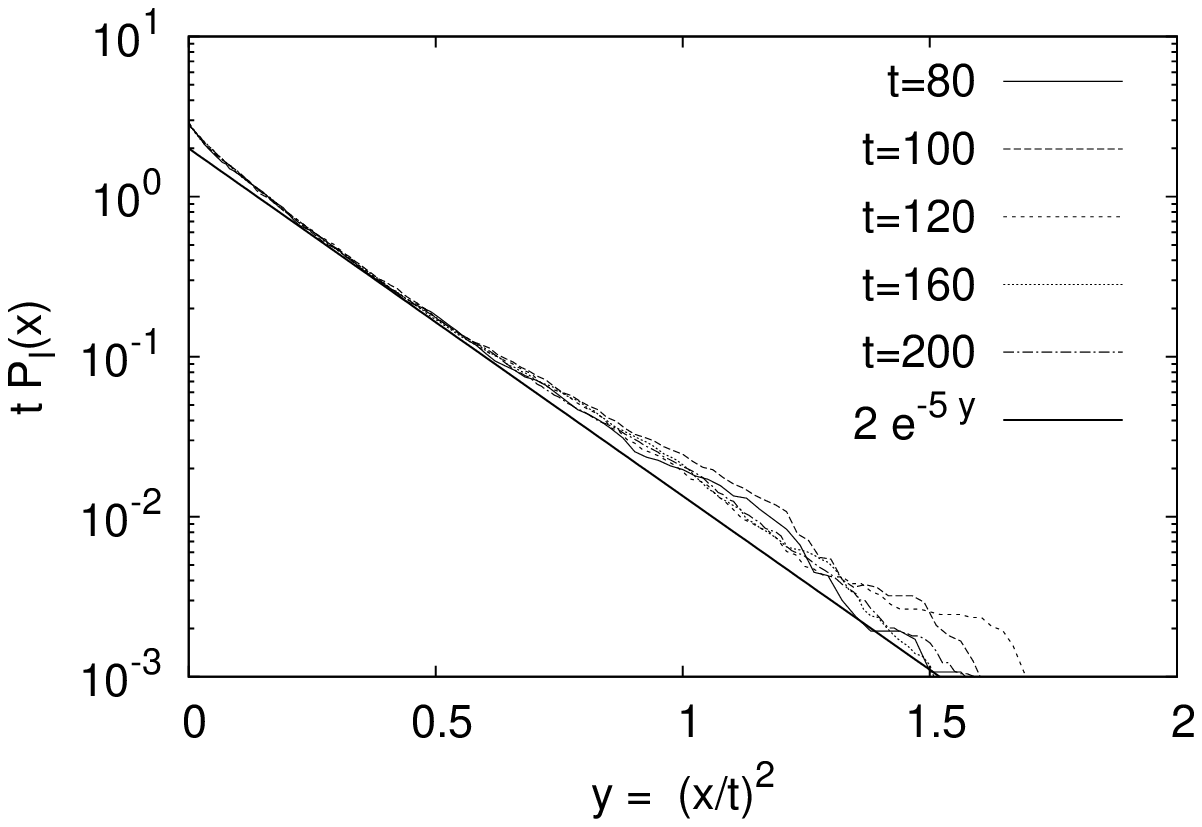}\hskip4pt
  \includegraphics[width=0.49\textwidth]{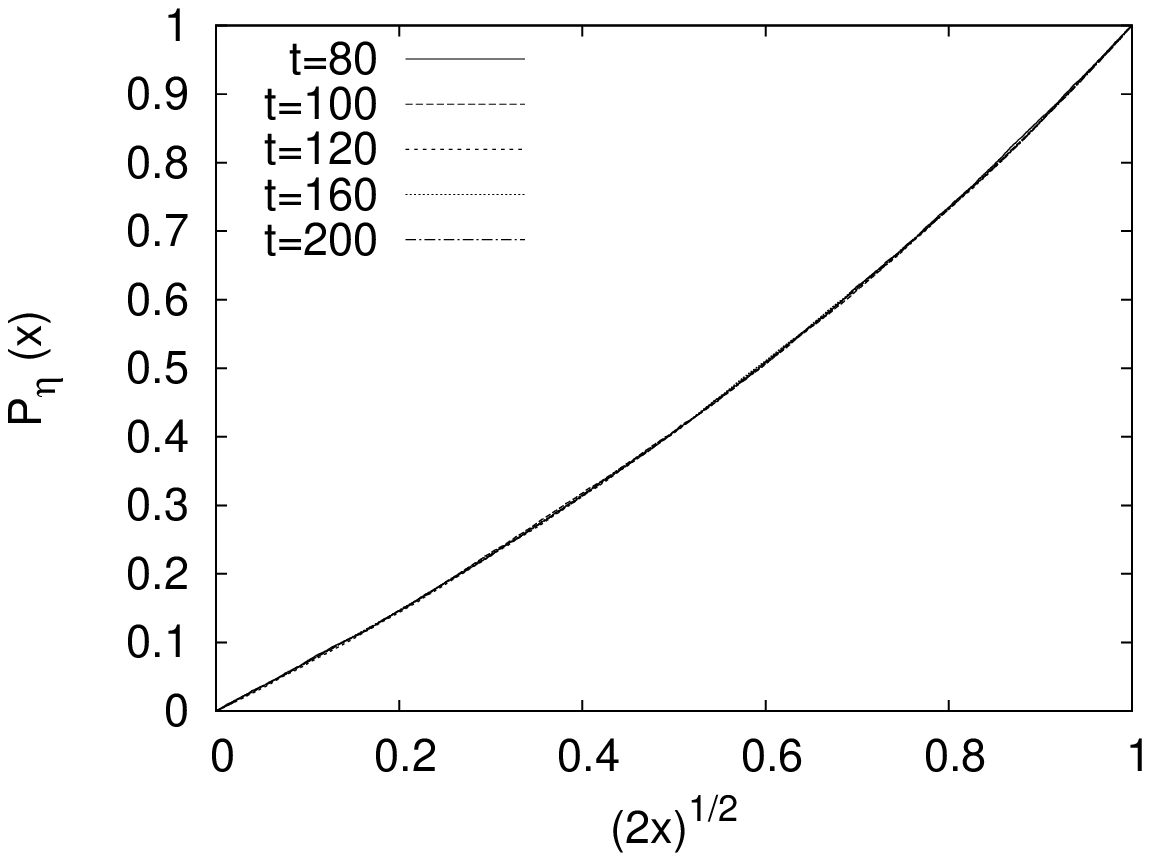}
    \hbox to\textwidth{\small\hfil(a)\hfil\hfil(b)\hfil}
  \caption{The distribution of free propagation length of partitions (a)
  and the cumulative distribution of splitting strengths (b) in the Markov matrices calculated in the polygonal description at $\alpha=1/e$, $\beta=(\sqrt{5}-1)/2$ and for different times $t$.}
  \label{fig:tp_split}
\end{figure}
\begin{figure}[!htb]
  \includegraphics[width=0.49\textwidth]{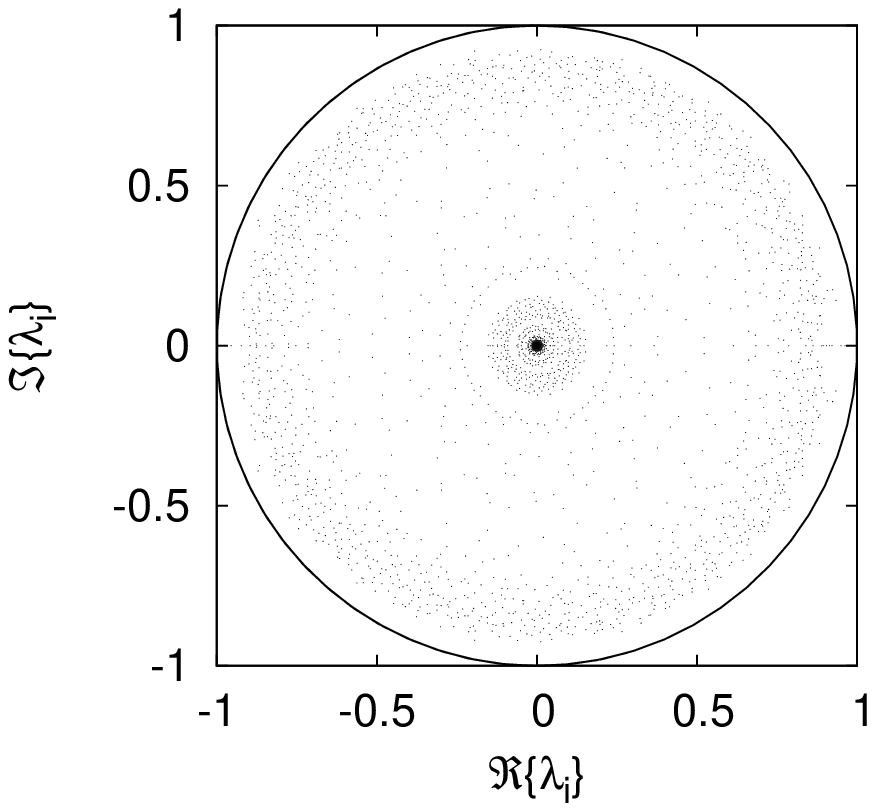}\hskip4pt%
  \includegraphics[width=0.49\textwidth]{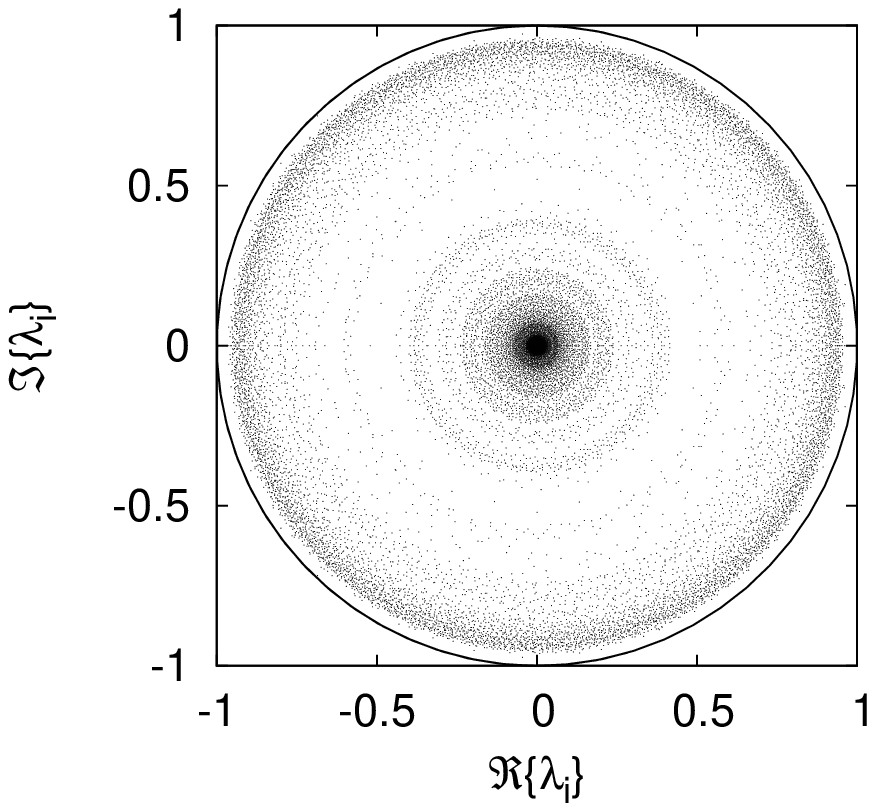} \\
  \includegraphics[width=0.49\textwidth]{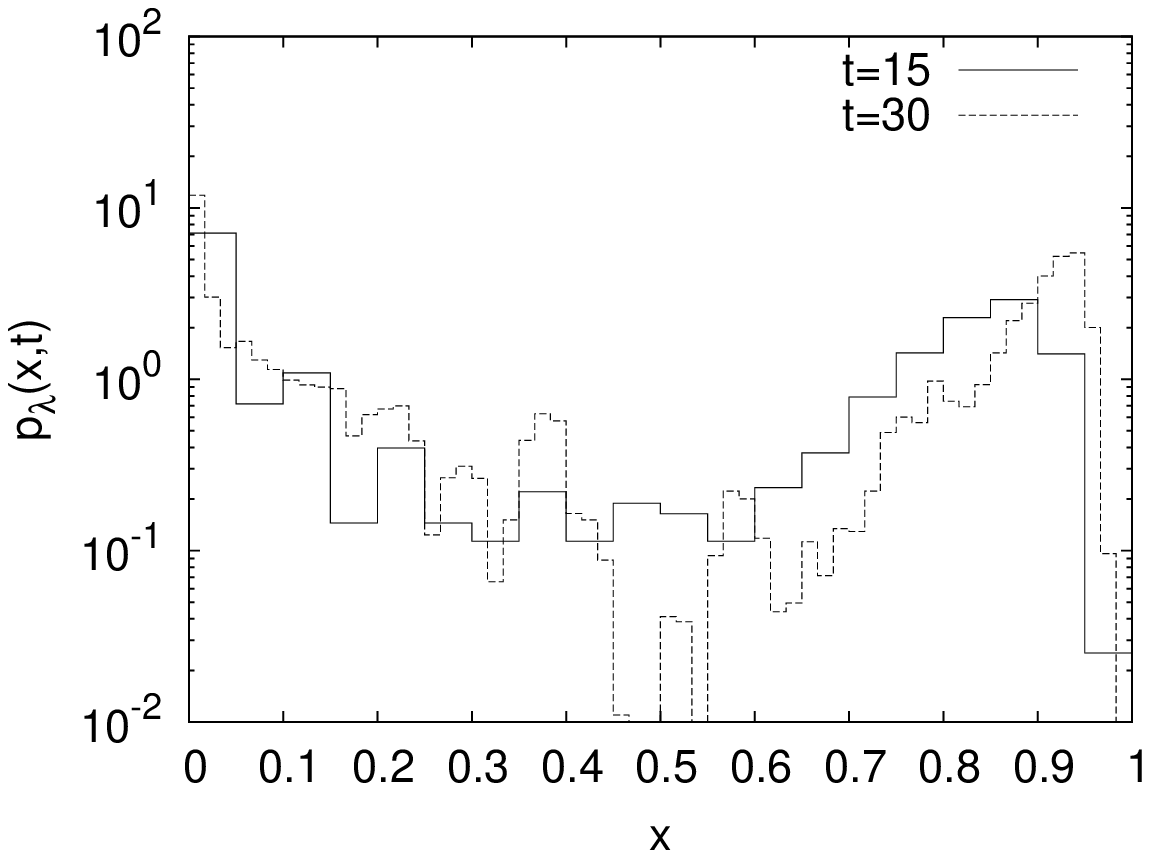}\hskip4pt
  \includegraphics[width=0.49\textwidth]{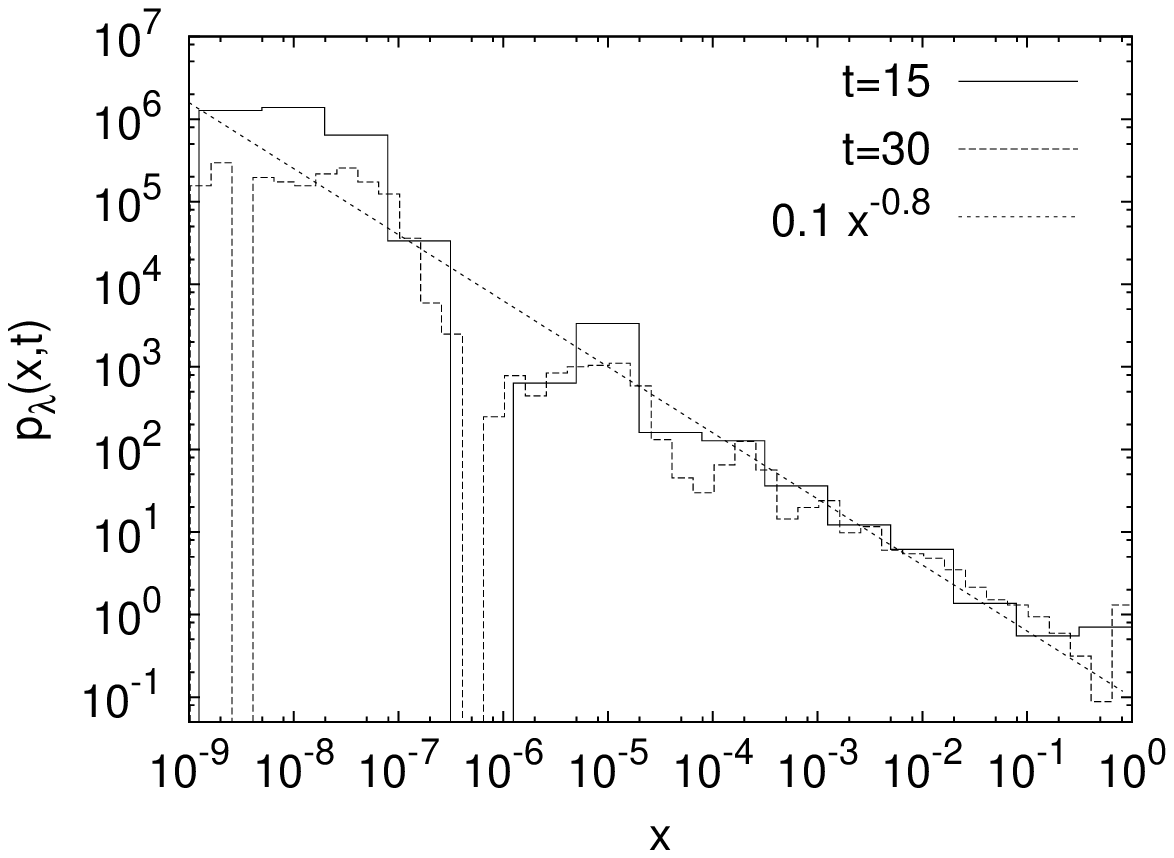}
  \caption{The eigenvalues $\lambda_i$ of the Markov matrix in the complex plane calculated in polygonal description at $\alpha=1/e$ and $\beta=(\sqrt{5}-1)/2$ for $t=15$ (top, left),$t=30$ (top, right) and the distribution of eigenvalue amplitudes $p_\lambda(x)$ in different representations (bottom).}
  \label{fig:tp_spect_distr}
\end{figure}
\begin{figure}[!htb]
  \includegraphics[width=0.48\textwidth]{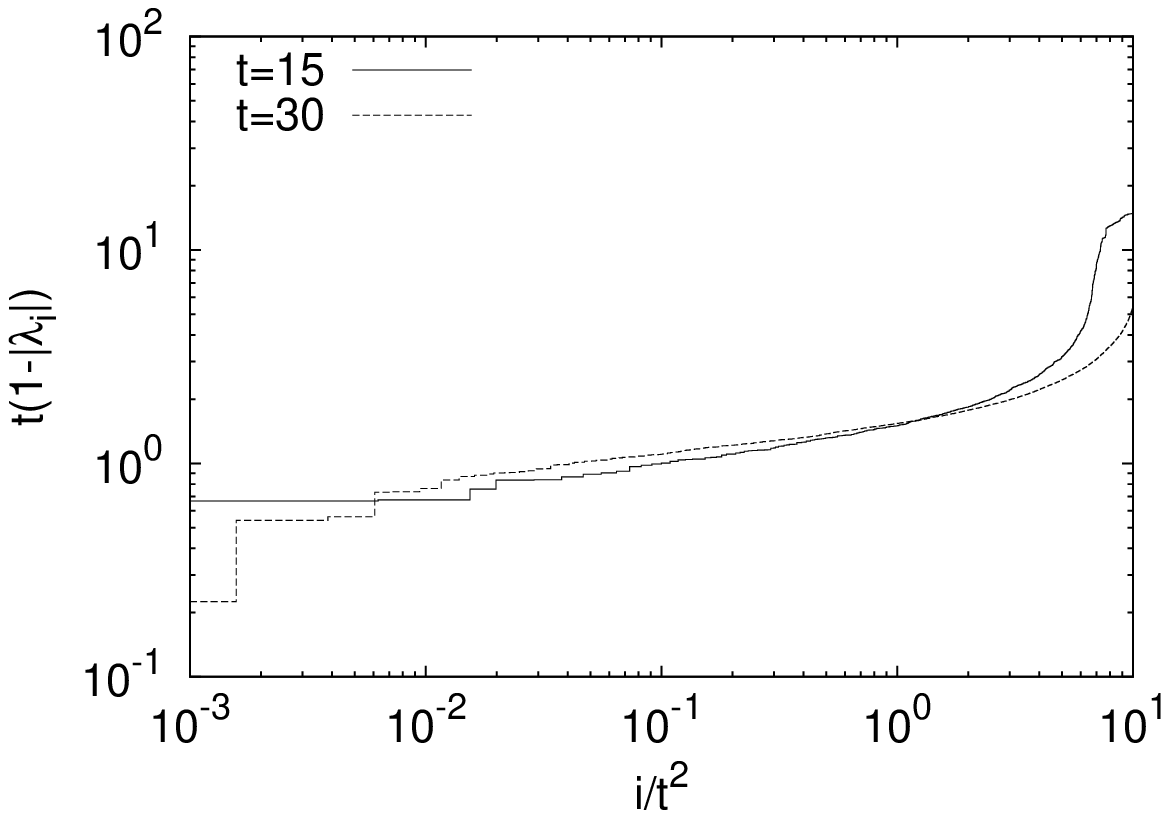}\hskip4pt
  \includegraphics[width=0.51\textwidth]{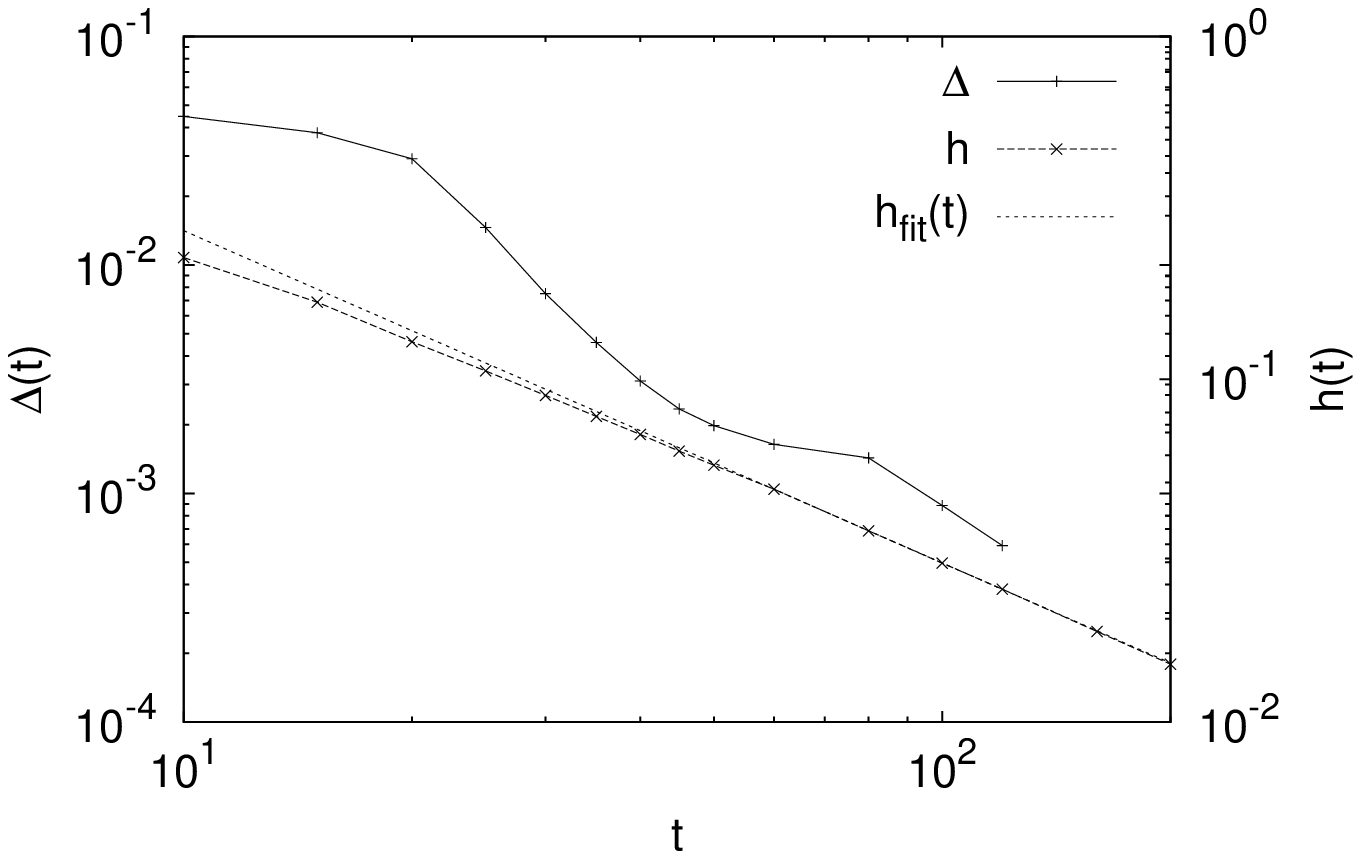}
    \hbox  to \textwidth{\small\hfil(a)\hfil\hfil(b)\hfil}
\caption{The largest amplitude eigenvalue for different times (a) and the spectral gap $\Delta$ together with the Kolmogorov-Sinai entropy $h$ as a function of time (b) in the Markov matrix in polygonal description at $\alpha=1/e$ and $\beta=(\sqrt{5}-1)/2$. The fit to the entropy curve is equal to $h_{\rm fit}(t)=2.529\, t^{-0.969}$.}
\label{fig:tp_gap_ent}
\end{figure}

Let us now study subsequent images of this set and their union up to
integer time $t$:
\beq
  \cD^{(t)} = \bigcup_{n=0}^{t-1} \phi^{(n)}(\cD).
  \label{eq:defD}
\eeq
A set of lines $\cD^{(t)}$ can be considered as a boundary set for a finite set of $N(t)$ polygons, let us denote it by $\cM^{(t)}$. It is obvious that $\phi : \cM^{(t)} \to \cM^{(t+1)}$. For any $t$, a set ${\mathcal M}^{(t)}$ defines a convenient partition of phase space to $N(t)$ elements on which we define a probabilistic process in terms of a Markov transition matrix $M^{(t)}$ of dimension $N(t)$.
Note that $\cD^{(t')} \subseteq \cD^{(t)}$ if $t' \le t$. It is obvious that each element $\cA\in \cM^{(t)}$ is mapped again onto a single element  $\cB \in \cM^{(t)}$, provided this element is bounded only by the lines from the set $\cD^{(t')}$ for $t' < t$. For each such $\cA$ the column of transition matrix reads $M^{(t)}_{\cA',\cA} = \delta_{\cA',\cB}$. Hence $M^{(t)}$ there acts as a simple permutation. Note that since the number of elements grows as $N(t) \propto t^3$, this happens for a majority of cases, namely only for $N'(t) \propto N(t+1)-N(t) \propto t^2$ elements $\cA\in \cM^{(t)}$ it may happen that their boundary also includes lines from the last image $\phi^{(t)}(\cD)$ and hence they are cut and split into several elements of $\cM^{(t)}$. Since the boundary set $\cD^{(t)}$ is a subset of ${\cD}^{(t+1)}$, it is obvious that each such polygon $\cA$ can be cut only once, and hence mapped to at most two polygons, say $\cB_1$ and $\cB_2$. The probabilities are given simply in terms of area measures of image polygons, hence such columns $\cA$ of the transition matrix read
\begin{eqnarray}
  M^{(t)}_{\cA,\cA'}
  &=& \frac{1}{\mu(\cA)}
  \int_{\cA}\dd x\int_{\cA'}\dd y\, \delta(y - \phi(x))\> \nonumber \\
  &=& \frac{1}{\mu(\cA)}
    \left (\delta_{\cA',\cB_1}\mu(\cB_1) +
           \delta_{\cA',\cB_2}\mu(\cB_2) \right)\>.
\label{eq:markovmatrix1}
\end{eqnarray}
To each split of a polygon we associate the relative splitting strength as
\beq
  \eta (\cA) := \min \{\mu(\cB_1), \mu(\cB_2)\}/\mu(\cA)\le \frac{1}{2}\>.
\eeq
We study the cumulative distribution of splitting strengths, denoted by $P_\eta(x)$. In Figure \ref{fig:tp_split}a we plot the $P_\eta(x)$ for different $t$. We notice that small splitting strengths are dominant and the probability density distribution has a square-root singularity at $\eta=0$.
The distribution converges with increasing $t$ to its limiting form, which is well fitted by the phenomenological formula
\beq
  N_\eta(x) = y + a y(1-y)(y + b) + \epsilon\>, \qquad y =\sqrt{2x}\>,
  \label{eq:eta_model}
\eeq
with absolute error $|\epsilon|< 5\cdot10^{-3}$ and constants $a\approx 0.1\pm 0.05$ and $b\approx 2\pm 0.3$. The first two central moments of the limiting distribution are
\beq
  \mean{\eta}= 0.2 \pm 0.002\>,\qquad \sigma_\eta = 0.155 \pm 0.004\>.
\eeq
In the following we are interested in dynamical and statistical properties of the Markov process for finite times $t$ and how they evolve with time $t$. The Markov chain is found to be ergodic and mixing with the spectrum denoted by $\{\lambda_i(t) \in \bC\}_{i=0}^{N(t)-1}$. In Figure \ref{fig:tp_spect_distr} we show two examples of spectra of finite Markov matrices and the distribution of eigenvalue magnitudes $|\lambda_i|$ which shall be later compared to Markov spectra with respect to an alternative partition coding. Further on, we study the evolution of the spectral gap
\beq
  \Delta(t) = 1- |\max_i \{ \lambda_i(t) \neq 1\}|
\eeq
and Kolmogorov-Sinai entropy
\beq
  h(t) = -\sum_{\cA \in \cM^{(t)}} \mu(\cA) \sum_{\cA' \in \cM^{(t)}}
  M^{(t)}_{\cA,\cA'} \log \left(M^{(t)}_{\cA,\cA'}\right).
  \label{eq:entropy}
\eeq
Because the triangle map is non-hyperbolic, we expect that in the limit $t\to\infty$ the Markov matrix does not have a spectral gap and entropy is equal to zero. This is supported by the numerical results shown in Figure \ref{fig:tp_gap_ent}. The entropy as a function of time decreases monotonically. Its asymptotic dependence fits well to an algebraic law
\beq
  h \sim C_{\rm entropy} t^{-\tau}\>,\qquad \textrm{where}\quad\tau = 0.97 \pm 0.03\>.
\eeq
Something similar can not be said for the gap, because its dependence on time is not so simple. Nevertheless numerical results suggest that the gap asymptotically decreases inversely proportional with time as
\beq
  \Delta \sim C_{\rm gap}\, t^{-1}\>.
\eeq

\section{A suitable binary coding}

The description in terms of polygons leaves a lot of freedom in the choice of the initial setup. Nevertheless, the dynamical properties of time asymptotics are shown to be only trivially dependent on the initial partition set. In the following we present a description of dynamics based on binary coding of trajectories. In certain chaotic systems, e.g. baker map \cite{arnold:book:89}, Lozi and H\'enon map (see  \cite{cvitanovic:pra:88}, \cite{dalessandro:ptp:91} ), such description is proved to be asymptotically exact. We shall provide evidence that this is the case for our non-hyperbolic system as well.

\subsection{Topological complexity of the triangle map}

Let $\La=\{\La_0,\La_1\}$ be the binary partition of $\bT^2$ into the half-open rectangles defined as
\beq
  \La_i=[i,{i+1\over 2})\times [0,1), \quad i=0,1\>,
\eeq
and we consider the natural coding associated to it. More formally, we let $\sigma: \bT^2\to \{0,1\}^{\bN}$ be the coding map defined by
\beq
  \sigma(q,p) = \{ \chi_{\La_1}(\phi^t(q,p))\}_{t=0}^\infty\>,
\eeq
where $\chi_\cA(q,p)= (1: (q,p)\in \cA;\, 0 : \textrm{otherwise})$ is the characteristic function of set $\cA$. One may wonder if $\La$ is a {\it generating} partition, in the sense that
$$
  \sigma(q,p) = \sigma (q',p') \Longleftrightarrow (q,p)=(q',p')\>.
$$
Since horizontal segments remain horizontal during the evolution, whereas  arbitrary small vertical segments are linearly stretched while they become asymptotically horizontal, one easily checks that
\beq
  \sigma(q,p) = \sigma (q',p') \Longrightarrow p=p'\>.
\eeq
Therefore, in order that ${\La}$ be generating it suffices that the sequence $(q_t)_{t\geq 0}$ defined in (\ref{eq:trm_dyn_t}) is dense in $\bT$ for all $(q,p)\in \bT^2$. In the non-generic case, $\alpha =0$ and $\beta \in \bR \setminus \bQ$, this property follows at once from a theorem of Weyl according to which if $P(x)$ is a polynomial of degree $m\geq 1$ with real coefficients and at least one of the coefficients is irrational then the sequence $P(n)$ is uniformly distributed mod 1 and thus dense, whereas in the generic case this is directly related to the validity of Conjecture \ref{ergo}.\par
\begin{figure}[!htb]
  \includegraphics[width=0.49\textwidth]{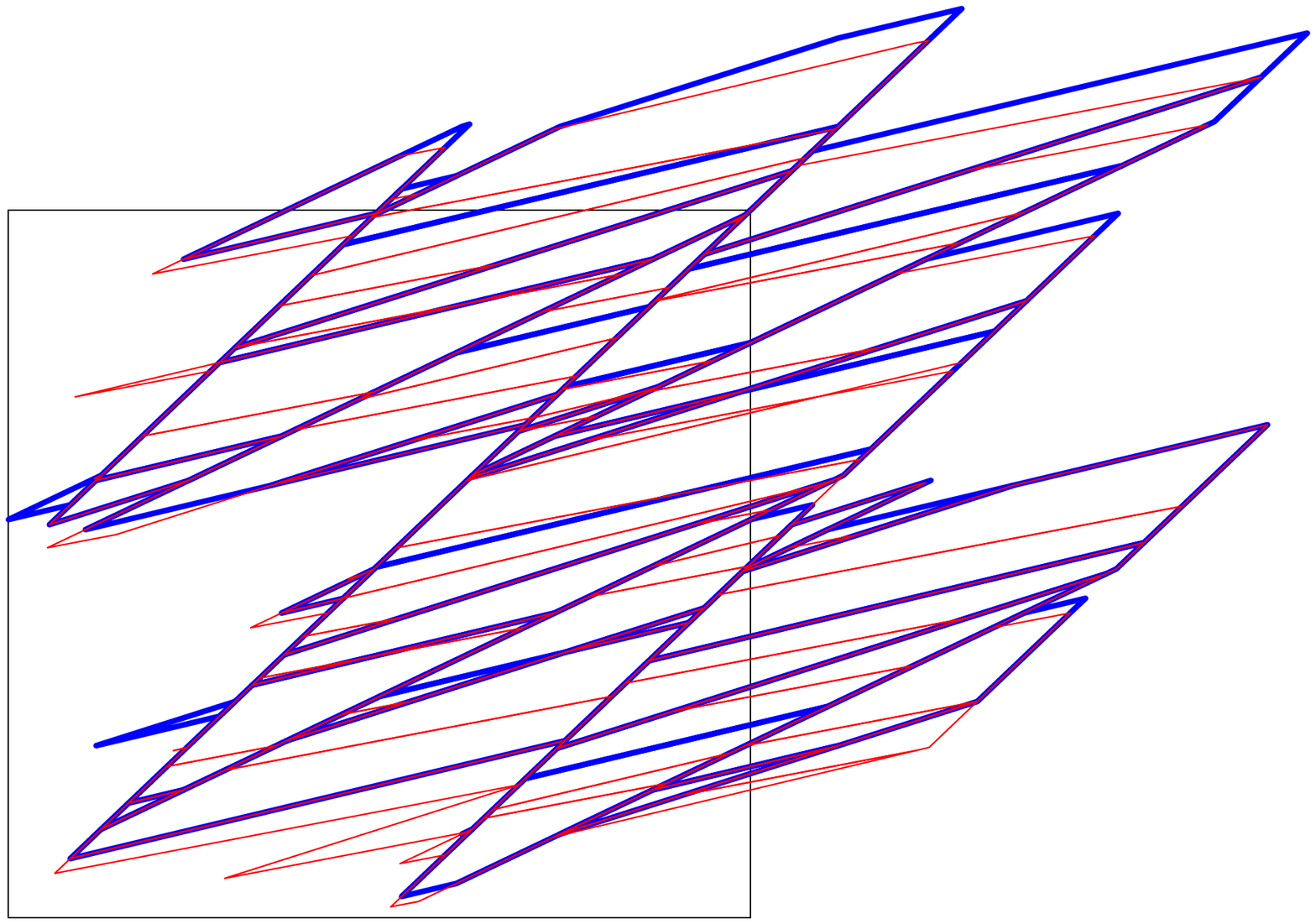}\hskip4pt%
  \includegraphics[width=0.49\textwidth]{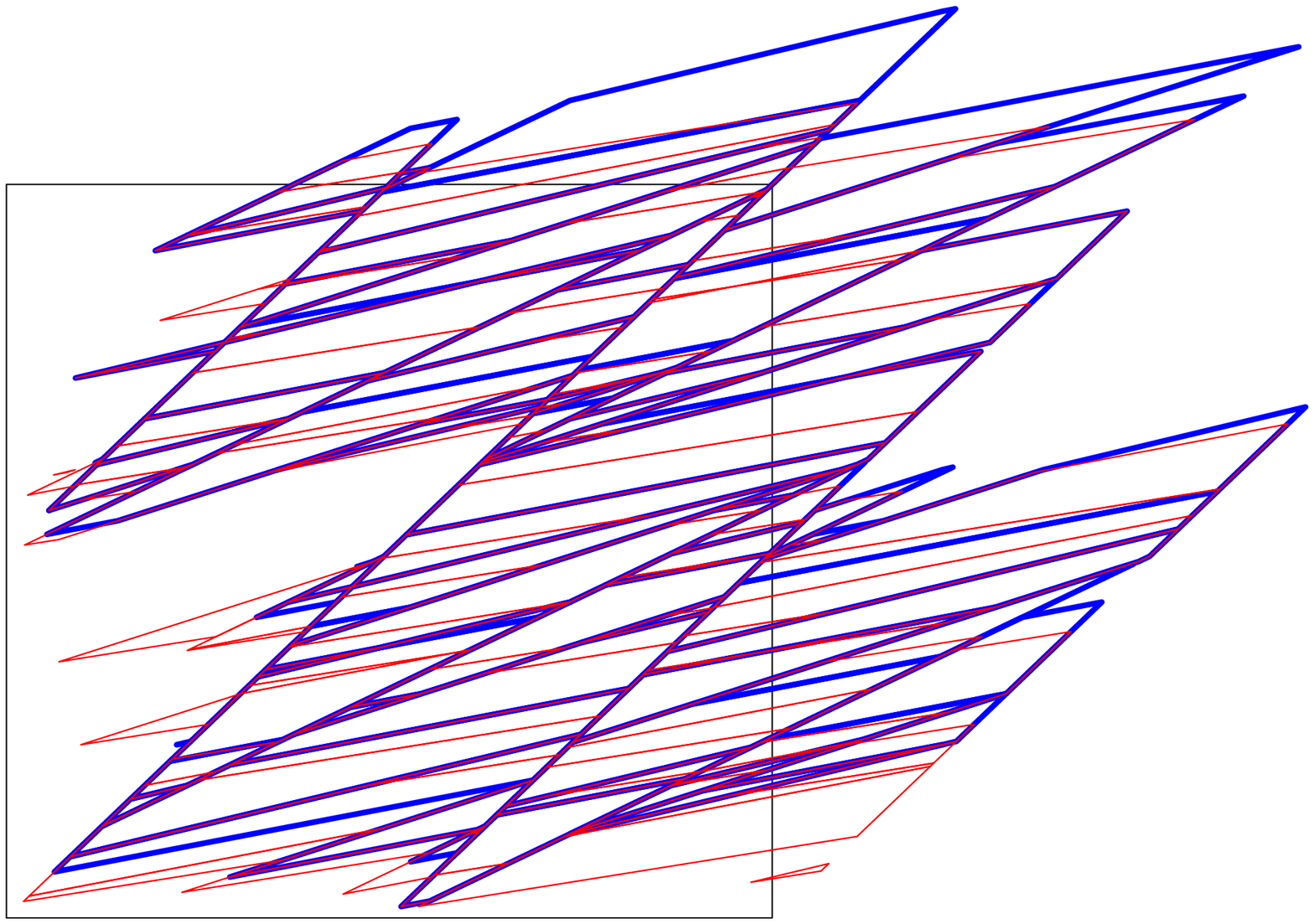}\\
  \includegraphics[width=0.6\textwidth]{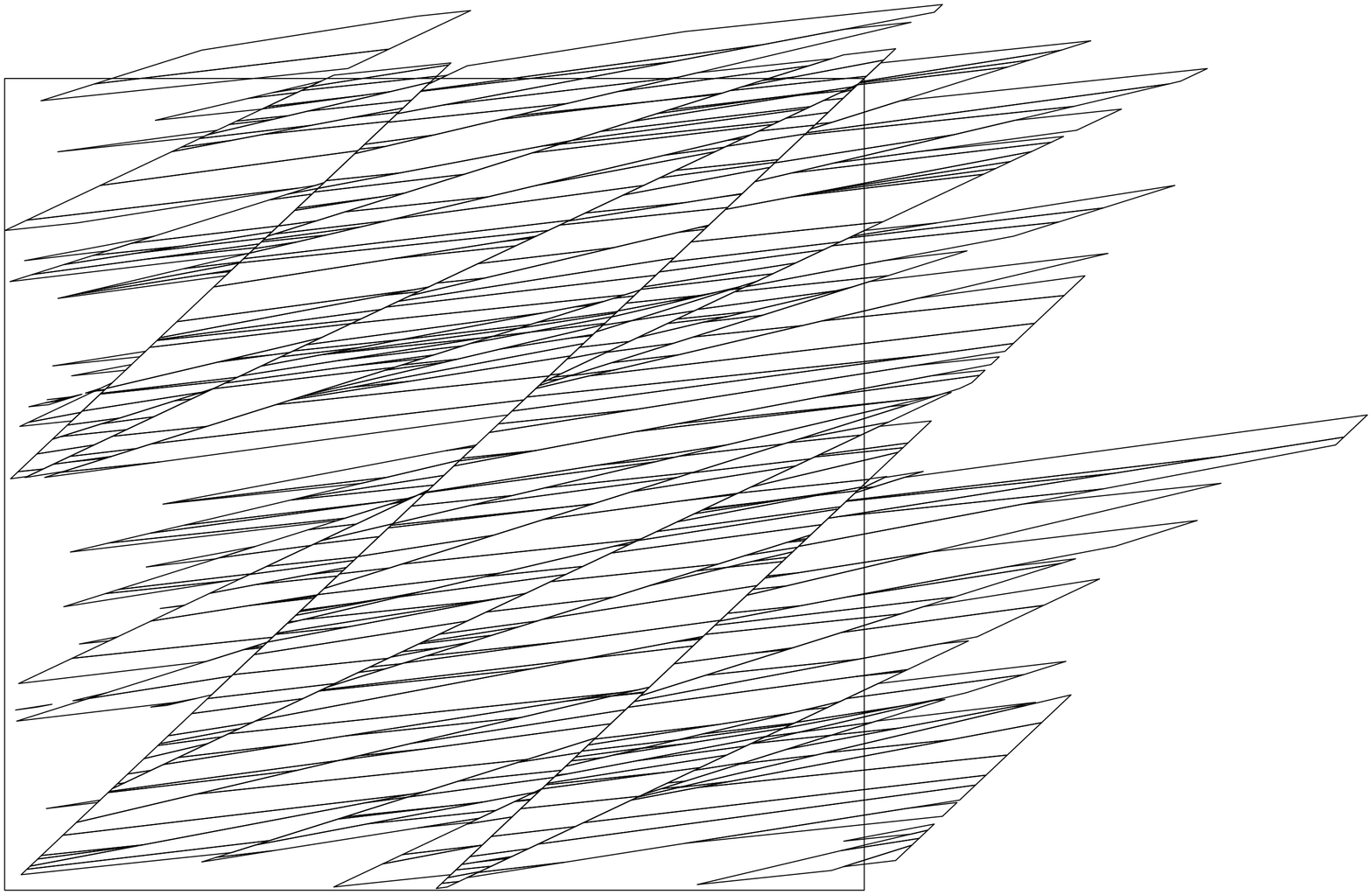}
\caption{In the upper two plots we show polygonal boundaries of the
partitions $\Lambda^{(t)}$ for two subsequent time steps, namely
${\mathcal D}^{(t)}$ (blue thick lines) and on top of them
${\mathcal D}^{(t+1)}$ (red thin lines), for $t=3$ (left plot) and
$t=4$ (right plot). Fundamental square $[0,1)^2$ is indicated by a
thin black line, and periodic boundary conditions are imposed. In the lower plot we show a finer partition, i.e. polygonal boundaries ${\mathcal D}^{(8)}$.} \label{fig:prop_parts}
\end{figure}
\begin{figure}[!htb]
  \includegraphics[width=0.5\textwidth]{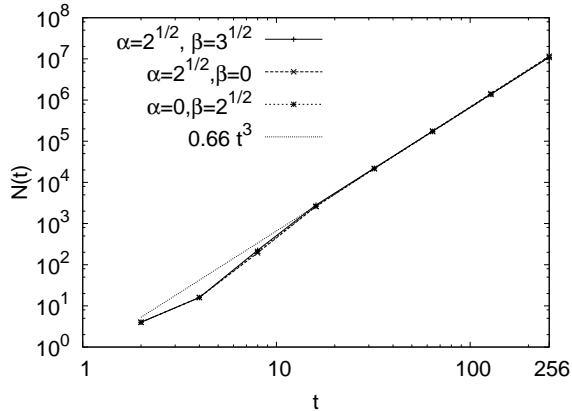}
\caption{The number of partition elements $N(t)$ for various values of parameters $\alpha,\beta$ as indicated in labels calculated by counting the binary codes corresponding to points uniformly sampled over the torus.}
\label{fig:piecebasis}
\end{figure}

We associate to an individual binary code $\omega=(\omega_0, \ldots ,\omega_{t-1})\in\{0,1\}^t$ a set of points $\La_\om$ defined as
\beq
  \La_\om
  = \left\{x\,:\, \phi^k(x)\in\La_{\om_k},\, 0\leq k<t\right\}
  = \bigcap_{k=0}^{t-1} \phi^{-k}\left(\La_{\om_k}\right)\>.
  \label{eq:Lambda_omega}
\eeq
We denote by $L_t \subseteq \{0,1\}^t$ the set of $\phi$-admissible words of length $t$, i.e.
\beq
  L_t=\{ \omega \in \{0,1\}^t \,: \, \La_\om \neq \emptyset\}\>.
\eeq
The partitions $\La^{(t)}$ based on the binary partition $\La$ is then defined as a collection of all non-empty $\La_\om$
\beq
  \La^{(t)} = \left\{\La_\om\right\}_{\omega \in L_t}\>.
\eeq
Note that {\em each} $\La_\om$ is a union of disjoint polygons with
horizontal parallel sides, whereas the other two opposite sides are
in general not parallel and each has a slope not smaller than $1/t$
(see also below). In Figure \ref{fig:prop_parts} snapshots of the
iterated partitions $\La^{(t)}$ for different values of $t$ are
shown, in fact we plot the set of polygonal boundaries of all
elements of $\La^{(t)}$, which is just ${\mathcal D}^{(t)}$ (see equation
(\ref{eq:defD})).
The cardinality of the set $L_t$, that is
\beq
  N(t)= \# \{ \om \in \{0,1\}^t \,: \, \La_\om \neq \emptyset\}\>,
\eeq
is called the {\it topological complexity function} of the pair $(\phi, {\La})$. In positive entropy systems the cardinality of the set is expected to  grow exponentially as
\beq
  N(t) = O(\exp (h\, t))\>,
\eeq
where $h$ is the topological  entropy. In the non-hyperbolic triangle map we are considering here  $h=0$ and therefore we could expect only an algebraic  growth. An interesting problem is that of estimating the asymptotic growth of $N(t)$ as $t\to \infty$ which is up to a constant factor equal to $N(t)$ discussed in Section \ref{sec:poly}. The numerical experiments (see Figure \ref{fig:piecebasis}) lead to the conjecture on existence of a positive constant $C$ independent of the value of irrational parameters $(\alpha, \beta)$ such that
\beq
  N(t)\sim C t^3\>,\qquad  C \doteq 0.66 \pm 0.02 \>.
  \label{eq:p_trm}
\eeq
A slightly weaker result was recently proved in \cite{bonanno:priv:06} and argues that for almost all $\alpha$ and $\beta$ one can find two positive constants $C_i$, $i=1,2$ such that $C_1 \, t^3\leq N(t) \leq C_2 \, t^3$. Moreover, a similar result is known in rational polygonal billiards \cite{cassaigne:aif:02}. Compatible with this law, at time $t$ about $t^2$ elements of the partition get cut into two new pieces, i.e. for about $t^2$ words $(\om_0,\ldots,\om_{t-1})$ we have
\beq
  \La_{\om_0,\ldots,\om_{n-1}}=\La_{\om_0,\ldots,\om_{n-1},0}\bigcup
  \La_{\om_0,\ldots,\om_{n-1},1}
\eeq
with both  $\La_{\om_0,\ldots,\om_{n-1},0}\neq \emptyset$ and $\La_{\om_0,\ldots,\om_{n-1},1} \neq \emptyset$.\par
\begin{figure}[!htb]
  \includegraphics[width=0.35\textwidth]{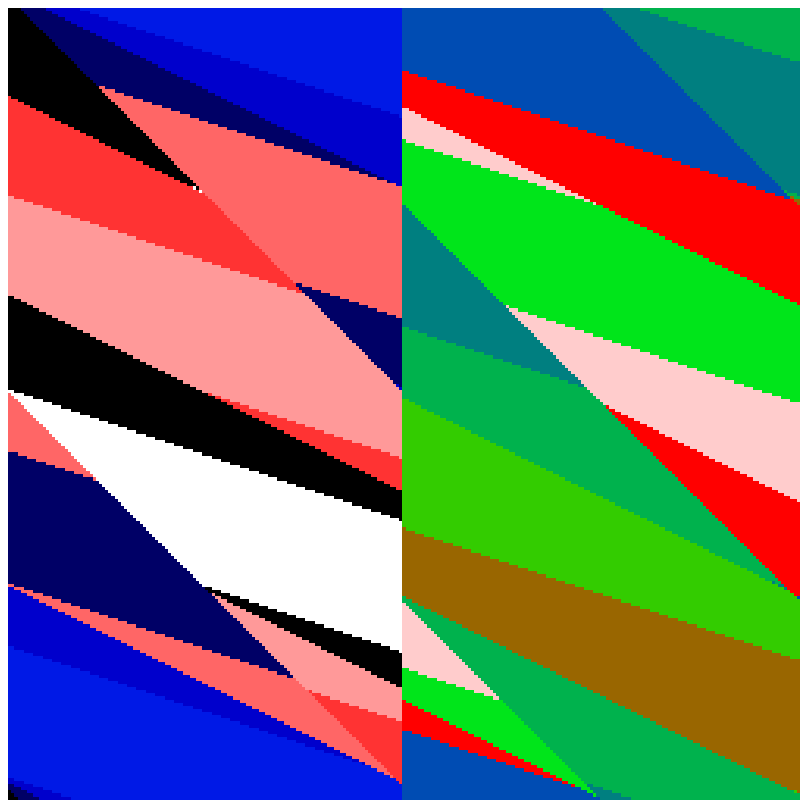}\hspace{1cm}%
  \includegraphics[width=0.5\textwidth]{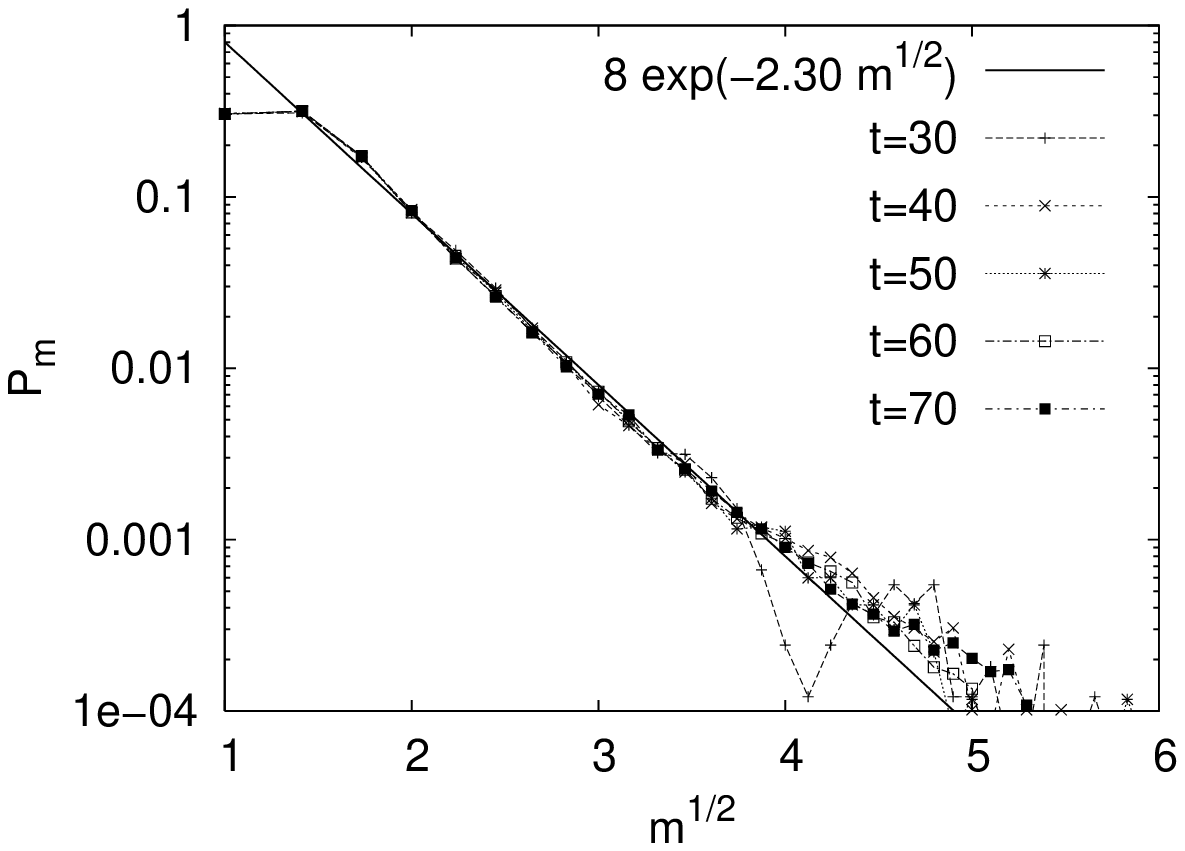}
  \hbox to\textwidth{\small\hfil(a)\hfil\hfil(b)\hfil}
\caption{The elements of the partition of the triangle map at time $t=4$ individually coloured (a) and the distribution $P_m$ of the number of polygons $m$ in the elements of the partition (b) obtained at different times $t$ indicated in the figure, both obtained at $\alpha=1/e$, $\beta=(\sqrt{5}-1)/2$.}
\label{fig:trm_part}
\end{figure}

\begin{figure}[!htb]
  \includegraphics[width=0.49\textwidth]{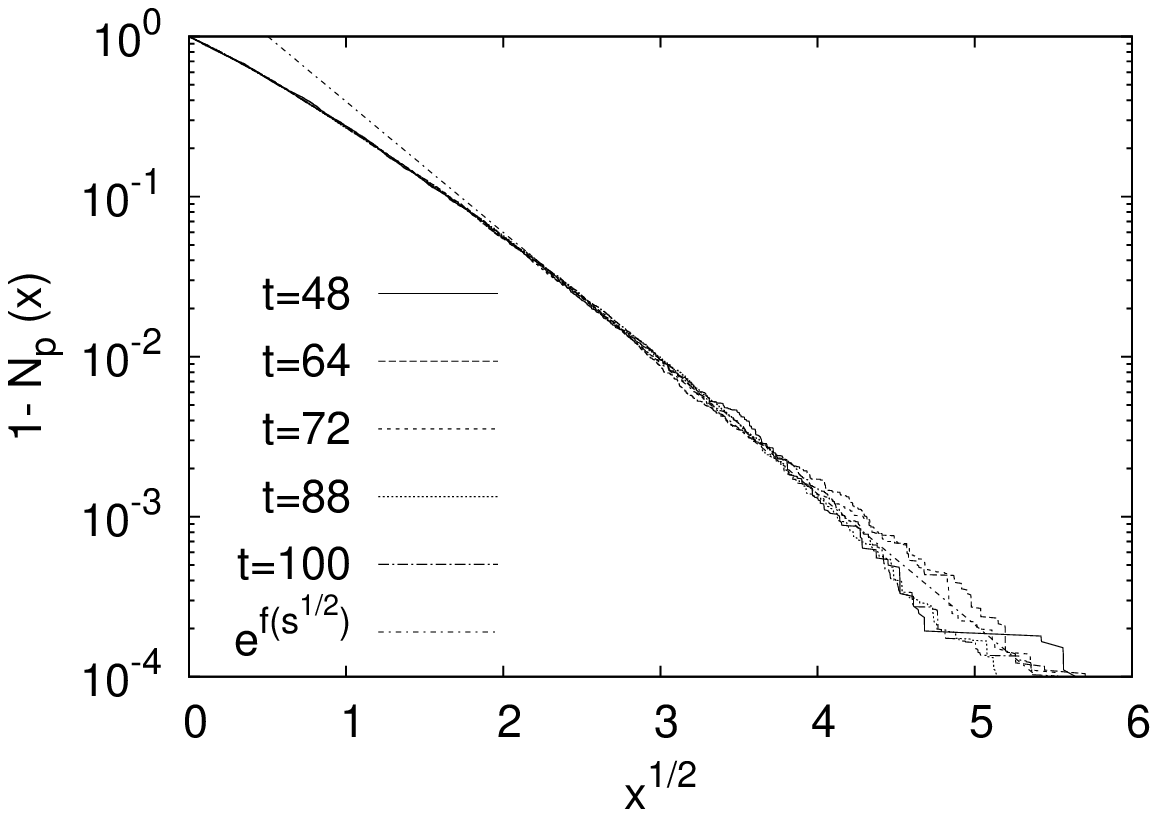}\hskip4pt%
  \includegraphics[width=0.49\textwidth]{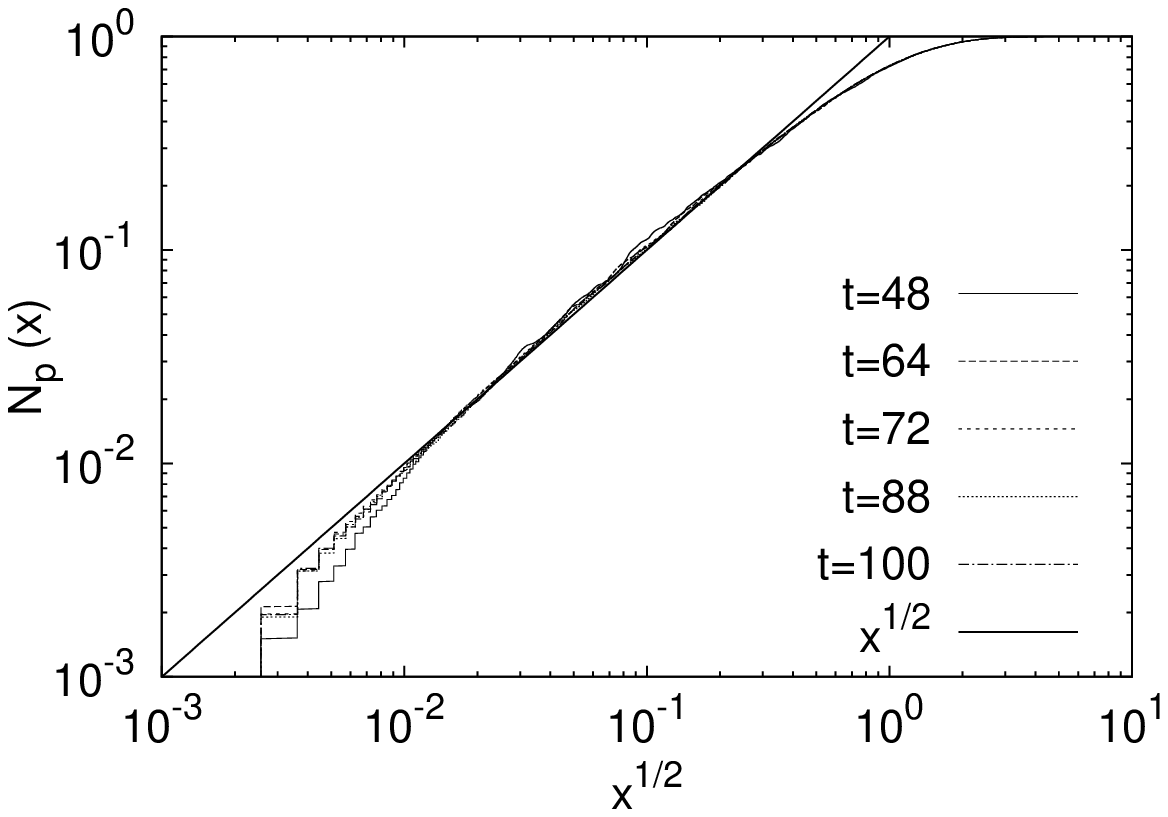}
\caption{The cumulative distribution of rescaled areas of partition elements $N_{\rm p}(x,t)$ in triangle map in two different representations calculated at $\alpha=1/e$ and $\beta=(\sqrt{5}-1)/2$ for different length of trajectories $t$. The fitted curve is $f(x)=-1.88098\, x + 0.945298$.}
\label{fig:trm_cum_size}
\end{figure}

The partition elements $A\in \La^{(t)}$ are of different sizes (measures) $\mu(A)$ and are decomposed of disjoint polygons, see Figure \ref{fig:trm_part}a. We calculate the (heuristic) probability $P_m$ for a partition element being composed of exactly $m \in \bN$ disjoint polygons. An example of the latter is presented in Figure \ref{fig:trm_part}b for the generic case. The typical number of polygons in a partition element is one or two, where the second is slightly more favoured. In addition
we see that the probability $P_m$ is decreasing approximately stretch-exponentially, $\log P_m = -|O(\sqrt{m})|$. The maximal number of polygons in a partition element appears to be growing approximately proportionally with $t$, but this result is not strongly statistically relevant. The presented properties about decomposition of partitions for the generic case do not change significantly in the non-generic case.
The areas of partition elements are decreasing with increasing $t$. We study their statistical properties using the cumulative distribution of areas
\beq
  N_{\rm a}(x,t) = \#\{\om \in L_t\, :\, x > \mu(\La_\om)\}/N(t)\>.
\eeq
and cumulative distribution of rescaled areas $N_{\rm p}(x,t) = N_{\rm a}(N(t)x,t)$. In Figure \ref{fig:trm_cum_size} we present $N_{\rm p}(x,t)$ in the generic case for different $t$.
We see that $N_{\rm p}(x,t)$ converges with increasing $t$ to an unique distribution $N_{\rm p}(x)$ with the following behaviour
\beq
   N_{\rm p}(x) = O(x^{\frac{1}{2}}) \quad \textrm{for}\quad x \ll 1\>,
\eeq
and
\beq
   N_{\rm p}(x)= e^{-|O(x^\frac{1}{2})|}\>\quad \textrm{for}\quad x \gg 1\>.
\eeq
Consequently the average area of partition elements in average decreases as $O(1/N(t))$ with increasing $t$, which is similar to the polygonal description.

\subsection{Transition probability matrix}

We are now again going to approximate the dynamics of the triangle map through a probabilistic model i.e {\it Markov chain}. The long range correlations in this system prevent (statistical) validity of this description on the long time scale. Nevertheless it gives an useful insight into the properties of the dynamics and the slow mixing decay which we numerically detected in the triangle map.\par
\begin{figure}[!htb]
  \includegraphics[width=0.49\textwidth]{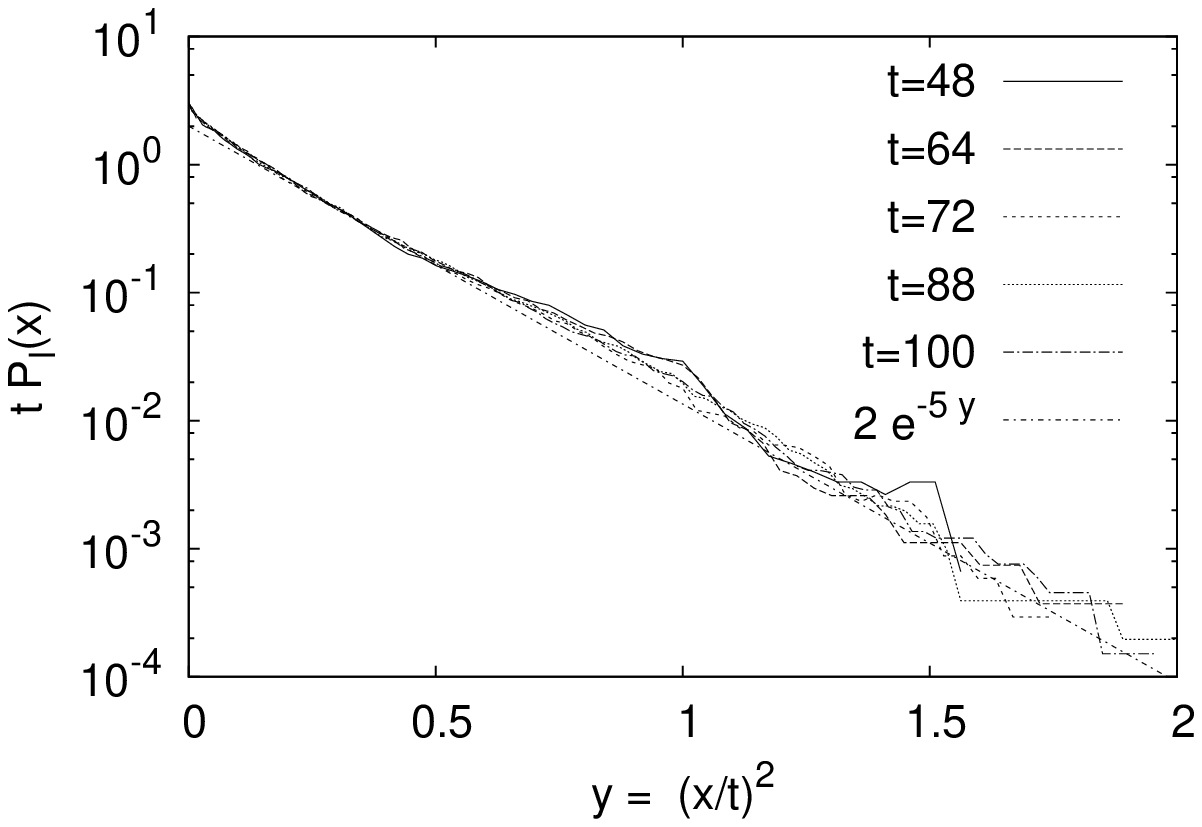}\hskip4pt%
  \includegraphics[width=0.49\textwidth]{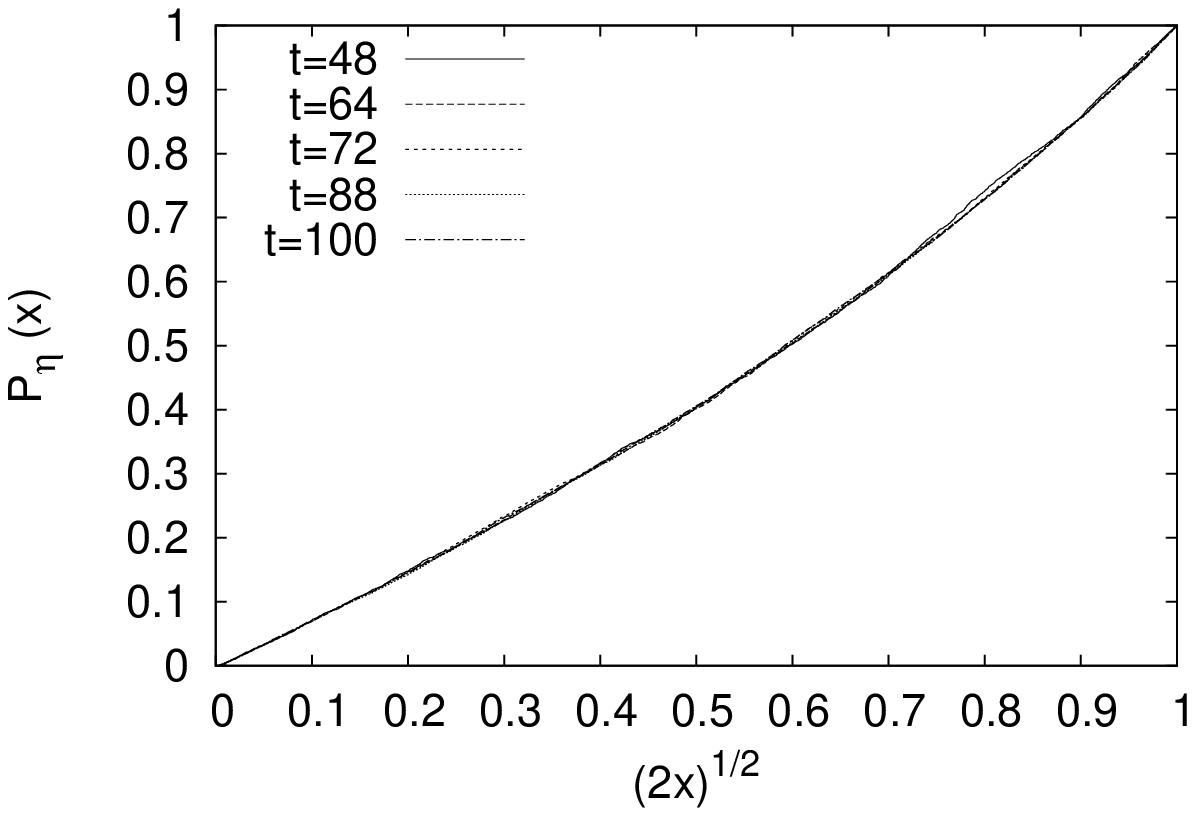}
  \hbox to\textwidth{\small\hfil(a)\hfil\hfil(b)\hfil}
\caption{Distribution of free propagation lengths of the binary partition $\La^{(t)}$ (a) and cumulative distribution of splitting strength $P_\eta (x,t)$ of the Markov matrix (b) calculated at $\alpha=1/e$ and $\beta=(\sqrt{5}-1)/2$ and different times $t$ as indicated in the figure.}
\label{fig:trm_free_split}
\end{figure}
\begin{figure}[!htb]
  \includegraphics[width=0.49\textwidth]{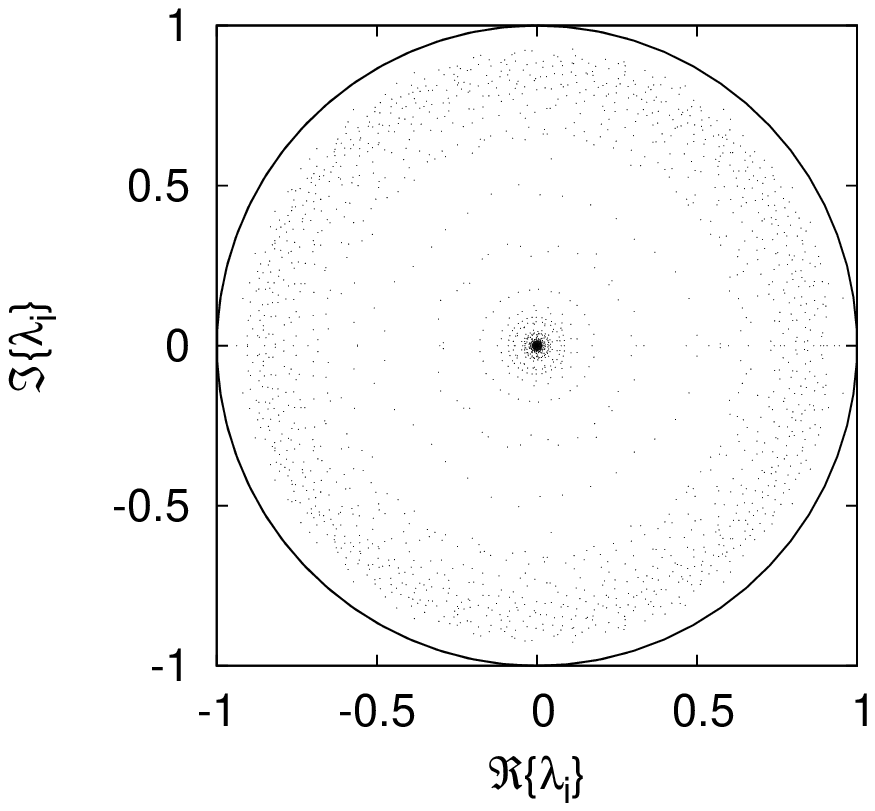}\hskip4pt%
  \includegraphics[width=0.49\textwidth]{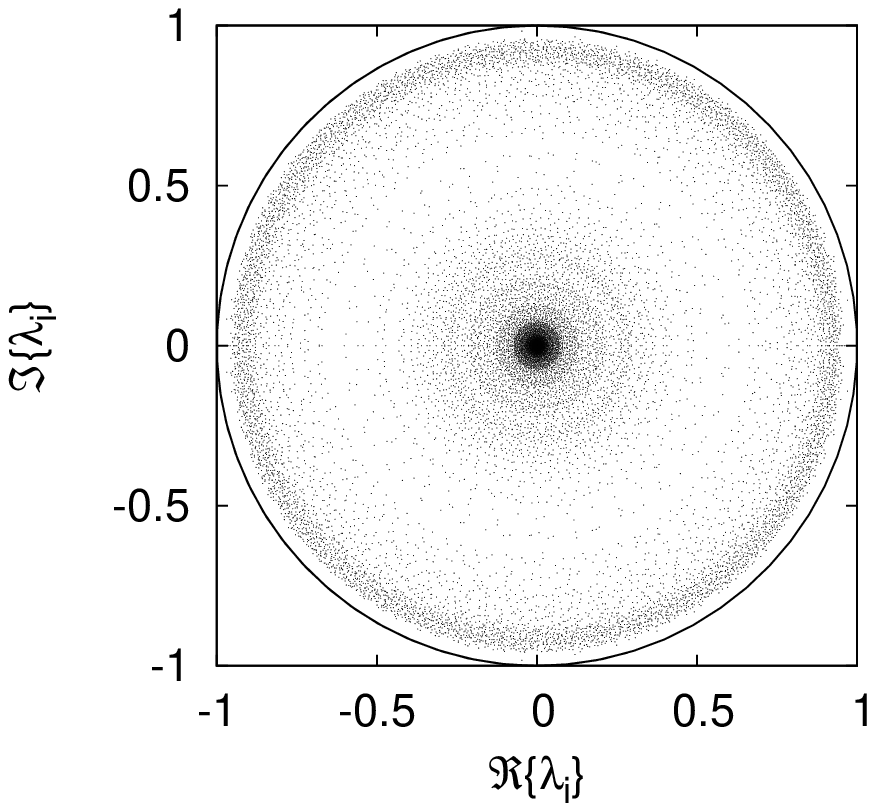}\\
  \includegraphics[width=0.49\textwidth]{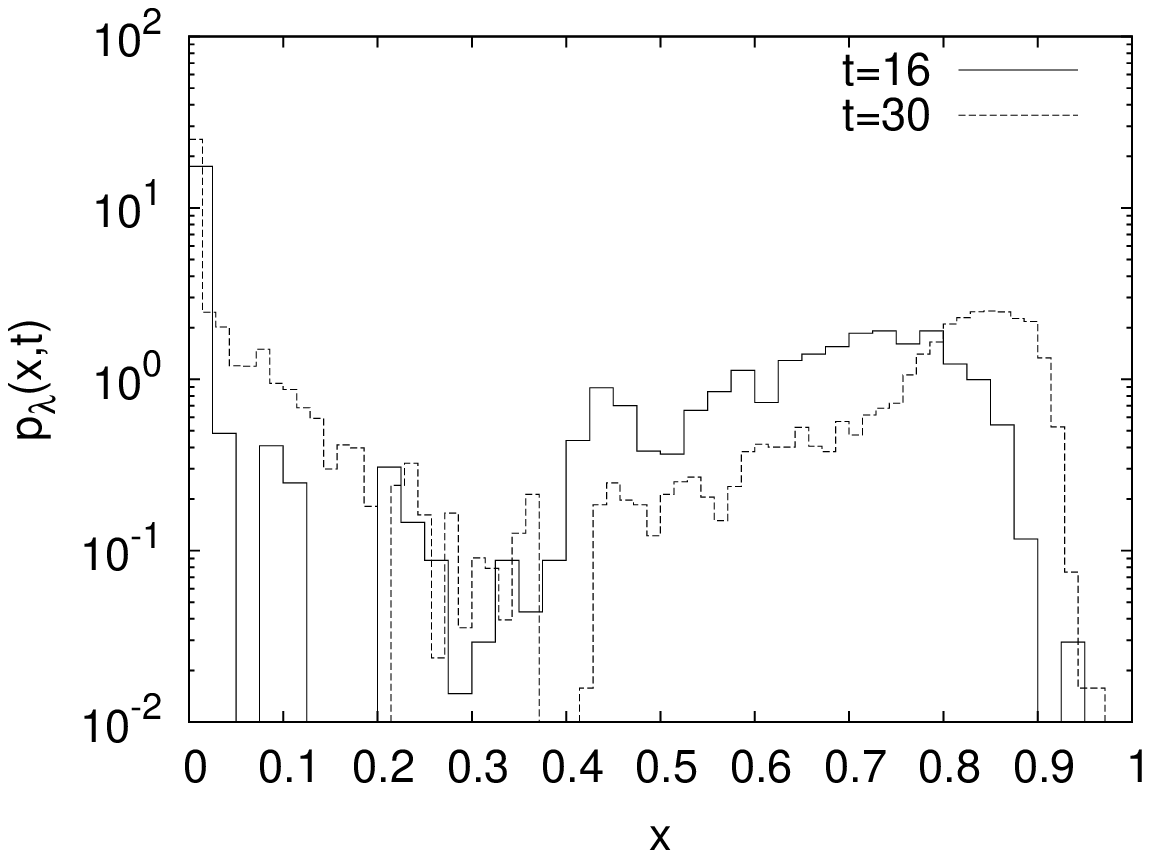}\hskip4pt%
  \includegraphics[width=0.49\textwidth]{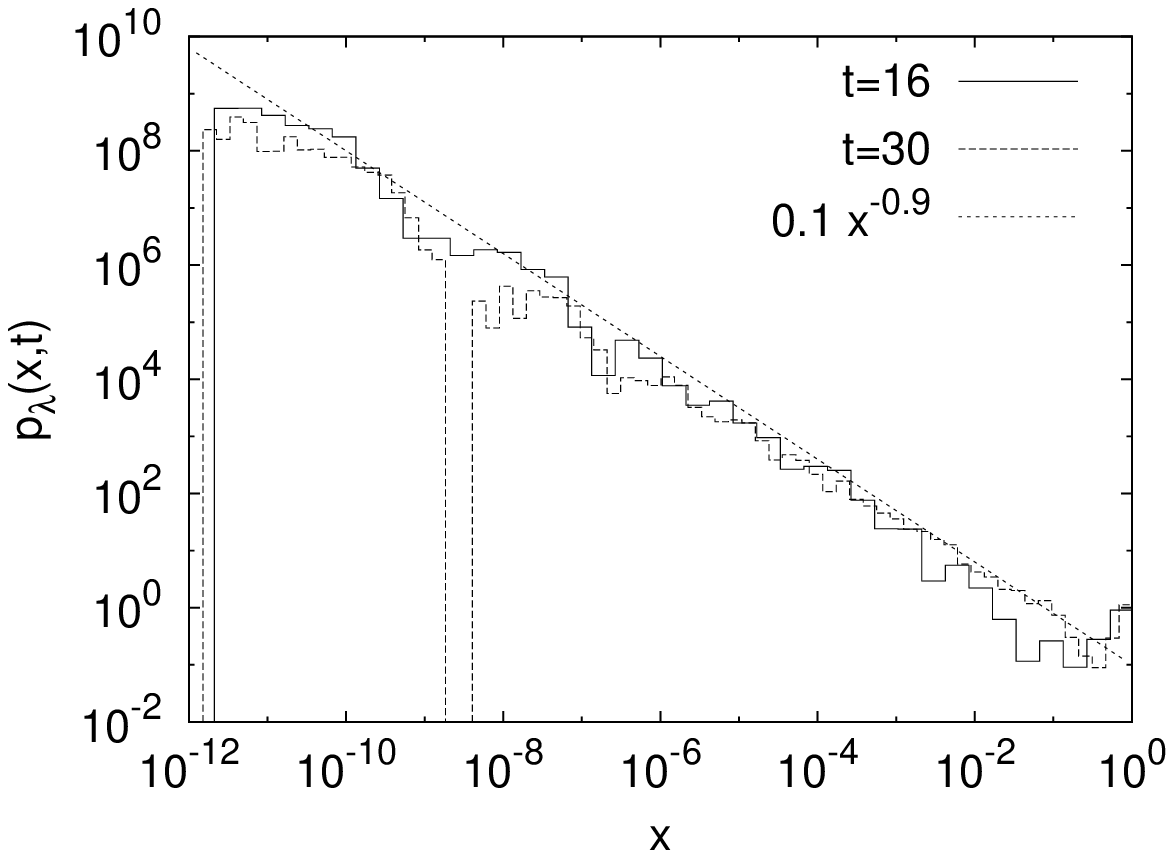}
  \caption{The eigenvalues $\lambda_i$ of the Markov matrix calculated using the binary description at $\alpha=1/e$, $\beta=(\sqrt{5}-1)/2$ for times $t=16,30$ (top) and the distribution of eigenvalue amplitudes in semi-log and log-log presentation (bottom).}
  \label{fig:bin_sp}
\end{figure}
\begin{figure}[!htb]
  \includegraphics[width=0.49\textwidth]{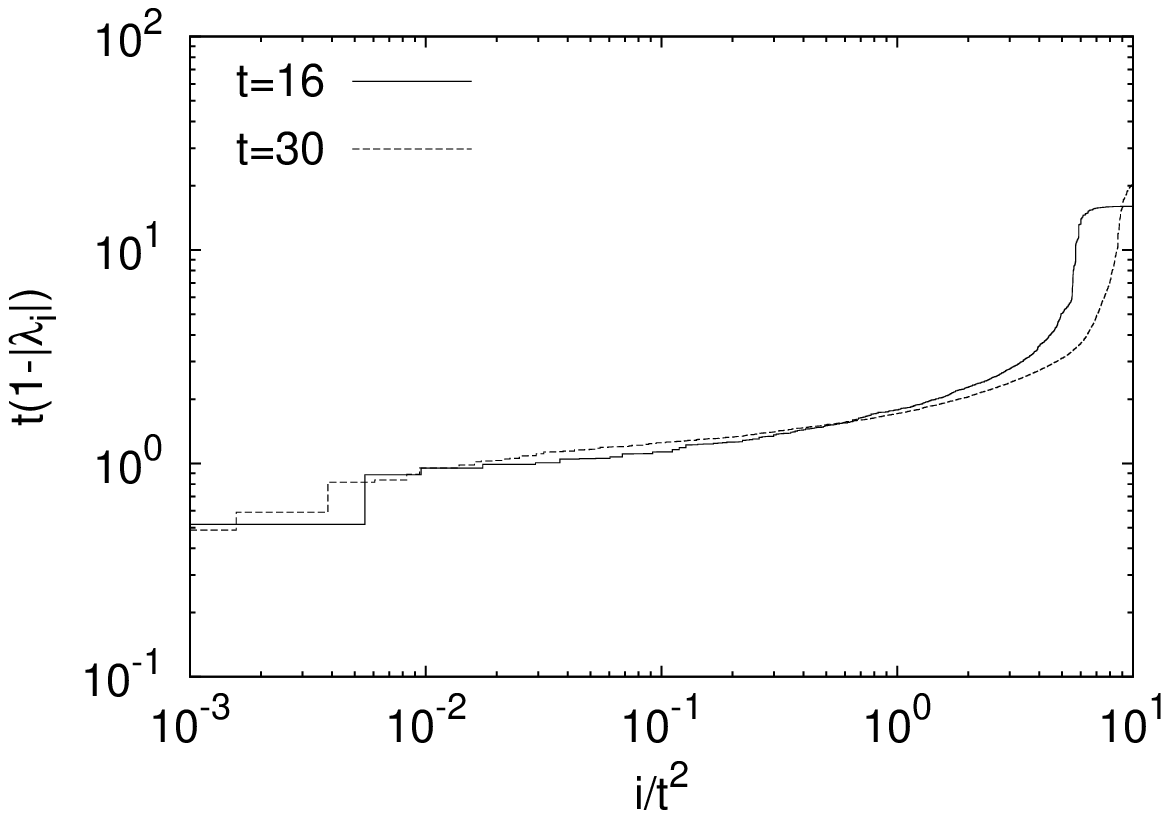}\hskip4pt%
  \includegraphics[width=0.49\textwidth]{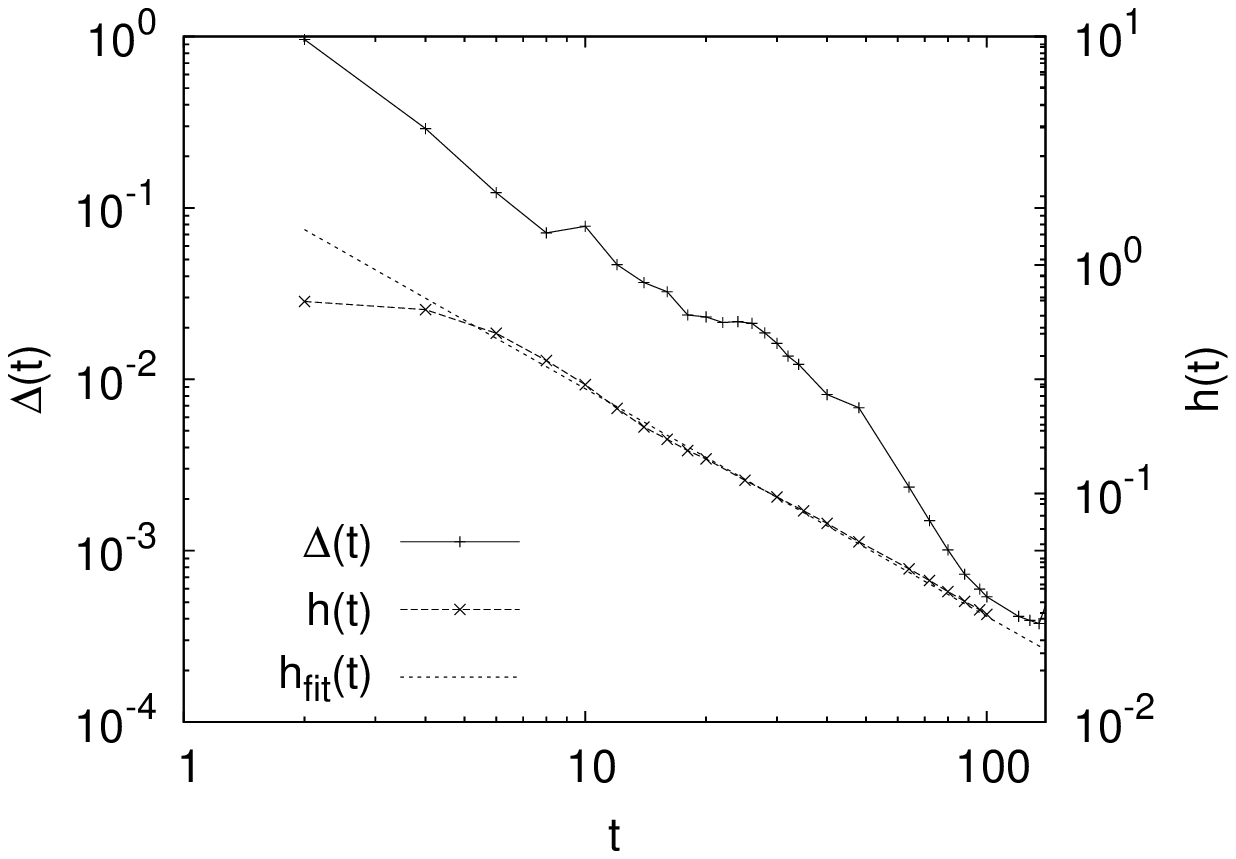}
  \hbox to \textwidth{\small\hfil(a)\hfil\hfil(b)\hfil}
\caption{The largest eigenvalues (a) and the  spectral gap with the entropy of the Markov matrix (b) at parameters $\alpha=1/e$, $\beta=(\sqrt{5}-1)/2$. The fitted curve correspond to $h_{\rm fit}(t)= -0.996524\, t + 1.04747$.}
\label{fig:sp_gap}
\end{figure}

Similar as above (\ref{eq:markovmatrix1}) we associate to the triangle map a Markov process \cite{meyn:book:95} build on transition probabilities
\beq
M_{\omega, \omega'} = {\rm Prob} \left(\phi(x\in \La_\om) \in \La_{\om'} \right)
\eeq
between elements of the partition $\La^{(t)}=\{\La_\omega:\, \omega \in L_t\}$. One has
\beq
  M_{\om, \om'}^{(t)}
  = \frac{1}{\mu(\La_\omega)}
  \int_{\La_\om}\dd x\int_{\La_{\om'}}\dd y\, \delta(y - \phi(x))\>.
\eeq
The transition probabilities determine the Markov matrix $M^{(t)}$ of our process. Because the cardinality of the partition $N(t)$ grows as $O(t^3)$ with increasing $t$, the matrix $M^{(t)}$ is sparse for large $t$ and therefore can be efficiently stored. The sparseness is increasing with increasing $t$. We recognise that the number of non-zero elements in $M^{(t)}$ is equal to
\beq
  \#\{M_{\om, \om'}^{(t)} \neq 0 : \, \om,\om' \in L_t \} = N(t+1)\>.
\eeq
By considering the Markov matrix of the $m$th iterate of the map $\phi$ the number of non-zero elements is $N(t+m)$. Note that a point $x\in \La_\om$ with $\om=(\om_0,\om_1,\ldots,\om_{t-1})$ is mapped to $\phi(x)\in\La_{\om'}$, where $\om'=(\om_1,\om_2,\ldots,\om_{t-1},\epsilon)$ and $\epsilon \in \{0,1\}$. Consequently each column of $M^{(t)}$ has at most two nonzero entries denoted by $\eta_\om \in (0,\frac{1}{2})$ and $1-\eta_\om$. We refer to $\eta_\om$ as {\it the splitting strength} of the element in $\La^{(t)}$ labeled by $\om$ and is defined as
\beq
  \eta_\om := \min_{\om'} M_{\om, \om'}^{(t)}\>.
\eeq
An individual element of the partition $\La^{(t)}$ is mapped by the Markov matrix $M^{(t)}$ usually to another element or its image is split between two elements. We calculate for each element in $\La^{(t)}$ {\it the free propagation length} denoted by $l(e)\>,\; e \in \La^{(t)}$ i.e. how long it can be evolved by $M^{(t)}$ without being split. This time is necessarily finite as the opposite would indicate that the map is non-ergodic. The distributions of free propagation lengths it is defined as
\beq
  P_l(x,t) = \frac{1}{N(t)} \sum_{e\in \La^{(t)}} \delta_{x,l(e)}\>,
\eeq
and shown in Figure \ref{fig:trm_free_split}a for different $t$. Properly scaled free propagation lengths reveal that the distribution $P_l(x,t)$ has an asymptotic form for large $t$
\beq
  P_l(x,t) = \frac{1}{t} P_l\left(\frac{x}{t}\right)\>,\qquad
  P_l(x) = e^{-|O(x^2)|}\>,
  \label{eq:free_prop_scale}
\eeq
We see that the topology of transitions in the Markov matrix strongly suppresses a long free propagation. The average free length scales with time as $\ave{l}\sim \const\,t$, which can be read from (\ref{eq:free_prop_scale}). The splitting strength strongly varies between partitions. This can be seen in Figure \ref{fig:trm_free_split}b, where we plot the cumulative distributions of splitting strength
\beq
  P_\eta (x,t) = \frac{\#\{ \eta_\om \in (0,x]:\om \in L_t \}}
                      {\#\{ \eta_\om \neq 0: \om \in L_t\}}
\eeq
for one set of parameters. We see that the variable $\sqrt{\eta_\om}$ is almost uniformly distributed and that the distribution of $\eta$ converges with increasing $t$ to a form well described by model (\ref{eq:eta_model}) with $a=0.195 \pm 0.005$ and $b=1.42 \pm 0.05$.
It seems that the limiting distribution is common for all generic cases with the first two central moments reading
\beq
  \mean{\eta} = 0.2 \pm 0.005\>,\qquad
  \sigma_\eta = 0.15 \pm 0.01\>.
\eeq
Without detailed analysis we could say that the the distribution of splitting strengths in the binary and polygonal description is almost identical.\\
The eigenvalue spectrum $\Sigma(t) = \{\lambda \in \bC:\,\det(M^{(t)} - \lambda \id)= 0 \}$ of the Markov matrix is confined to the unit circle. Particularly interesting is the spectral gap $\Delta$ between the unit-circle and the second largest eigenvalue. Examples of spectra obtained for different times are shown in Figure \ref{fig:bin_sp}. We notice a persistent eigenvalue $\lambda=1$ corresponding to the invariant distribution. Other eigenvalues inside the unit-circle move with increasing time $t$ toward the unit circle. This can be clearly observed in distribution of eigenvalue amplitudes
\beq
  p_\lambda(x,t)
  = \frac{1}{N(t)} \sum_{\lambda \in \Sigma(t)} \delta (x - |\lambda|)\>,
\eeq
and in particular by the spectral gap $\Delta(t)$ as a function of time $t$. The general form of the distribution of eigenvalues amplitudes do not change with parameters $\alpha$ and $\beta$. An example is shown in \ref{fig:bin_sp}. The concentration of eigenvalues increases near to the unit circle and around the origin of the complex plane. Around the origin the distribution has an algebraic singularity, approximately $p_\lambda(x,t)\sim \const\, x^{-1}$, similarly as in the case of the Koopman operator.
From the Figure \ref{fig:sp_gap}a we can read that the compact "cloud" of largest eigenvalues in average converges to the unit circle as $O(t^{-1})$. The outermost point defines the spectral gap of the system, which is plotted as a function of time in Figure \ref{fig:sp_gap}b. We see that the gap decreases with increasing $t$ algebraically, nevertheless its functional dependence on time $t$ is unclear. The gap in the {\em binary coding} is smaller than in the {\em polygonal description}. This behaviour is generally expected by coarse graining the partition of Markov matrices, which in our case means grouping polygons in binary coded partition elements. The entropy of the Markov matrix in the binary encoding can be written in the following way
\beq
  h(t) = - \sum_{\om \in L_t} \mu(\La_\om)
  \left[\eta_\om \log \eta_\om  + (1-\eta_\om) \log (1-\eta_\om) \right]\>.
  \label{eq:bin_entropy}
\eeq
The entropy is bounded from above as
\beq
  h(t) \le \log(2)\, A_{\rm split} \>,\qquad
  A_{\rm split} =\sum_{\om \in  L_t} H(\eta_\om) \mu(\La_\om) \>,
\eeq
where $A_{\rm split}$ is the total area of elements of the partition $A_{\rm split}$ that are split by the action of the map $\phi$. The number of split partition elements is $N(t+1)-N(t)$ and their total average area is $A_{\rm split}\approx (N(t+1)-N(t))/N(t) \sim \const\, t^{-1}$. Therefore we conclude that $h(t) \sim \const\, t^{-1}$ and this is numerically verified in Figure \ref{fig:sp_gap}b.
We calculate the matrix elements $M_{\om', \om}$ at given $t$ by a Monte-Carlo integration using $N_{\rm mc}$ points uniformly sampled over the torus so that $N(t)/N_{\rm mc} < 10^{-3}$. We empirically find that it suffices to have the relative error of the gap in observed time windows smaller than $5\%$.\par
Next we consider a class of random Markov matrices $\{M\in \bR_+^{n\times n} : \sum_j M_{i,j}=1\}$ of dimension $n \gg 1$, which mimics known statistical properties of the Markov matrices obtained in the generic case of the triangle map system. We are interested in the asymptotics as $n\to\infty$. The individual matrix $M$ in this class is a perturbed permutation matrix defined as
\beq
  M = R\, (\id +  D)\>,\qquad M\in \bR_+^{n\times n}\>,
  \label{eq:markov_model}
\eeq
where $R$ is a permutation matrix \cite{heersterman:book:90} associated to a random indecomposable permutation $(r_i)_{i=0}^{n-1}$, with matrix elements $P_{ij} = \delta_{i,r_j}$, and $D$ represent the perturbation in the expression (\ref{eq:markov_model}) given by
\beq
  D_{ij} = \xi_i (\delta_{i,d_j} -\delta_{i, j})\>,\qquad \xi_i \in [0,1]\>,
\eeq
with $(d_i)_{i=0}^{d-1}$ being a random permutation. There is only $m=\#\{\xi_i\neq 0\} \ll n$ non-zero coefficients $\xi_i$, with an average $\mean{\xi}=\frac{1}{m}\sum_i \xi_i$. We estimate the average gap between eigenvalues and the unit circle in these class of Markov matrices using the standard perturbation method \cite{courant:book:89}. The indecomposible permutation matrix $R$ is similar to down-shift permutation matrix $S$
\beq
  S = O^T\, R\, O\>,\qquad
  S_{i,j} = \delta_{i- j\; {\rm mod}\; n, 0}\>,
\eeq
where $O$ is just a product of transpositions. The matrix $S$ can be diagonalized and its eigensystem is given by
\beq
  S\, F =  F\,\Lambda \>,\quad
  \Lambda= \diag \left(e^{\ii \frac{2\pi}{n} j}\right)_{j=0}^{n-1}\>,\quad
  F_{j,k} = \frac{1}{\sqrt{N}} e^{-\ii \frac{2\pi}{n} jk}\>.
\eeq
Knowing that the eigensystem of the permutation matrix is $R\,V = V\,\Lambda$ with $V = O\, F=[v_i]_{i=0}^{n-1}$, we find that amplitudes of eigenvalues $\lambda_i$ of matrix $M$ in the first order perturbation are
\beq
  |\lambda_i| \doteq \sqrt{1 + 2\Re \{v_i^\dag D v_i\} + |v_i^\dag D v_i|^2}\>,
\eeq
and in the leading order depend only on the coefficients $\xi_i$
\beq
  |\lambda_i| \approx 1 - \frac{1}{n} \sum_j \xi_j
  = 1 - \frac{m}{n} \mean{\xi}\>.
  \label{eq:ruelle_res}
\eeq
Following this formula (\ref{eq:ruelle_res}) the effective spectral gap for the Markov matrices of the triangle map should scale with time asymptotics as
\beq
  \Delta = 1 - \max_i |\lambda_i| \sim \frac{3}{2t}\>,
\eeq
where $n=N(t)\sim C t^3$, $m=N(t+1)-N(t)\sim 3C t^2$ and, because the elements in the rows of matrix $D$ can be equally large we take $\mean{\xi} = \frac{1}{2}$. The results do not give the dependence of the gap in the polygonal and binary encoding of dynamics, but rather a correct time scaling of the distribution of the group of largest eigenvalues in both descriptions.

\section{Stochastic models}

The deterministic model is found to be a hard system to analyse, and in particular to establish rigorous results. In order to avoid some of the most difficult technical problems, but still to obtain intuitively correct understanding of the dynamics of triangle maps we turn to stochastic models. The stochasticity is introduced in the position of the discontinuity of the triangle map (the cut), which is from analytical point of view, the most problematic.

\subsection{Random triangle map}

We introduce a stochastic version of the triangle map, called {\it random triangle map}, by randomising the position of the discontinuity and thereby preserving essential properties of the deterministic case e.g. the non-hyperbolic nature of the dynamics. The random triangle map is defined as the iteration formula
\beqa
  p_{t+1} &=& p_t + \alpha\ \sgn(\xi_t - q_t ) + \beta \quad \mod~1\>,\\
  q_{t+1} &=& q_t + p_{t+1} \quad \mod~1\>,
\eeqa
where $\xi_t \in [0,1]$ is a u.d. random variable. In this system we again introduce the binary coding for a single realisation of the random sequence $\Gamma=\{ \xi_t\}_{t\in\bZ^*}$ as it was introduced for the deterministic variant.\par
\begin{figure}[!htb]
  \includegraphics[width=0.48\textwidth]{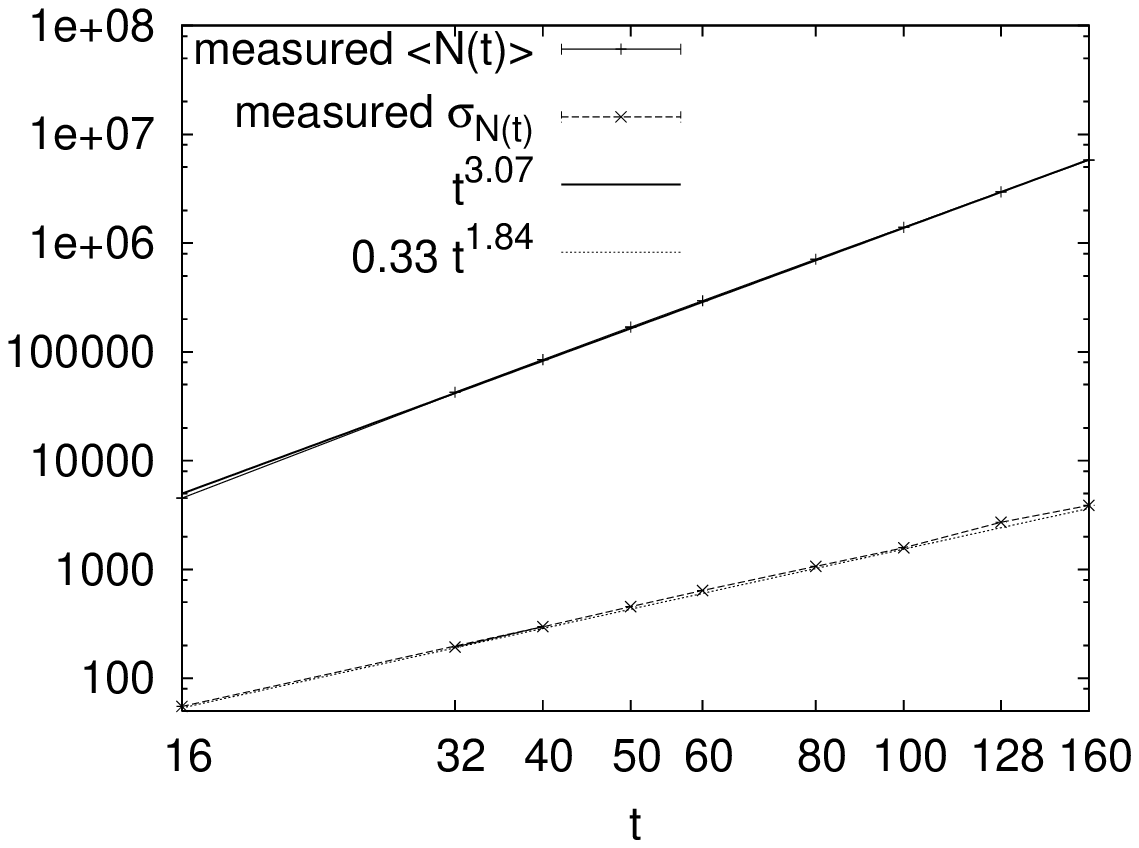}\hskip4pt%
  \includegraphics[width=0.5\textwidth]{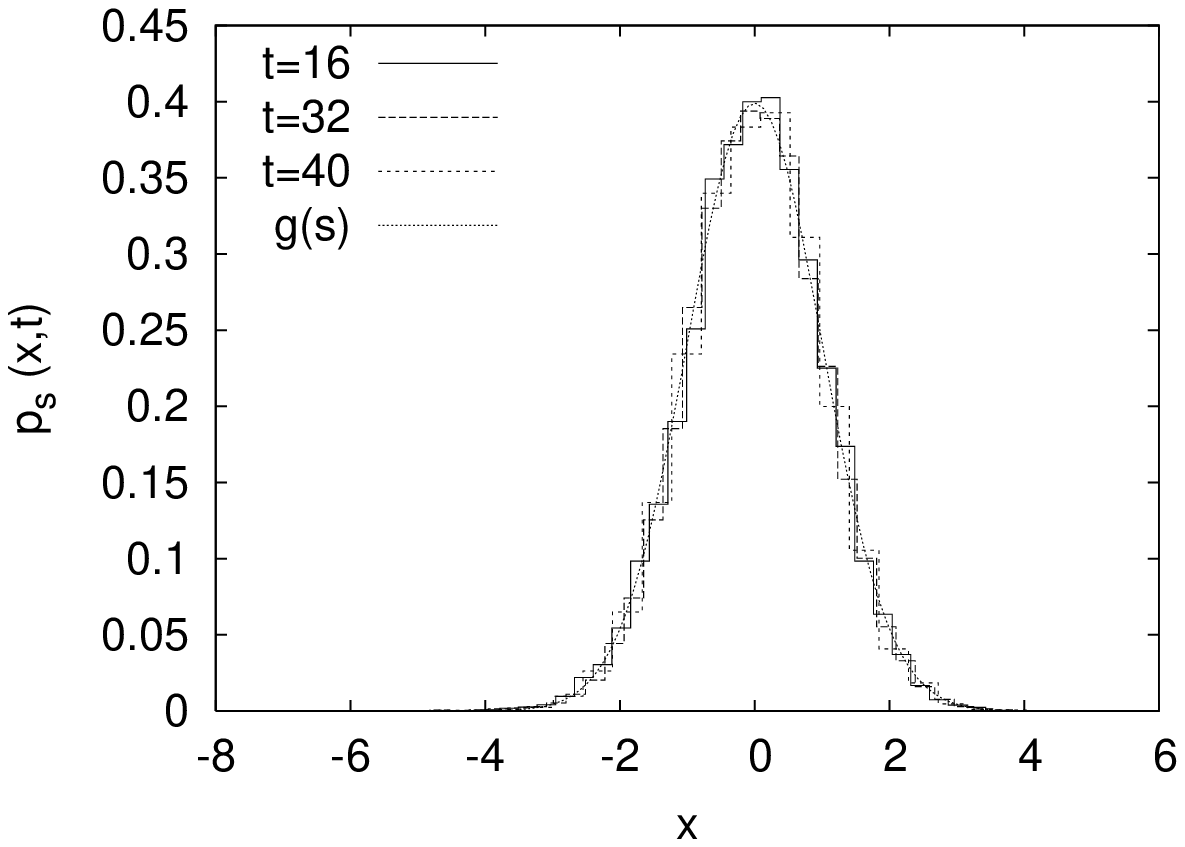}\\
  \hbox to  \textwidth{\small\hfil(a)\hfil\hfil(b)\hfil}
\caption{The average number of the elements of the partitions $\mean{N}(t)$ as a function of time $t$ (a) and distribution of the normalised number of elements of the partitions $P_{\rm s} (x,t)$ (b) in the generic case of random triangle map at $\alpha=1/e$ and $\beta=(\sqrt{5}-1)/2$.}
\label{fig:rtrm_pt}
\end{figure}
\begin{figure}[!htb]
  \includegraphics[width=0.49\textwidth]{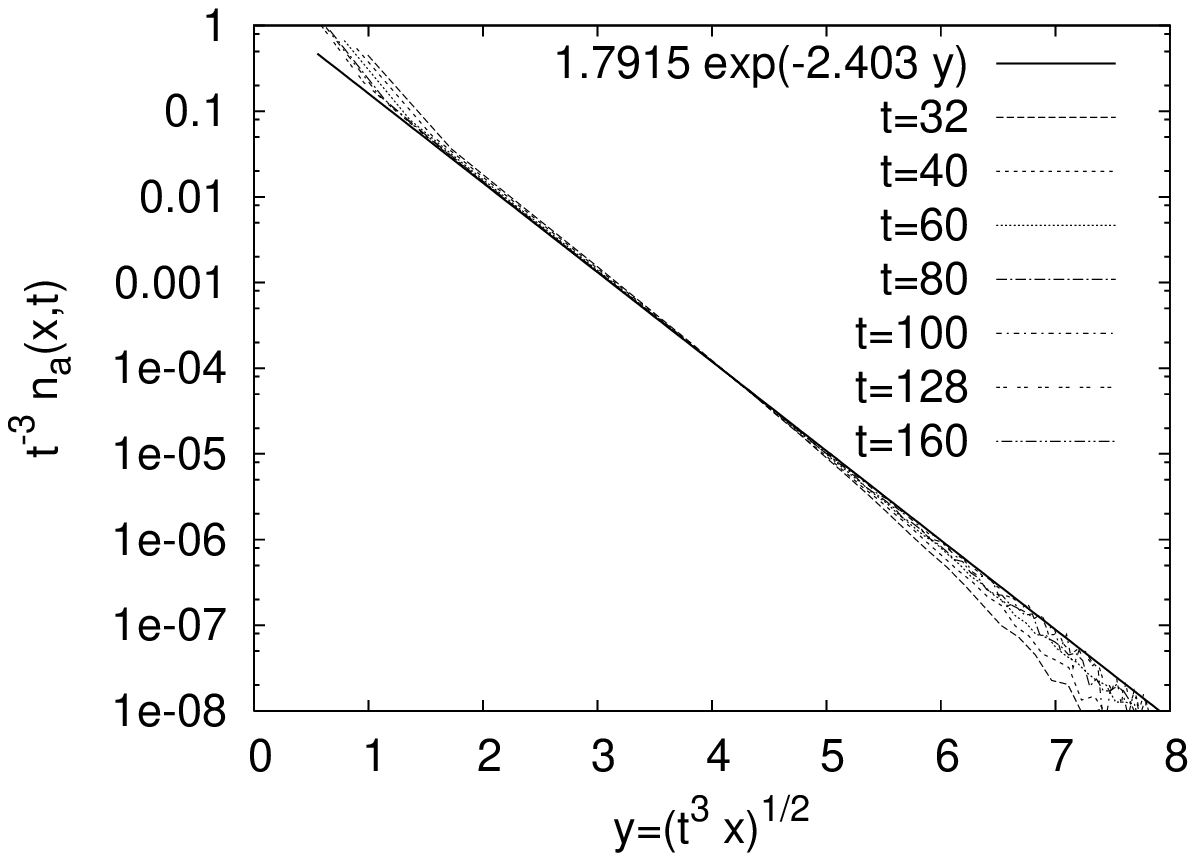}\hskip4pt%
  \includegraphics[width=0.49\textwidth]{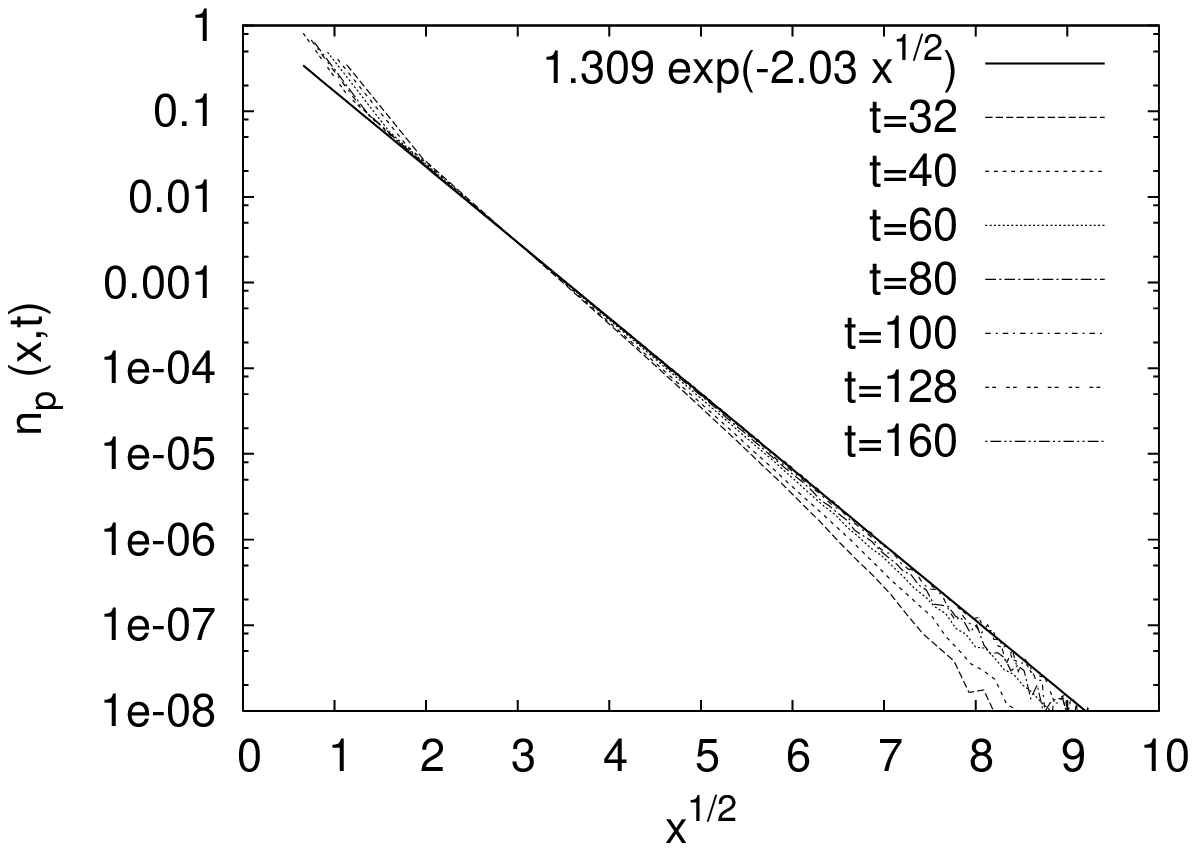}
  \hbox to\textwidth{\small\hfil(a)\hfil\hfil(b)\hfil}
\caption{The average distribution of areas of the elements of partitions $n_{\rm a}(x,t)$ (a) and of rescaled ares of the elements of partitions $n_{\rm p} (x,t)$ (b) in the generic case of triangle map at $\alpha=1/e$ and $\beta=(\sqrt{5}-1)/2$.}
\label{fig:rtrm_part_distr}
\end{figure}

We study the behaviour of partitions $\La^{(t)}(\Gamma)$ with time $t$ across realisations of the random sequence $\Gamma$. The number of partitions $N(t, \Gamma)$ in binary coding at some $t$ varies strongly with the realisation of the cuts $\Gamma$. Nevertheless we find a very simple behaviour of its mean $\mean{N}(t)$ and standard deviation $\sigma^2_N (t)$ in the generic case:
\beqa
  \mean{N}(t) := \ave{N(t,\Gamma)}_\Gamma \sim  C_1 \, t^\rho\>, \quad
  \sigma^2_N (t) := \ave{N(t,\Gamma)^2}_\Gamma - \mean{N}(t)^2 \sim C_2 \,t^{2r}\>,
  \label{eq:p_rtrm}
\eeqa
with $C_{1,2}$ being constants and $\ave{\cdot}_\Gamma$ denoting the average over all realisations of $\Gamma$. The numerically results are depicted in Figure \ref{fig:rtrm_pt}a, where by fitting the data to the power law we find exponents $\rho\doteq 3.07 \pm 0.02$ and $r\doteq 1.84 \pm 0.01$. We may heuristically expect $\rho=3$ as this is somehow the largest power we observe theoretically in sequences of such simple maps.
We conclude that the number of elements of the partitions is asymptotically proportional to $t^3$ similarly as in the deterministic variant. In addition the relative deviation $\sigma_N/\mean{N} = O(t^{r-\rho})$, where $\rho>r$, meaning that presented behaviour for the mean value is asymptotically stable. Furthermore we investigate the distribution of $N(t, \Gamma)$ and we find a very good agreement with the Gaussian distribution for all $t$
\beq
  p_{\rm s}(x,t)
  = \ave{\delta\left(x-s(t,\Gamma)\right)}_\Gamma
  = \frac{1}{\sqrt{2\pi}} \exp\left(-\half x^2\right)\>,
\eeq
for the rescaled number of elements of the partitions $s(t,\Gamma) = (N(t,\Gamma)-\mean{N}(t))/\sigma_N(t)$. This is demonstrated in Figure \ref{fig:rtrm_pt}b. The form of the distribution and the same scaling of $\mean{N}$ and $\sigma_N$ with time $t$ makes us to suspect that $N(t,\Gamma)$ should be, for $t\gg 1$, approximately represented as a sum of $\sim t$ independent stochastic variables. Next we take a look a the average distribution of areas of the elements of the partitions in the binary coding. We study the average distribution of areas of elements of the partitions
\beq
  n_{\rm a}(x, t) =
  \ave{\sum_{\La \in \La^{(t)}(\Gamma)}
  \delta \left(x -\mu(\La)\right)}_\Gamma\>,
\eeq
and average distribution of rescaled areas
\beq
  n_{\rm p} (x, t) =
  \ave{\sum_{\La \in \La^{(t)}(\Gamma)}
  \delta\left(x - N(t,\Gamma)\mu(\La)\right)}_\Gamma\>.
\eeq
over the set of partitions. These two average distributions are not equivalent due to changes to the number elements of the partitions between different realisations of the cuts $\Gamma$. In generic case shown in Figure \ref{fig:rtrm_part_distr} we find that these distributions have similar functional dependence, which seems to have a simple asymptotics at large times reading
\beq
  \lim_{t\to\infty} t^{-3} n_{\rm p} (t^3 x, t)\>,\, n_{\rm a} (x, t)
    \approx \exp(-|O(x^\frac{1}{2})|)\>,
\eeq
The obtained dependence coincides with the dependence in the deterministic case using the binary coding, but due to the computational complexity of the problem we cannot establish the nature of the decay more precisely.\par
\begin{figure}[!htb]
%from olimp, average over m = 4000 using extrapolation of grids Ng=500 and 1000
  \includegraphics[width=0.49\textwidth]{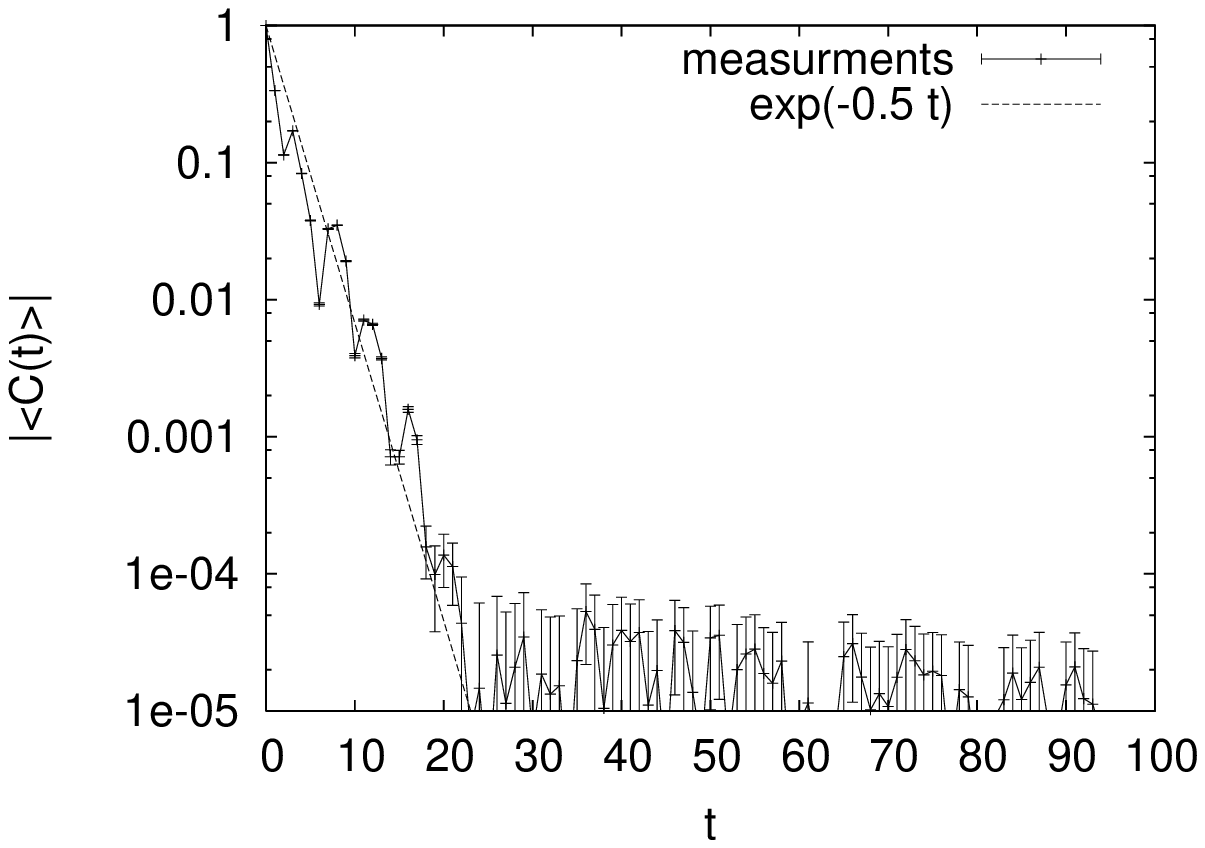}\hskip4pt%
%from olimp, average over m = 20180 using extrapolation of grids Ng=1000 and 2000
  \includegraphics[width=0.49\textwidth]{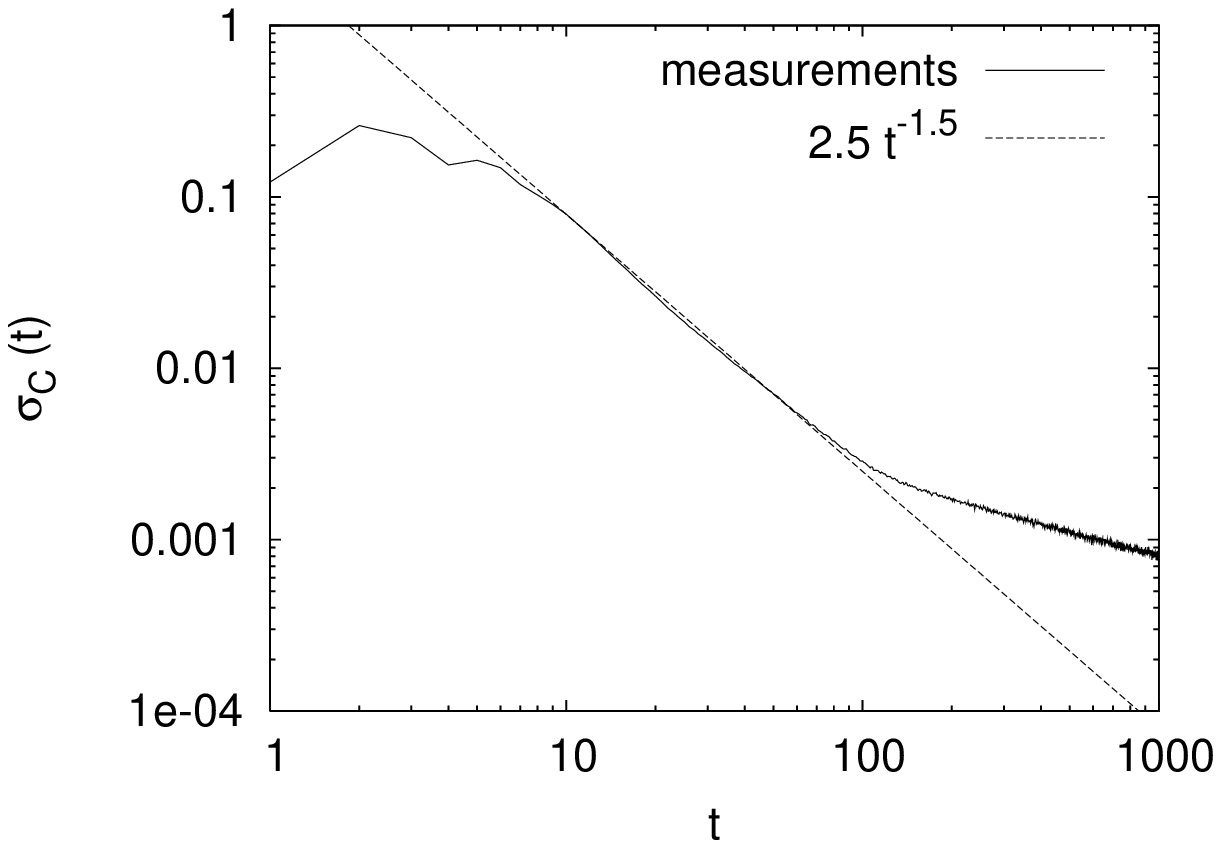}\\
%from asgard, average over m =11216 using Np=10^7 points
  \includegraphics[width=0.49\textwidth]{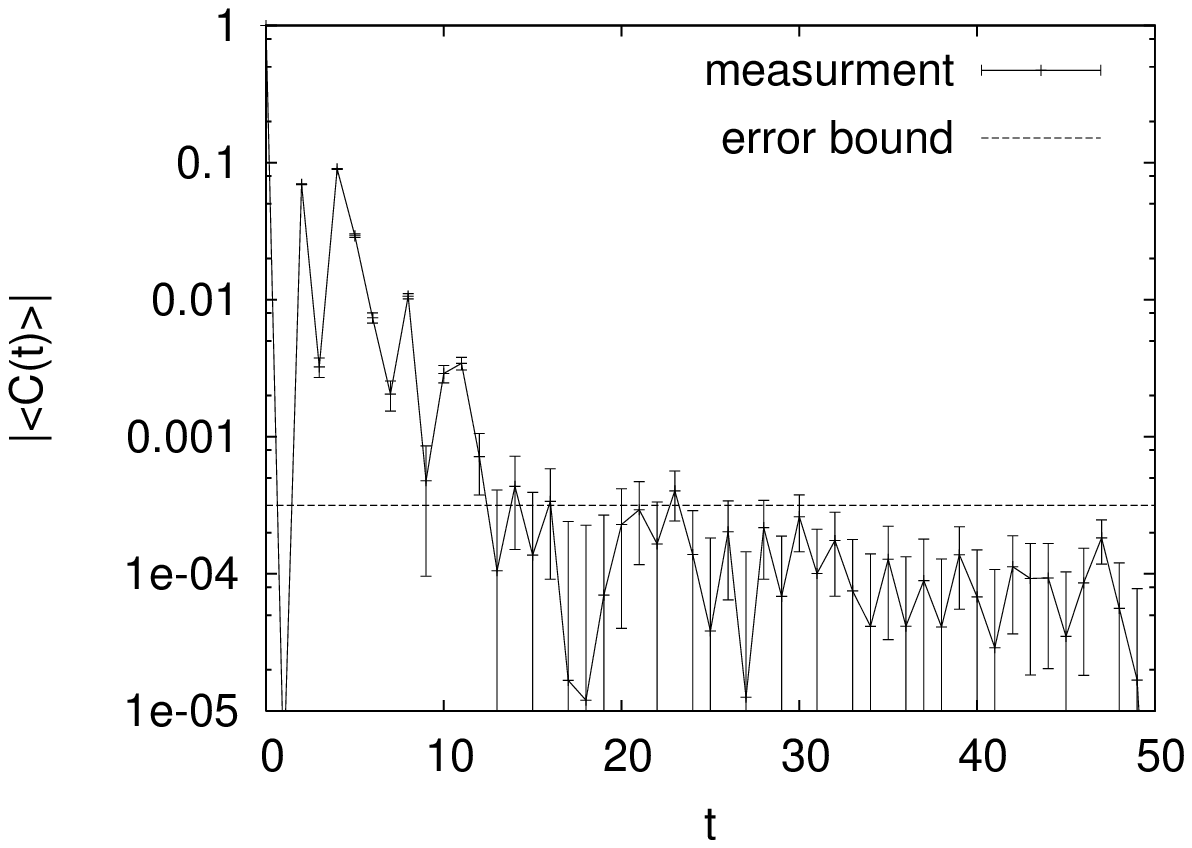}\hskip4pt%
  \includegraphics[width=0.49\textwidth]{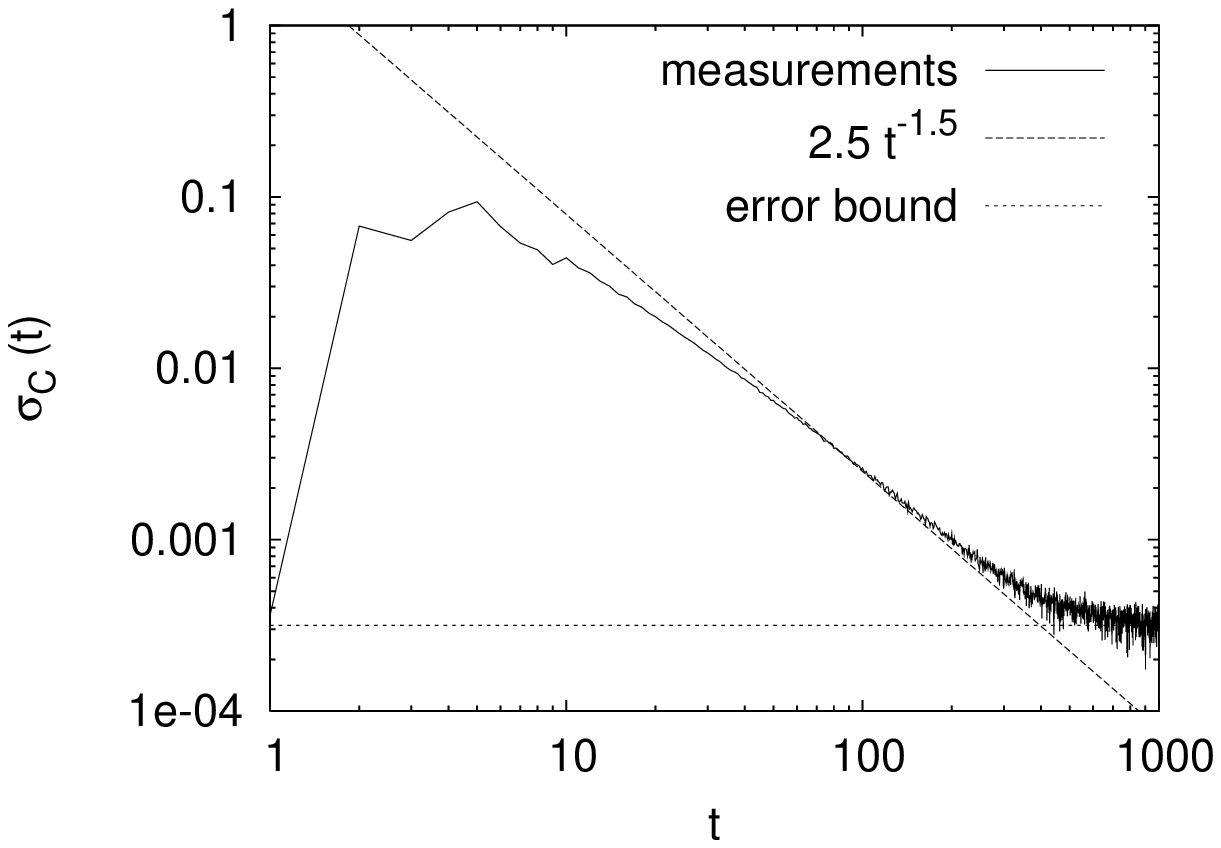}
  \hbox to \textwidth{\small\hfil(a)\hfil\hfil(b)\hfil}
\caption{The average correlation $\mean{C}(t)$ (a) and its standard deviation $\sigma_C(t)$ (b) for observables $f(q,p)= \sin(2\pi q) + \sin(2\pi p)$ (top) and $f(q,p)=\theta(q)$ (bottom) in random triangle model for $\alpha=1/e$ and $\beta=(\sqrt{5}-1)/2$. The horizontal line in figures at the bottom represents the value of the statistical error of the method.}
\label{fig:C_gen_rtrm}
\end{figure}

\begin{figure}[!htb]
%calcolo3, average over m = 1600 using extrapolation of grids Ng=500 and 1000
  \includegraphics[width=0.49\textwidth]{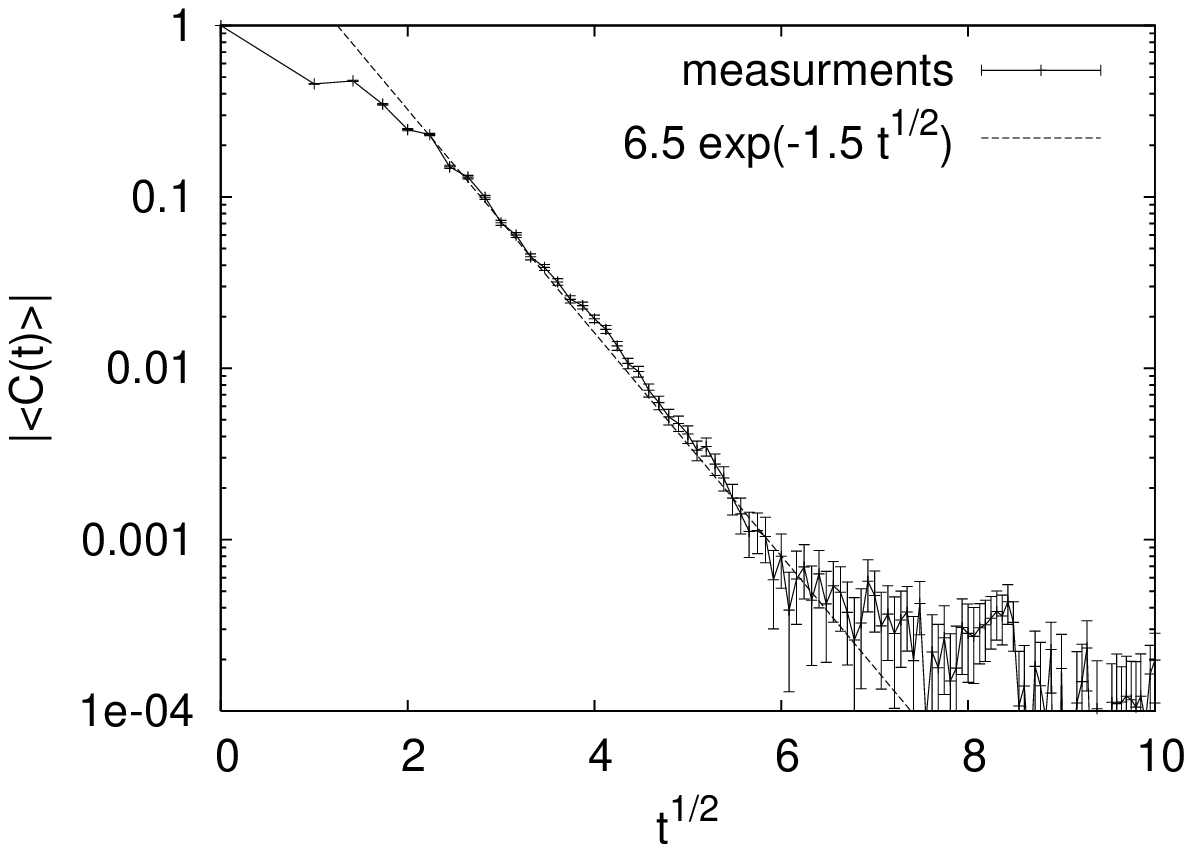}\hskip4pt%
  \includegraphics[width=0.49\textwidth]{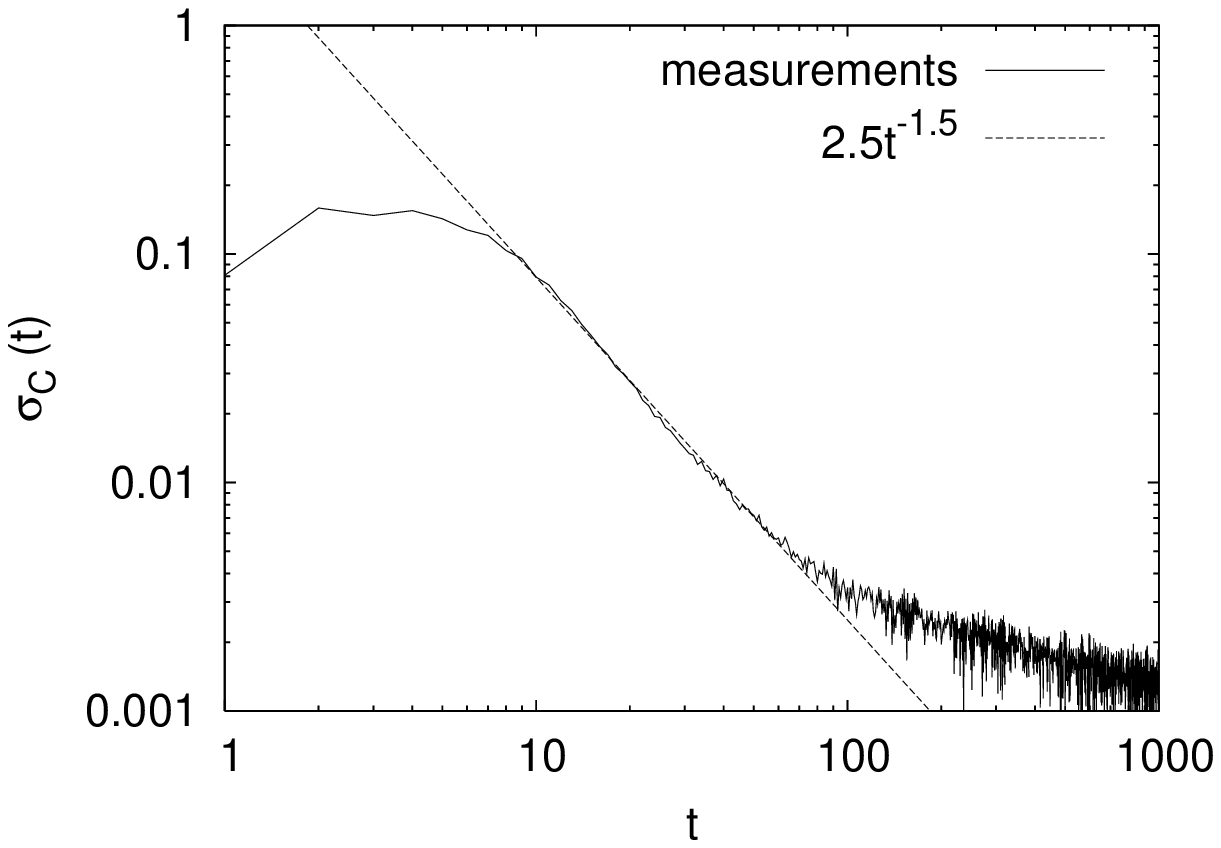}
\hbox to \textwidth{\small\hfil(a)\hfil\hfil(b)\hfil}
\caption{The average correlation $\mean{C}(t)$ (a) and its standard deviation $\sigma_C(t)$ (b) for observables $f(q,p)=\theta(q)$ in random triangle model at $\alpha=1/e$ and $\beta=0$ calculated by extrapolating results obtained by applying Simpson quadrature rule with $N_x=N_y = 500$ and averaging over 1600 realisations.}
\label{fig:C_nongen_rtrm}
\end{figure}

Next we examine the mixing property in the stochastic model by studying the auto-correlation function of an observable $f:\bT^2 \to \bR$ with zero phase-space integral $\int_{\bT^2}\dd q\dd p\, f(q,p) = 0$ reading
\beq
  C(t, \Gamma)[f]
  = \int_{\bT^2} \dd q\, \dd p\, f(q, p) f(q_t(q,p,\Gamma), p_t(q,p,\Gamma))\>,
\eeq
where a trajectory with a given $\Gamma$ and initial point $(q,p)$ is denoted by \linebreak $(q_t(q,p,\Gamma), p_t(q,p,\Gamma))$ for $t\in\bZ^*$. In particular, we focus on the first two central moments of the correlations w.r.t. the average over $\Gamma$ i.e. $\mean{C}(t):= \ave{C(t, \Gamma)}_\Gamma$ and $\sigma_C^2(t) := \ave{C(t,\Gamma)^2}_\Gamma-{\mean{C}(t)}^2$. We performed numerical calculations of the two moments using different parameters $(\alpha,\beta)$. The results for the generic case of parameters are shown in Figure \ref{fig:C_gen_rtrm}, whereas for the non-generic case in Figure \ref{fig:C_nongen_rtrm}.
In Figures \ref{fig:C_gen_rtrm}a and \ref{fig:C_nongen_rtrm}a we see that the average autocorrelation $\mean{C}(t)$ in the generic and non-generic case decays exponentially
\beq
  \mean{C}(t) = \exp(-|O(t)|)\>,
\eeq
and stretch-exponentially
\beq
  \mean{C}(t) = \exp(-|O(t^{\frac{1}{2}})|)\>,
\eeq
for large times $t$, respectively. The decay of average correlation is faster than algebraic, which is expected in stochastic systems. In all tested cases we obtain in the random triangle map the same power law decay of the standard deviation of the autocorrelation reading
\beq
  \sigma_C (t) = O(t^{-\frac{3}{2}})\>.
\eeq
This is the same power-law decay as the one found for the correlation in generic case of the deterministic triangle map. This behaviour can be clearly seen in figures \ref{fig:C_gen_rtrm}b and \ref{fig:C_nongen_rtrm}b. In Subsection \ref{sec:heur_corr} we clarify this behaviour by showing that the dispersion of correlation $\sigma_C^2 (t)$ is inversely proportional to $N(t)$, which in our case scales as $O(t^3)$.
\subsection{Random triangle map acting on one-dimensional sets}
\label{xy}
Here we would like to show that the model of random triangle map can be even  simplified and still possesses some interesting features of the (random) triangle map.\par
In the previous subsection we have analysed distributions of polygon/partition attributes within a stochastic model. The following random model indeed mimics the deterministic triangle map (\ref{eq:trm_map}) using the polygonal description, if we assume random positions ($x$ coordinates) of the cuts, relative to the positions of a given polygonal piece. In this subsection we show that such analysis can be performed analytically for a special case of a set of degenerate polygons, namely the line segments. We derive explicit results for the distribution of the length of line segments as a function of time.\par
Let us take initial set to be a vertical line of certain length. Then after time $t$, the image is composed of many short lines, each of which has slope $s=1/t$. So, the correct scaled slope variable is a constant $\chi_s = s t = 1$ which is consistent with the scaling of the general case above (\ref{eq:scalz}). Therefore we may continue investigating just one of the quantities, e.g.  $\delta_x:=h \le 1$, while the others are related to it, e.g. $\delta_y= h/t$. Let $P(h,t)\dd h$ be the probability that at time $t$ a randomly chosen piece has a horizontal length between $h-\dd h/2$ and $h+\dd h/2$. Vertical size of this piece is $h/t$, so the total probability that this piece does not get cut in one iteration is $1 - h$. Here all pieces are taken with the statistical weight $1$, so $P(h,t)$ represents a distribution of the number of such pieces. Using elementary probability it is easy to derive the following dynamic evolution integro-difference equation
\beq
  P(h,t+1) = (1-h) P(h,t) + \int_h^{1} \dd h'\, P(h',t)\>.
  \label{eq:dyn_scal}
\eeq
We are interested in a possible asymptotic scaling solutions of the above equation. Indeed, using the ansatz
\beq
  P(h,t) = t \psi(t h)\>,\qquad \int_0^\infty \dd t\,\psi (t) = 1\>,
\eeq
for large $t$, the equation (\ref{eq:dyn_scal}) is in the leading order in $1/t$ equivalent to an ordinary integro-differential equation
\beq
  \psi(\chi) + \chi\psi'(\chi) + \chi\psi(\chi)
  = \int_\chi^\infty \dd x\,\psi(x)\>,
\eeq
which has a unique solution $\psi(\chi) = \exp(-\chi)$ satisfying the appropriate boundary conditions, $|\psi(0)| < \infty, \psi(\infty)=0$. Note that $\psi(\chi)$ is the scaled number density, $\psi(\chi)=\dd N_x(\chi)/\dd\chi$, so we have the cumulative distribution,
\beq
  N_x(\chi) = 1 - \exp(-\chi)\>,
  \label{eq:res_x}
\eeq
The first remarkable result of this simplified analysis is the correct scaling variables $\chi_y = t^2 \delta_y = t^2 h,\chi_x = t \delta_x = t h$. In the spirit of the presented simplified analysis we can also express the effective area of a polygon $a$ to be proportional to $\delta_y \delta_x = \chi_x\chi_y/t^3$. Then the formula (\ref{eq:res_x}) immediately yields  the cumulative distribution of the areas reading
\beq
 N_a(a) \approx 1- \exp(O(\sqrt{a/t^3})) \>.
 \label{eq:res_a}
\eeq
The second notable results is that the distributions (\ref{eq:res_x}) and (\ref{eq:res_a}) are essentially identical, or very close for the areas distribution,
to the numerical results in the deterministic triangle map reported in Figure \ref{fig:tp_part_cum}.

\subsection{Relation for the correlations decay}\label{sec:heur_corr}

In the following we present a heuristic derivation of the correlation decay in (random) triangle map using only its basic properties. Let us take two sets $\cA,\cB\subset \bT^2$ of positive measure and consider the correlations of characteristic functions of the sets,
\beq
  C_{\cA,\cB}(t) =
  \mu({\mathcal B}\cap \phi^{(t)}({\mathcal A})) - \mu({\mathcal B})\mu({\mathcal A})\>.
  \label{eq:cfab}
\eeq
The total number of pieces to which ${\mathcal A}$ is cut up to time $t$,  with average area $\ave{a}(t)$, see equations (\ref{eq:scala}), (\ref{eq:p_trm}) and (\ref{eq:p_rtrm}), is in all studied cases estimated by
\beq
  N_{\mathcal A}(t) = \frac{\mu(\cA)}{\ave{a}(t)} \sim \const\, t^3 \>
  \qquad \textrm{for}\quad
  t \gg 1\>.
\eeq
The number of pieces of $\cA$ which map to a test set ${\mathcal B}$, and hence contributes to the correlation function (\ref{eq:cfab}), is
\beq
  N_{\cA,\cB}(t) \approx \frac{\mu(\cA)\mu(\cB)}{\ave{a}(t)}\>.
\eeq
Assuming that pieces (partitions) are uniformly and pseudo-randomly distributed over the set $\cB$, the expected standard deviation of $N_{\cA, \cB}$ is $\sqrt{N_{\cA, \cB}}$, and an estimated value of the fluctuation of the correlation function reads
\beq
  \sigma_{C_{\cA, \cB}}(t)
  = \sqrt{N_{\cA, \cB}(t)} \ave{a}(t)
  = \sqrt{\mu(\cA)\mu(\cB)\ave{a}(t)} = O(t^{-\frac{3}{2}})\>,
\eeq
for $t\gg 1$, where we implicitly use that $\ave{C_{\cA, \cB}(t)}$ is negligible in comparison to $\sigma_{C_{\cA, \cB}}(t)$. This is precisely the scaling law which is observed in numerical explorations of the random triangle map. We believe that this elucidates the law of decay of correlations in the (deterministic) triangle map.

\section{Conclusions}

We presented both theoretical and numerical analysis of a two-parameter family of nonhyperbolic dynamical system  related to the classical motion in a polygonally shaped billiard. We formulated several conjectures, and provided strong numerical evidence for their validity, concerning the ergodic behaviour of this family for different sets of parameters values. In particular, we believe that the system is strongly mixing (and thus ergodic) at least in a situation named {\em generic}, where both parameters are mutually incommensurate irrationals so that in particular no periodic orbits exist. We also found a kind of mixing behaviour in the case where only the parameter $\alpha$ controlling the strength of the cut is irrational whereas $\beta$ is equal to zero.  However, the two cases exhibit different mixing behaviours, diffusion properties and periodic orbit structure. \par
The dynamical properties are studied by means of two different
symbolic encoding schemes based on propagating an initial set of
polygons and following the two portions of phase space separated by
the discontinuity lines, referred to as the polygonal and the binary
description, respectively. Using these schemes we numerically
deduced  several asymptotic scaling properties - for some of which
we provided a semi-heuristic analytical explanation. In particular
we focused on the spectrum of the Markov transition matrix (in
either of the two symbolic descriptions) near to the unit circle and
on its spectral gap in relation to the correlation decay.  The
average convergence of the group of largest eigenvalues to the unit
circle as the size of the matrix increases is quantitatively
explained by perturbative analysis of a simple random model. We have
also briefly studied the Koopman operator associated to the map on a
finite dimensional Hilbert space obtained by truncating the Fourier
basis. There we found interesting scaling properties of
the spectrum and fidelity with respect to the dimension of the
space.\par

Finally, we have introduced a stochastic version of the triangle map
which allowed us to obtain some theoretical asymptotic scalings of
the distribution of degenerate polygons, thus yielding the
$t^{-3/2}$ decay of correlations observed numerically in the generic
case.\par

This paper presents a bulk of numerical results,
along with sparse analytical arguments, on the properties of the
triangle map. From the mathematical point of view several important
problems remain to be solved. Some of them, in particular the
dynamical mechanism underlying the observed strongly mixing
behaviour (for some parameter values) of this piecewise parabolic
system is a challenging problem in ergodic theory whose solution may require the
development of new ideas and techniques.

\section*{Acknowledgments}
MH thanks for the support by the Department of Mathematics at the University of Bologna, Italy.  MH and TP acknowledge support from research grants P1-0044 and J1-7347 of Slovenian Research Agency.

% inserting bibliography

\end{document}